\documentclass[fleqn,usenatbib,useAMS]{mnras}
\usepackage{amsmath,amstext,amssymb,amsgen,amsopn,amsfonts,theorem}
\usepackage[T1]{fontenc}
\usepackage{ae,aecompl}
\usepackage[pdftex]{graphicx}
\usepackage{xspace}
\usepackage{framed}
\usepackage{gensymb}
\usepackage[normalem]{ulem}
\usepackage{pdflscape}

\makeatletter
\def\mr@ignsp#1 {\ifx\:#1\@empty\else #1\expandafter\mr@ignsp\fi}%
\newcommand{\multiref}[1]{\begingroup
\xdef\mr@no@sparg{\expandafter\mr@ignsp#1 \: }%
\def\mr@comma{}%
\@for\mr@refs:=\mr@no@sparg\do{\mr@comma\def\mr@comma{,}\ref{\mr@refs}}%
\endgroup}
\makeatother

\newcommand{\hypref}[2]{\ifx\href\asklfhas #2\else\href{#1}{#2}\fi}
\newcommand{\Secref}[1]{Section~\multiref{#1}}
\newcommand{\secref}[1]{Sec.~\multiref{#1}}
\newcommand{\Appref}[1]{Appendix~\multiref{#1}}

\newcommand{\Figref}[1]{Figure~\multiref{#1}}
\newcommand{\figref}[1]{Fig.~\multiref{#1}}
\renewcommand{\eqref}[1]{(\multiref{#1})}

\newcommand{\eq}[1]{\begin{align}#1\end{align}}

\newcommand{\kms}{\,km\,s$^{-1}$}
\newcommand{\masyr}{\,mas\,yr$^{-1}$}

\title[Discovery of the Pal~5 stream perturbations]{A sharper view of Pal 5's tails: Discovery of stream perturbations with a novel non-parametric technique}

\author[D. Erkal, S.E. Koposov and V. Belokurov]
  {Denis Erkal$^1$\thanks{derkal@ast.cam.ac.uk}, Sergey E. Koposov$^1,^2$\thanks{skoposov@cmu.edu} and Vasily Belokurov$^1$\thanks{vasily@ast.cam.ac.uk} \\
  $^1$Institute of Astronomy, Madingley Road, Cambridge, CB3 0HA, UK \\ $^2$McWilliams Center for Cosmology, Department of Physics, Carnegie Mellon University, 5000 Forbes Avenue, Pittsburgh, PA 15213, USA}

\begin{document}

\label{firstpage}

\maketitle

\begin{abstract}

Only in the Milky Way is it possible to conduct an experiment which
uses stellar streams to detect low-mass dark matter subhaloes. In
smooth and static host potentials, tidal tails of disrupting
satellites appear highly symmetric. However, perturbations from dark subhaloes, as well as from GMCs and the Milky Way bar, can induce density fluctuations that destroy this symmetry. Motivated by the
recent release of unprecedentedly deep and wide imaging data around
the Pal~5 stellar stream, we develop a new probabilistic, adaptive and
non-parametric technique which allows us to bring the cluster's tidal
tails into clear focus.  Strikingly, we uncover a stream whose density
exhibits visible changes on a variety of angular scales. We detect
significant bumps and dips, both narrow and broad: two peaks on either
side of the progenitor, each only a fraction of a degree across, and
two gaps, $\sim2^{\circ}$ and $\sim9^{\circ}$ wide, the latter
accompanied by a gargantuan lump of debris. This largest density
feature results in a pronounced inter-tail asymmetry which cannot be
made consistent with an unperturbed stream according to a suite of
simulations we have produced. We conjecture that the sharp
peaks around Pal 5 are epicyclic overdensities, while the two dips
are consistent with impacts by subhaloes. Assuming an age of 3.4 Gyr
for Pal 5, these two gaps would correspond to the characteristic size
of gaps created by subhaloes in the mass range of $10^6-10^7 M_\odot$
and $10^7-10^8 M_\odot$ respectively. In addition to dark
substructure, we find that the bar of the Milky Way can plausibly
produce the asymmetric density seen in Pal 5 and that GMCs could cause
the smaller gap.

\end{abstract}


\section{Introduction}

Palomar 5 is so diffuse that it was once mistaken for a low surface
brightness galaxy by \citet{wilson1955} who ``rediscovered" the
globular and called it the Serpens Dwarf, the name that appears
surprisingly fitting today, after the detection of the conspicuous
S-shaped tails attached to the cluster
\citep[][]{pal5disc}. Naturally, both the low stellar density of the
satellite and the prominence of its associated stellar stream are
tell-tale signs of the ongoing disruption by the Galactic tides. Over
the years, Pal 5's tidal tails grew and are currently traced across
several tens of degrees on the sky \citep[see
  e.g][]{rockosi2002,odenkirchen2003,gd_pal5}. Thus, Pal 5 has quickly
become a poster child for Milky Way accretion.  To date, the cluster's
role as a possible powerful and precise Galactic accelerometer has
been emphasized by many \citep[see e.g.][]{ibata_et_al_pal5}, but it
remains modelled by few \citep[e.g.][]{dehnen2004,kuepper_et_al_pal5}.

\citet{dehnen2004} who presented the very first - but nonetheless
impressively comprehensive - study of the Pal 5 disruption,
established with certainty several key aspects of the satellite's
accretion: the cluster's orbit, its mass and size, and the importance
of disk shocks in driving the mass loss. However, while getting many
observables right, such as the shape of the stream track and the
overall behavior of the debris density along the tails, no model in
the \citet{dehnen2004} suite could match the level of asymmetry
between the star counts in the leading and the trailing tail of the
cluster, as displayed in e.g. their Figure 16 and Figure 4 of
\citet{odenkirchen2003}. It was then concluded that the observed
asymmetry ought to be due to the processes not captured by the
simulations. The authors point out that the most likely phenomenon -
not included in their numerical setup - which could produce such a
small-scale density enhancement in one of the tails is the interaction
of the stream with a low mass substructure. They offer three examples
of such perturbers: giant molecular clouds, spiral arms and dark
matter subhaloes.

The irrefutable detection of small-scale density perturbations in the
Pal 5 tails as presented by \citet{odenkirchen2003} and emphasized by
\citet{dehnen2004} called for an explanation. This inspired
\citet{capuzzo2005} to revisit the numerical experiments of
\citet{combes1999} who had predicted that globular cluster tails ought
to contain low-level stellar clumps. In their simulations,
\citet{capuzzo2005} not only confirmed the presence of ubiquitous
small-scale substructure in tidal tails, but also provided an
intuitive justification of their existence: the stars in the clumps
move slightly slower compared to the rest of the surrounding debris in
the tail. The deceleration of the stars in the clumps was the clue
which helped \citet{kuepper_et_al_2008} establish the genesis of the
overdensities: the orbits of the stripped stars in the reference frame
of the progenitor are oscillatory, composed of the guiding center
circular orbit and the epicyclic ellipse. The two vertices of the
ellipse are crossed at the lowest speed, thus leading to episodic
bunching of stars along the tidal tail. Although the presence of these epicyclic overdensities was initially shown for progenitors on circular orbits, further efforts showed that it is also present on orbits with a wide range of eccentricities \citep[e.g.][]{kuepper_et_al_2010}. 

Now, could the clump in the Pal 5's trailing tail simply be an example
of an epicyclic overdensity as described above? This seems unlikely as
the simulations of \citet{dehnen2004} actually produce epicyclic
``feathering'' of the cluster's tails as is obvious from e.g. Figure
6: from panel to panel, the bunching appears either enhanced or
reduced depending on the orbital phase of the progenitor
considered. Remarkably, however, the epicyclic clumping is almost
undetectable in the snapshot corresponding to Pal 5's current location
as shown in their Figure 18, thus begging for a conclusion that the
apocentre - near which Pal 5 is currently situated - is not the
optimal location for the detection of epicyclic overdensities. Most
importantly, however, the overdensities produced are always at a very
similar level in the leading and the trailing tails, especially so
close to the progenitor. This is exactly the point highlighted by
\citet{dehnen2004}: whether the epicycles are strong or not, the
density profile of the two tails should be symmetric (see their Figure
16). Note also, that such a (approximate) symmetry of the tidal tails'
density profiles might simply be a direct consequence of the symmetry
of the Hill's surface \citep[see e.g.][]{bt2008}.

Thus, given that the epicyclic clumps - a generic feature of the
globular cluster tidal streams - cannot seem to explain the sizeable
overdensity in the trailing tail of Pal 5, it is prudent to attempt to
establish the actual mechanism behind the apparent asymmetry. Our work
is motivated not only by the fact that this conundrum has remained
unsolved for more than 10 years, but also by the two recent advances
in the tidal tail studies. First, the Pal 5 stream has been mapped by
\citet{ibata_et_al_pal5} to an unprecedented depth allowing for a
robust determination of the minute details of the debris
distribution. Armed with this remarkable dataset, we use a powerful
new and robust non-parametric algorithm to determine the stream track
and the associated density variation. We confirm with very high
confidence, both the detection of a non-monotonic star count evolution
along the trailing tail and the asymmetry between the leading and
trailing portions of the stream. In addition, we also find a gap-like
feature in the leading tail and evidence of two epicyclic
overdensities near the progenitor.  At first glance, these findings appear to be in tension with \citet{ibata_et_al_pal5}
who analyzed the same dataset and found that there are no statistically significant gaps in Pal 5. However, their search was
performed on scales smaller than $1$ degree while the search for gaps in this work is focused on larger scales, guided
by predictions of the expected distribution of gap sizes from subhaloes in \citep{number_of_gaps} which brings us to the second advance in studies of tidal streams.

There is now a much better understanding of the impact of the
massive perturber fly-bys on the structure of the tidal tails
\citep[see
  e.g.][]{carlberg_2009,yoon_etal_2011,carlberg_2012,carlberg_2013,three_phases,subhalo_properties,sanders_bovy_erkal_2015,bovy_erkal_sanders_2016,number_of_gaps}
- the primary mechanism put forward by \citet{dehnen2004} to explain
the unruly star counts in the Pal 5 tails. While it had been known
that a fly-by leads to a density depletion around the projected impact
point \citep[see e.g.][]{ibata_et_al_2002,johnston_et_al_2002}, it has
now been established that the induced stream gap is always accompanied
by density hikes on either side \citep[see e.g.][]{carlberg_2012,
  three_phases}. Moreover, stream gaps go hand in hand with stream
wobbles: small-scale perturbations visible in {\it all} phase-space
projections of the debris track
\citep[see][]{three_phases,subhalo_properties,bovy_erkal_sanders_2016}. Finally,
notwithstanding the degeneracy between the age of the gap, the mass of
the perturber and its speed, there exists a distinct characteristic
gap size for subhaloes with different masses as shown in
\citet{number_of_gaps}, with smaller subhaloes tearing smaller
holes. However, there also exists a lower bound to the size of the
gap. This minimum size emerges because lighter subhaloes on average
impart a smaller velocity kick onto the stream stars. Therefore, it
takes longer for the density in the gap to drop to detectable levels
hence widening it to sizes comparable to those of the gaps induced by
more massive subhaloes. For example, according to
\citet{number_of_gaps}, it is not feasible to expect DM subhaloes with
masses of $10^7 M_{\odot}$ to produce deep gaps less than 5$^{\circ}$
wide in a Pal 5-like stream, with most detectable gaps produced by
these subhaloes opening to $\sim10^{\circ}$.

Interestingly, the flyby of dark subhaloes is not the only conceivable
mechanism that can produce small-scale perturbations in the
stream. Naturally, exactly the same generic features described above
are also expected from the gaps torn by giant molecular clouds
\citep{amorisco_et_al_2016_gmcs}. In addition, in
\cite{kohei_rotating_bar} it was shown that the rotating bar of the
Milky Way can reshape the stream drastically since different portions
of the debris approach their pericenter at different times and hence
experience a different force from the bar. The influence of the bar
was also studied in \cite{price_whelan_et_al_chaotic_fanning} in terms
of the chaos it can create. Sending some of the tidal debris on
chaotic orbits can dramatically affect the stream appearance, leading
to substantial perturbations of the stream track
\citep[e.g.][]{pearson_et_al_pal5}, as well as stream fanning
\citep{price_whelan_et_al_chaos}.

In this work, through a series of numerical experiments involving
N-body simulations of the Pal 5-like cluster disruption as well as the
approximate stream models based on modified Lagrange Cloud Stripping
\citep[mLCS,][]{gibbons_et_al_2014}, we will demonstrate that the
observed small-scale disturbances of the Pal 5 tails are consistent
with an impact by two low-mass substructures. If the stream features
are indeed caused by the passage of dark subhaloes, we argue that the
features in the leading and trailing tails are most likely caused by
subhaloes in the mass range $10^6-10^7 M_\odot$ and $10^7-10^8
M_\odot$ respectively.  Such subhaloes have long been predicted in
$\Lambda$CDM and their detection would represent a stunning
confirmation of the theory.  However, unfortunately, with the data in
hand, we cannot distinguish the higher mass flyby from the influence
of the Milky Way bar. In addition, we cannot distinguish the lower
mass flyby from a GMC flyby. The complications introduced by the bar
and GMCs also suggest that searches for gaps which focus on streams at
larger radii will yield detections that can be interpreted more
straightforwardly and hence may be more fruitful.

This Paper is organized as follows. In \Secref{sec:data} we discuss
the deep CFHT photometry published by \citet{ibata_et_al_pal5}. In
\Secref{sec:stream_model} we present a novel non-parametric model
which we use to extract the stream track and to measure the stellar
density variation and the evolution of its width. In
\Secref{sec:unperturbed_pal5} we give a review of how tidal streams
form in static and smooth potentials to highlight the discrepancy with
the observed features. Next, in \Secref{sec:asymmetric_mechanisms}, we
study the effect of an impact by substructure and that of the Milky
Way's rotating bar and show that they can both contribute to the
features seen in Pal 5. In \Secref{sec:other} we discuss several other
mechanisms such as the internal Pal 5 rotation, chaos, and
perturbations by other globular clusters, as well as how these can be
distinguished. The results are compared against expectations in
\Secref{sec:discussion}. Finally, we conclude in
\Secref{sec:conclusion}.

\section{Data}
\label{sec:data}

\begin{figure}
\includegraphics{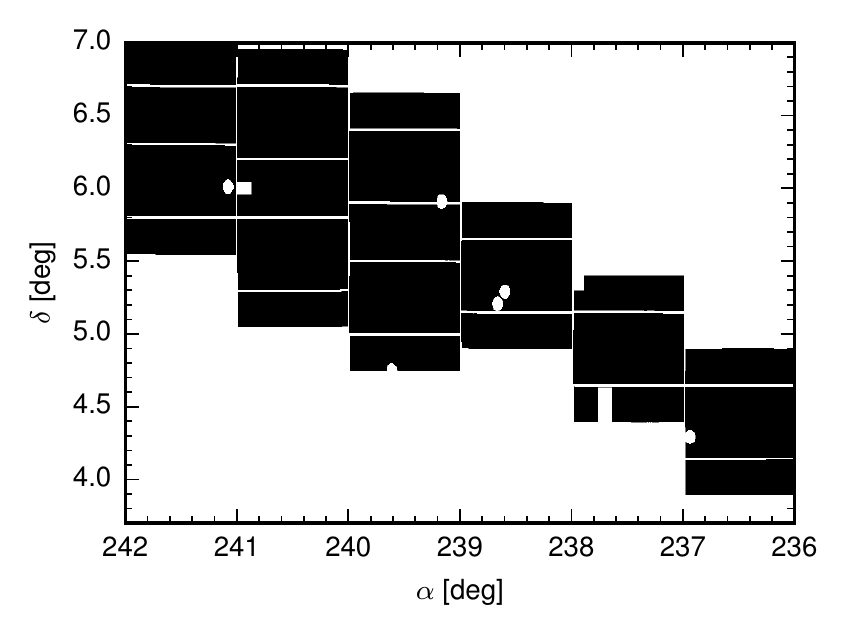}
\caption{A portion of the footprint derived for the CFHT dataset
  presented in \citet{ibata_et_al_pal5}. The footprint shows
  noticeable gaps between the CCD chips as well as holes due to masked
  bright stars.}
\label{fig:foot_plot}
\end{figure}

In this study, we use the catalog produced from CFHT observations by
\citet{ibata_et_al_pal5} which is publicly available through the
VizieR
web-site\footnote{\url{http://vizier.u-strasbg.fr/viz-bin/VizieR-3?-source=J/ApJ/819/1}}. For
details of the observations and the data reduction we refer the reader
to the original paper. However, before we can proceed with the
analysis, several additional processing steps are required. The most
important one is the determination of the survey footprint, i.e. the
area of the sky containing data of sufficiently good quality. This
step is essential as the CFHT catalogs do not yield a continuous
coverage of the stream. To establish the footprint, we consider all
individual telescope pointings and all CCDs in the mosaic used for the
catalog creation, while removing all objects located closer than 20
pixels from the edges of each CCD. The remaining objects from all CCDs
are then used to construct the combined footprint. To make sure that
the data quality is as uniform as possible throughout the footprint,
all regions of the data within 3\arcmin\ of the known bright stars
(i.e. those brighter than $V_T=8$ in Tycho-2) are
removed. Additionally, several regions where the source density is
noticeably lower compared to the typical levels are masked
out. Figure~\ref{fig:foot_plot} illustrates a portion of the footprint
constructed. Note the multiple apparent CCD chip gaps and holes caused
by bright stars. The footprint of the entire dataset was set up on a
HEALPix \citep{Gorski2005} grid at a high resolution of $N_{\rm
  side}$=65536 (pixel size of $3\arcsec\times 3\arcsec$). Finally, in
addition to excising some of the problematic portions of the data as
described above, one group of pointings at $\alpha \sim $
245\degr\ which is disconnected from the rest of the survey was also
excluded.

With the footprint in hand we can construct a detailed density map of
the stars in and around the Pal~5 stream. Although we would ideally
prefer to work with the unbinned data, in practice when dealing with a
complex footprint, rather than modelling the set of unbinned positions
of the objects on the sky, it is more convenient to describe the
Poisson number counts, $H_j$, in HEALPix pixels
($\alpha_{pix,j},\delta_{pix,j}$). If the pixels are small enough relative to the relevant length scales in the problem then
the information content is the same as in the unbinned stellar
distribution. Specifically in this study, we use stellar densities
calculated inside pixels with $N_{\rm side}$=4096 which have a size of
$\sim 50\arcsec\times 50\arcsec$, which is approximately seven times
smaller than the width of the Pal~5 stream on the sky.

\subsection{Color-magnitude mask}

\begin{figure}
\includegraphics{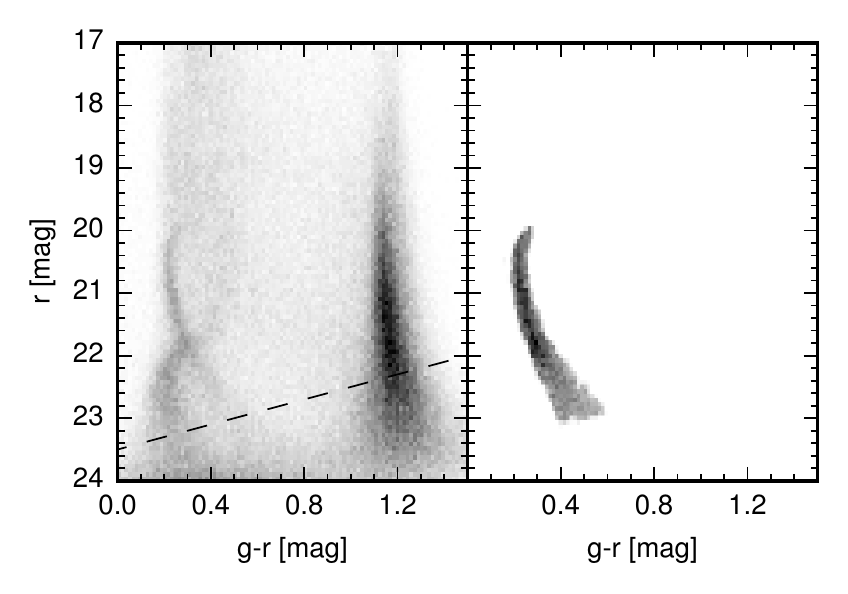}
\caption{Extinction corrected color-magnitude density distribution
  (a.k.a. Hess diagram) of stars in the CFHT Pal~5 dataset published
  by \citet{ibata_et_al_pal5}.  Left panel shows all objects in the
  catalog within 10 degrees of the Pal~5 globular cluster, while the
  right panel gives the subset of stars inside the optimal mask which
  will be used in this work. Note that the Hess diagram on the left
  shows not only the stellar main sequence of the Pal~5 (with a
  turn-off at $r\sim20.5$) but also that of the leading tail of 
  Sagittarius stream with a turn-off at r $\sim 23$. The dashed line show
  the $g=23.5$ magnitude limit adopted for this work.}
\label{fig:cmd_plot}
\end{figure}

Matched filters have been used with great success to maximise the
signal to noise when constructing density maps of low surface
brightness substructure in the stellar halo \citep[see
  e.g.][]{rockosi2002,odenkirchen2003,Grillmair2009}. The standard
matched filter approach proceeds by weighting each star by the ratio
of the target (typically a single stellar population) stellar density
in color-magnitude space, $P_{str}(g-r,r)$, to the density of the
background stars, $P_{bg}(g-r,r)$. While this weighting scheme
produces the optimal signal to noise, its disadvantage is that the
weighted densities are not Poisson distributed. Cumbersomely, the
distribution of the stellar weights is typically very asymmetric as
some stars (e.g. those along the red giant branch) have very high
weights, but their incidence on the sky is extremely low. Therefore,
to mitigate the above drawbacks of the matched filter approach we use
a simpler, more convenient method. Namely, in this paper, candidate
Pal 5 stars are selected using boolean masks based on the ratio of
$P_{str}(g-r,r)/P_{bg}(g-r,r)$, i.e. weights equal to 1 when
$P_{str}(g-r,r)/P_{bg}(g-r,r)$ is greater then some threshold, and 0
below the threshold. This approach yields signal-to-noise levels that
are only marginally lower compared to those obtained with a matched
filter, while preserving the Poisson distribution of the stellar
densities. The exact value of the threshold which maximises the
signal-to-noise for $P_{str}(g-r,r)/P_{bg}(g-r,r)$ can be easily found
from Monte-Carlo simulations.

Figure~\ref{fig:cmd_plot} shows the extinction
corrected\footnote{Please note that the magnitudes listed in the CFHT
  catalog are already corrected for the Galactic dust reddening. For
  details, see \citet{ibata_et_al_pal5}} density of stars in the
color-magnitude space (Hess diagram), with stars inside the optimal
color-magnitude mask shown in the right panel of the Figure. Note that
the mask goes down to a magnitude limit of $g\sim23.5$ and
$r\sim23$. While the main analysis of this paper was carried out using
a g$<$23.5 magnitude cut, we have verified that all conclusions of
this paper remain unchanged if a more conservative cut of g$<$23 is
chosen.  We also note that \citet{ibata_et_al_pal5} measured a small
but detectable distance gradient along the stream of 0.009 mag per
degree. Here, we choose to ignore the possible distance variation
along the stream. Our color-magnitude diagram (CMD) mask does not
include the red giant branch and thus is not very sensitive to the
small shifts along the magnitude direction. However, this should not
impact our analysis since the maximum offset in magnitude at the edges
of our dataset would be $\sim$ 0.13\,mag, which is significantly lower
that the width of our mask in r-magnitude.

\begin{figure*}
\includegraphics{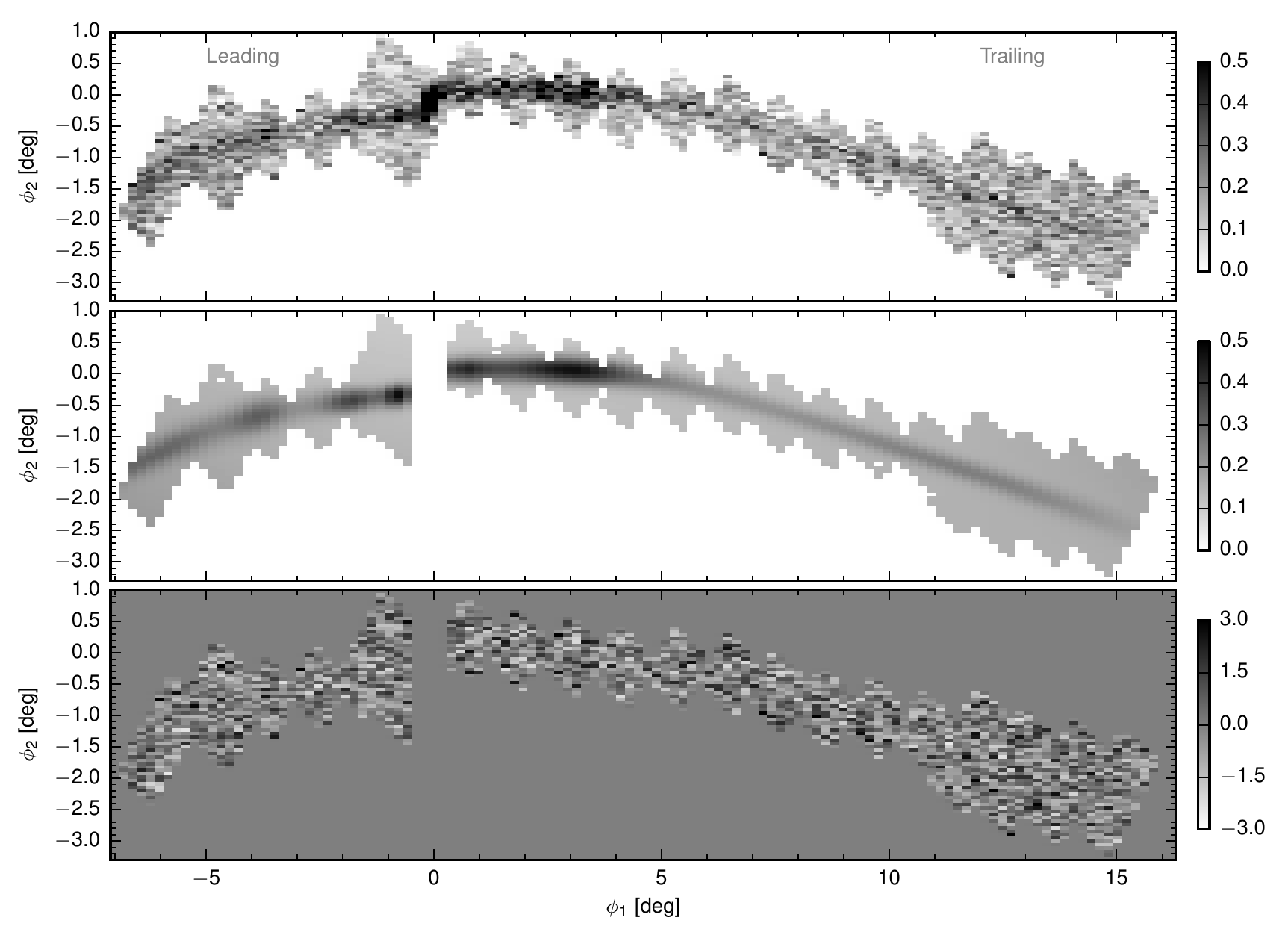}
\caption{{\it Top}: Two-dimensional (in stream-aligned coordinates,
  $\phi_1 $ and $\phi_2$) on-sky density distribution of the Pal~5
  stream candidate stars selected to lie within the optimal color-mask
  (see Figure ~\ref{fig:cmd_plot}). The density is given in
  units of the number of stars per square arc-minute.  {\it Middle}:
  Density distribution of stars predicted by our best-fit (maximum
  likelihood) model. The region near the progenitor,
  $-0\fdg4<\phi_1<0\fdg2$, has been excised as the data from this region was
  not used in the fits. {\it Bottom}: Two-dimensional map of the
  stellar density residuals with respect to our model scaled by the
  expected standard deviation in each pixel.}
\label{fig:model2d}
\end{figure*}

\subsection{Stream coordinate system} \label{sec:coordinate_system}

Throughout the paper, similarly to e.g. \cite{Koposov2010}, we use a
rotated coordinate system, $(\phi_1,\phi_2)$, which is approximately
aligned with the stream: the $\phi_1$ axis is along the stream, and
the $\phi_2$ axis is perpendicular to the stream. The pole of this
coordinate system is at $(\alpha_p, \delta_p)= (138\fdg95,
53\fdg78)$. The zero-point of the coordinate system, $(\phi_1,\phi_2)
=(0\degr,0\degr)$, lies at the crossing of the above great circle and
the $\alpha=229\degr$ great circle. The position of Pal~5 in this
coordinate system is $(\phi_1,\phi_2) \sim (-0\fdg07,-0\fdg13)$. For
convenience, the transformation matrix is also provided in
\Appref{sec:coo_appendix}.

\section{Non-parametric Stream Modeling} \label{sec:stream_model}

\begin{figure*}
\includegraphics{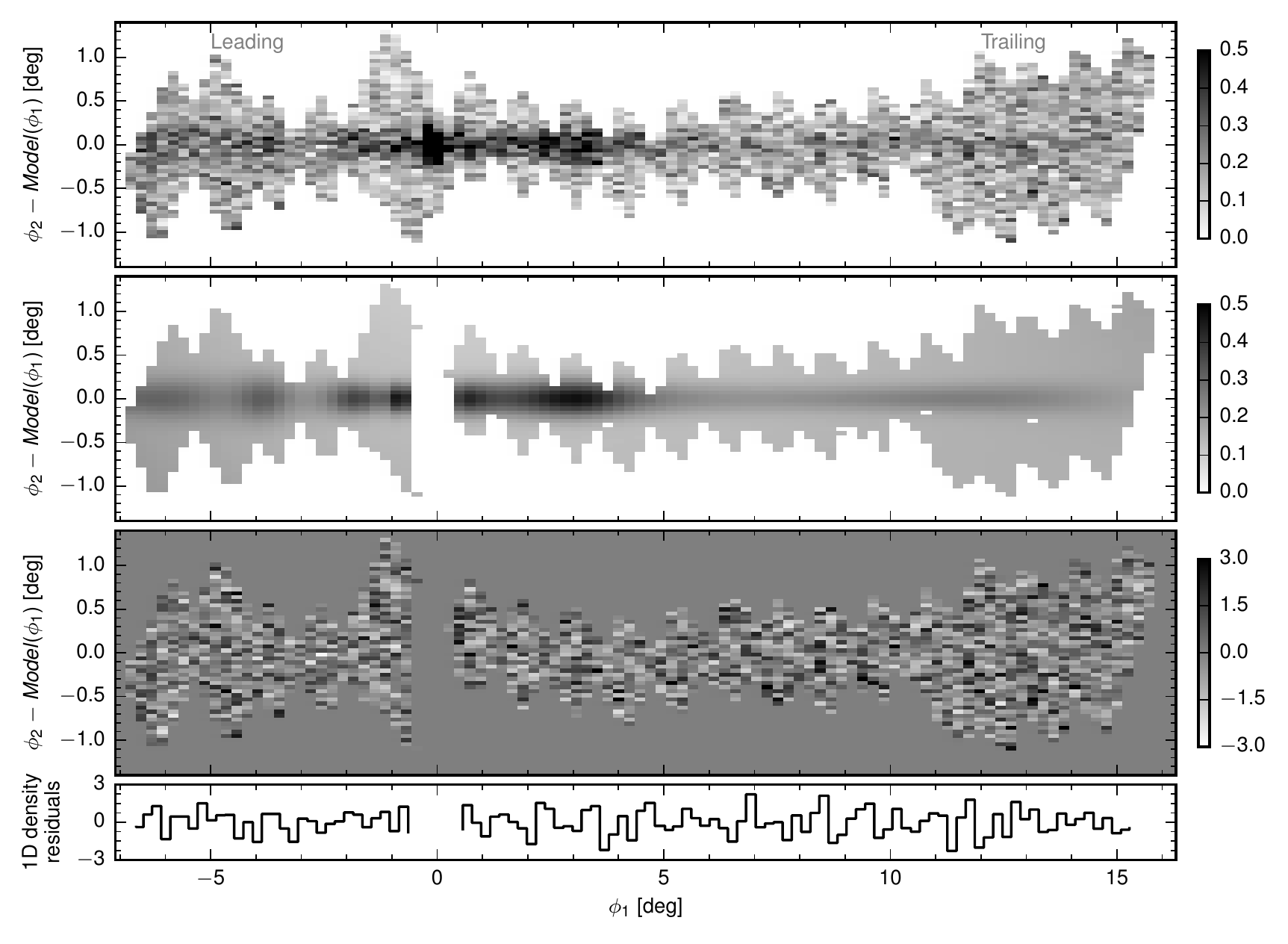}
\caption{{\it Top, second and third panels}: Observed, modelled and
  residual density distributions of the Pal~5 candidate stars, similar
  to Figure~\ref{fig:model2d}, but in a coordinate system where the
  best-fit stream track, $\hat{\phi}_2=\phi_2-{\Phi_2}(\phi_1)$, has
  been subtracted, i.e.  the stream should follow a horizontal
  straight line $\hat{\phi}_2=0$. {\it Bottom}: One-dimensional
  distribution of density residuals for the data within 0.2 degree of
  the stream track, where the residuals have been normalized by the
  standard deviation.}
\label{fig:model2dsub}
\end{figure*}

When measuring the properties of tidal tails in large area photometric
datasets, the familiar challenge is the absence of an appropriate
analytical model which would adequately describe the behavior of the
stellar density in the stream. Stellar streams are more than just
Gaussian tubes of stars following great circles on the sky. The stream
track deviates from a naive great circles due to the Galactic parallax
\citep[see e.g.][]{Eyre2010} or the precession in the aspherical
gravitational potential of the host \citep[see
  e.g.][]{ibata_et_al_2001,helmi_2004,johnston_et_al_2005,belokurov_sgr_precess,stream_width}. The
star counts along the stream can have small-scale bumps and dips
\citep[see e.g.][]{kuepper_et_al_2010, yoon_etal_2011, Ma2012}, while
the across-stream profile can become broader or narrower due to a
range of effects, including distance variation, epicyclic feathering,
interactions with small substructure and differential orbital plane
precession
\citep{Koposov2010,three_phases,amorisco_2014,stream_width,amorisco_et_al_2016_gmcs}.

Motivated by the complex picture of stellar streams which has recently
emerged from the literature, we develop here a novel, automated,
flexible, non-parametric/semi-parametric model. Our stream model is
based on cubic splines which are used to extract the stream's trajectory, the density variation along its track
and the across-stream profile evolution. The cubic splines are also used to model the background stellar density. The main innovative features
of our algorithm are as follows. First, the model complexity is set
independently for each stream dimension. Second, the flexibility of
the model is driven by the data and is determined automatically within
a probabilistic framework. Finally, because such a model by
construction would require a significant number of free parameters, we
place a particular emphasis on making sure that the model has
computable derivatives. This ensures that the likelihood gradient can
be obtained by automatic differentiation software, which allows us to
use efficient gradient-based methods for both optimization and
sampling of the posterior.

\subsection{Cubic Splines}

The key ingredient of our algorithm, the part of the model which
guarantees its flexibility, is the cubic splines. Here, we give a
brief summary of the spline properties that are relevant to our
particular case. The cubic splines are piecewise polynomial functions
described on each interval of the real line as:
\begin{align}
F(x|\{a_j,b_j,c_j,d_j,x_j\}) = \sum \limits_{0\leq j\leq n-1} {\bf I}_j(x) \,\left(  a_j+\right. \nonumber \\ 
\left.+b_j\,(x-x_j) + c_j\,(x-x_j)^2 + d_j\,(x-x_j)^3 \right) ,
\end{align}
where ${x_j}$ are the locations of the spline nodes and ${\bf I}_j(x)$
are the indicator functions of $x_j<x<x_{j+1}$. At the nodes, the
shape of the spline model changes, but with a high degree of
smoothness. The polynomial coefficients in the spline, $\{a_j, b_j,
c_j, d_j\}$, are set in such a way that the function $F(x)$ and its
first and second derivatives are continuous across the whole interval
considered. With this condition and the requirement that the second
derivatives of $F(x)$ at the edges of interval are zero, $F''(x_0)=0$
and $F''(x_{n-1})=0$ (the so called natural splines), the coefficients
of the polynomials are fully determined by the values of the spline at
the nodes, $\{y_j=F(x_j)\}$. Importantly, the polynomial coefficients
can be obtained by a simple tri-diagonal matrix inversion operation
from the values of $x_j$ and $y_j$ \citep[see
  e.g.][]{Bronshtein2007,Press2007}. Thus, with the location of the
nodes ${x_j}$ fixed, the function $F({x|y_j})$ has an easily computable
derivative with respect to the values at the nodes ${y_j}$.

\subsection{Constructing a flexible model of Pal 5's tails}
\label{sec:model_def}

In this section we demonstrate how, by using cubic splines as a
building block, it is possible to assemble a flexible Pal~5 stream
model which describes the data with high fidelity. In what follows it
is assumed that the Pal~5 stream has a Gaussian cross-section with a
variable width along the stream. The stream density is also variable
and the track is allowed to deviate from a great circle. The
Gaussianity of the stream profile is clearly an over-simplification,
as has been demonstrated in numerical experiments. Simulated tidal
tails typically show more complicated behavior with wisp-like features
known as epicyclic feathers \citep[see
  e.g.][]{combes1999,capuzzo2005,kuepper_et_al_2010,amorisco_2014}. Nonetheless,
as demonstrated below, we find that our model provides a satisfactory
description of the data in hand, devoid of any noticeable
inconsistencies.

As described in Section~\ref{sec:data}, the data consists of the set
of HEALPix pixels on the sky at locations $(\phi_{1,i},\phi_{2,i})$
and the number counts of the CMD-masked stars in those pixels, $H_i$
(the average number of stars in the pixel is $\langle H_i\rangle \sim
0.1$ ). The number counts in each pixel are Poisson distributed, with
the expected number of stars in the pixel given by the density
function $\lambda (\phi_1,\phi_2)$:
$$H_i \sim {\rm Poisson}(\lambda(\phi_1, \phi_2)) .$$
The expected number of stars per pixel, $\lambda (\phi_1,\phi_2)$, is
the sum of the contributions of the Galactic background/foreground
stellar density, $B(\phi_1,\phi_2)$, and the stellar density of the
stream stars, $S(\phi_1,\phi_2)$:
$$\lambda(\phi_1,\phi_2)  = B(\phi_1,\phi_2) + S(\phi_1, \phi_2).$$
Fortunately, in our case, the model for the background distribution
can be made extremely simple as the structure we are studying is
very elongated in the $\phi_1$ direction while being very narrow in
$\phi_2$. Furthermore, the CFHT footprint does not extend beyond
several degrees from the stream track. Therefore, the following
background model is adopted such that it changes linearly in
$\phi_2$:
$$ \log B(\phi_1, \phi_2) = B_0(\phi_1)+B_1(\phi_1) \phi_2.$$
However, the average dependence of the background on $\phi_1$,
$B_0(\phi_1)$, and the background slope in $\phi_2$, $B_1(\phi_1)$, are
modeled non-parametrically by cubic splines with nodes at
${\phi_{b0,j}}$, the node values $b_{0,j}$ and ${\phi_{b1,j}}$,
$b_{1,j}$ respectively. As noted in \cite{ibata_et_al_pal5}, the Sagittarius
stream crosses the leading tail of Pal 5 ($\phi_1 < 0^\circ$) and its presence can also be seen in the CMD shown in \Figref{fig:cmd_plot}.
The background model can account for this contamination since the Sagittarius stream is broad \citep[$\sim 10^\circ$,][]{sag}
compared to the width of Pal 5. Furthermore, since our model only fits stars within the CMD mask shown in the right panel of
\Figref{fig:cmd_plot}, this excludes the vast majority of stars from the Sagittarius stream so we do not expect a large contamination.

The model for the surface brightness of the stream is postulated to
have a Gaussian cross-section with the width and the density varying
along the stream:

$$ S(\phi_1, \phi_2) = I(\phi_1)\, \exp \left( - \frac{1}{2} 
\left( \frac{\phi_2 - \Phi_2(\phi_1)}{\Sigma(\phi_1)}\right)^2  \right).
$$
The track of the stream on the sky, $\Phi_2(\phi_1)$, is represented
by a cubic spline with nodes at $\phi_{\Phi,j}$ and values at the
nodes, $\Phi_{2,j}$. Likewise, the logarithm of the central surface
brightness of the stream, $\log I(\phi_1)$, is represented by a cubic
spline with nodes at ${\phi_{I,j}}$ and values at the nodes of $\log
I_j$, while the logarithm of the width of the stream, $\log
\Sigma(\phi_1)$, is represented by a spline with nodes at
$\phi_{\Sigma,j}$ and values at the nodes, $\log \Sigma_{j}$.  We note
that the width and surface brightness of the stream, as well as the
background density, are parametrized in log-space in order to enforce
non-negativity.

As mentioned earlier, the locations and the number of the spline nodes
is set {\it independently} for each component of the model. In other
words, the background density, the surface brightness of the stream,
its width and the track on the sky can all have a different total
number of interpolating nodes placed at different $\phi_1$ locations
(see Section~\ref{sec:nodes_def} for more details).

The model as described above has been coded in Python using the {\sc
  Theano} \citep{Theano2016} module. This module generates and
compiles the C++ code from the computation graph, thus providing a
high performance likelihood evaluation. Moreover, importantly, the
module has a symbolic differentiation functionality, and therefore
also delivers the gradient of the likelihood function with respect to
the model parameters.

\begin{table*}
\begin{tabular}{lc}
\hline
Background nodes $\phi_{b0,j}$ [deg] & $-7.1, 0., 8.2, 16.3$\\
Background slope nodes $\phi_{b1,j}$ [deg] & $-7.1, 0., 16.3$\\
Stream track nodes $\phi_{\Phi,j}$ [deg] &$ -7.1, -3.75, -0.4, -0.1, 0.05, 0.2, 4.23, 8.25, 16.3$\\
Stream density nodes $\phi_{I,j}$ [deg] & $-7.1, -5.425, -4.59, -3.75, -2.91, -1.66, -1.03, -0.82, -0.4$
\\& $-0.1, 0.2, 0.7, 0.96, 3.23, 4.23, 14.96, 16.3$\\
Stream width nodes $\phi_{\Sigma,j}$ [deg] &$ -7.1, -3.75, -0.4, 0.2, 8.25, 16.3$\\
\hline
\end{tabular}
\caption{The list of spline interpolation nodes used in  the Pal~5 model.}
\label{tab:nodes}
\end{table*}

\subsection{Automated spline node selection} \label{sec:nodes_def}

The main reason for choosing the spline parametrisation for the stream
modelling is its flexibility compared to e.g. a simple
polynomial. Importantly, the spline framework provides a
straightforward functionality to increase the model complexity through
the addition of extra nodes. Accordingly, there needs to be an
algorithm that dictates how to choose the number and the location of
the interpolation nodes. For example, one possible solution is to
place the nodes in $\phi_1$ as densely as possible -- that would
produce the most flexible, albeit over-fitting, model. Instead, we
seek a method which can deliver a model whose complexity is data
driven, i.e. we would like to select the simplest model which
describes the data well, in line with the {\it Occam's razor}
reasoning.

One way to implement such a principle is to start from an initial node
placement, $\{x_i\}$, evaluate the likelihood of the data given the
node positions, $\mathcal{P}(\mathcal{D}|\{x_i\})$, add a trial new
node $x_*$, and then evaluate the likelihood of the data with the
candidate node added $\mathcal{P}(\mathcal{D}|\{x_i\},x_*)$. Ideally,
the decision whether to include the new node $x_*$ would be based on
the set of criteria which can test the predictive performance of the
model, e.g. the cross-validation \citep{Gelman2014}. However, for
computational reasons, in this work we choose to use the Akaike
Information Criteria (AIC) \citep{Akaike1974}. Specifically, we start
the procedure with 4 interpolation nodes defined for each of the
stream track, width and stream density models (these interpolation
nodes are located at the edges of our data, $\phi_1= -7\fdg1$ and
$\phi_1 =16\fdg3$, and on either side of Pal~5, $\phi_1 = 0\fdg2$ and
$\phi_1=-0\fdg4$), and 3 interpolation nodes at $\phi_1=-7\fdg1,
0\degr, 16\fdg3$ for the background model. Then, new interpolation
nodes are trialled one by one at the locations between the existing
nodes. Namely, the candidate nodes are placed at 25\%, 50\%, and 75\%
of the distance between the previously chosen positions. At the next
step, the maximum likelihood is calculated together with the AIC of
the model. The new candidate node giving the highest likelihood is
chosen if the AIC of the corresponding model is better than the AIC
without the new node. If the AIC of the best candidate new node is
higher compared to the previous iteration, the node search procedure
is stopped.

When testing this procedure on artificial datasets, we have found it
to perform well, with the exception of when the data exhibit
deviations of borderline statistical significance (i.e. $4-5\sigma$)
with a characteristic size much smaller than the distance between
existing nodes. In this case, the procedure described above could miss
such a feature.  From one point of view, this is a desired quality of
the algorithm as it provides a natural regularisation for the model,
suppressing small-scale overfitting. However, because we know that
stream densities in particular can have low-amplitude small angular
scale fluctuations (i.e. due to epicyclic effects, uneven mass loss or
interaction with perturbers), it may be appropriate for the algorithm
to explore small scales more efficiently. Therefore, for the stream
density part of the model only (i.e. $I(\phi_1)$), we extend the node
placement algorithm with 3 additional steps. First, we split all node
intervals that are longer than 2 degrees, i.e. we insert $\lfloor
\frac{x_{j+1}-x{j}}{2} \rfloor$ equidistant nodes inside every
interval $(x_{j},x_{j+1})$ that is longer than 2 degrees, irrespective
of the AIC change. This ensures that the small-scale density
deviations are explored appropriately. Then, the node placement
algorithm is run again to insert new nodes based on the AIC
changes. As a final step, we iteratively remove the (excessive) nodes
that do not decrease AIC. We have found that with this modification,
the algorithm is noticeably more sensitive to small-scale structures
down to $\sim$ 0.1 degrees. Appendix~\ref{sec:sim_compare_appendix}
provides an illustration of the algorithm's performance on simulated
stellar streams. According to these tests, the method correctly
recovers the debris density evolution including small-scale behavior
such as epicyclic overdensities.

For the CFHT Pal~5 dataset, starting with 18 initial interpolation
nodes, the procedure described above leads to a model with 39 nodes in
total, as specified in Table~\ref{tab:nodes}. Note that for some of
the parameters such as the stream track and the density along the
stream, the number of nodes is significantly higher than for the width
of the stream or for the Galactic background, reflecting the higher
information content of the data for these stream dimensions. We also
remark that even if the total number of nodes assigned by the
algorithm is higher than required to describe the data (i.e. at the
onset of over-fitting), but the full covariance information from the
posterior samples between the values at the nodes is preserved, we
expect any further inference based on our measurements, such as
e.g. constraints on the gravitational potential from the stream track
fitting to be unbiased.

\subsection{Fitting the data}
\label{sec:posterior}

Here we describe the choice of the model parameter priors adopted as
well as the details of the posterior sampling. A uniform prior is
chosen for the stream track, $\phi_{\Phi,j} \sim U(-3,3)$, while the
background parameters, $b_{0,j}$, $b_{1,j}$, and the logarithm of the
stream density, $\log I_j$, have improper uniform priors. The prior on
the stream width is log-uniform, such that it would lie within the
range of 0.01 and 0.5 degrees:
$$\log \Sigma_j \sim U(\log (0.01), \log(0.5)),
$$
where in practice, instead of the uniform probability distribution, we
use its smooth approximation with two logistic functions.

With the priors defined and the likelihood function specified in
Section~\ref{sec:model_def}, we proceed to fitting the model and
sampling the posterior. The posterior is first optimized to find the
maximum a posteriori parameters using a Quasi-Newton L-BFGS-B
algorithm \citep{Zhu1997}. Because of high dimensionality of the
problem, multiple fits are always run from a number of different,
over-dispersed sets of starting points to ensure that the solution is
not trapped in a local minimum.  Finally, in order to properly explore
the covariance between different parameters, the posterior is sampled
using the Markov Chain Monte-Carlo (MCMC). Due to the high number of
the model parameters (39) as well as the substantial time required for
a single likelihood evaluation, it was crucial to use the Hamiltonian
Monte-Carlo sampling rather than e.g. Metropolis-Hastings sampler. The
former boasts much better scaling with the number of problem
dimensions $d$ \citep[$d^{\frac{5}{4}}$ vs $d^2$; see][]{Neal2012} by
avoiding random-walk behaviour.

The Hamiltonian Monte-Carlo (HMC) sampler was implemented by the
authors in Python, following the recommendations given in
\citet{Neal2012}. In our realization, each HMC fit starts from the
tuning step, where we determine the relative leapfrog step sizes for
different parameters, $\epsilon_i$, in order to achieve acceptance
rates in the range of $\sim$ 50\%-90\%. The total number of leapfrog
steps in one HMC trajectory was always fixed to 100. This is informed
by the test runs that showed that this number of steps produces chains
with short auto-correlation times ($\lesssim$ 1-3). In the final run,
each chain is advanced for 7500 iterations with first 2500 iterations
thrown away for the burn-in. To assess the convergence we ran many
chains in parallel, ensuring that the \citet{Gelman1992} $\hat{R}$
statistic is below 1.05 for all of the model parameters.

The estimates of the model parameters from the Markov Chain
Monte-Carlo runs, namely the medians and the 16\% and 84\% percentiles
from 1D posteriors are given in Tables~\ref{tab:meas_track},
\ref{tab:meas_width}, \ref{tab:meas_dens}. We have also made the MCMC
chains for all parameters publicly available\footnote{\href{https://zenodo.org/record/151912}{https://zenodo.org/record/151912}
} as these are
required to determine other useful statistics, such as covariance
matrices between parameters.

\begin{table*}
\begin{tabular}{llllllllll}
\hline
 $\phi_1$ [deg] & $-7.10$  & $-3.75$  & $-0.40$  & $-0.10$  & $0.05$  & $0.20$  & $4.23$  & $8.25$   & $16.30$ \\
 $\Phi_j$ [deg] & $-0.154$ & $-0.115$ & $-0.231$ & $-0.115$ & $0.063$ & $0.098$ & $0.085$ & $-0.075$ & $0.062$ \\
 $\Phi_j$(16\%) & $-0.203$ & $-0.135$ & $-0.254$ & $-0.138$ & $0.043$ & $0.079$ & $0.074$ & $-0.090$ & $0.006$ \\
 $\Phi_j$(84\%) & $-0.106$ & $-0.096$ & $-0.208$ & $-0.091$ & $0.083$ & $0.117$ & $0.095$ & $-0.060$ & $0.118$ \\
\hline
\end{tabular}
\caption{Stream track measurements at the interpolation nodes (medians) together with 16\%, 84\% percentiles.}
\label{tab:meas_track}
\end{table*}
\begin{table*}
\begin{tabular}{lllllll}
\hline
 $\phi_1$ [deg]        & $-7.10$  & $-3.75$  & $-0.40$  & $0.20$   & $8.25$   & $16.30$  \\
 $\log [\Sigma_j/1\, deg]$       & $-2.073$ & $-1.764$ & $-2.172$ & $-2.103$ & $-2.230$ & $-1.836$ \\
 $\log \Sigma_j$(16\%) & $-2.377$ & $-1.878$ & $-2.248$ & $-2.163$ & $-2.366$ & $-2.355$ \\
 $\log \Sigma_j$(84\%) & $-1.756$ & $-1.651$ & $-2.096$ & $-2.041$ & $-2.094$ & $-1.358$ \\
\hline
\end{tabular}
\caption{Stream width measurements at the interpolation nodes (medians) together with 16\%, 84\% percentiles.}
\label{tab:meas_width}
\end{table*}
\begin{table*}
\scriptsize
\setlength{\tabcolsep}{3pt}
\begin{tabular}{llllllllllllllllll}
\hline
 $\phi_1$ [deg]   & $-7.10$ & $-5.42$ & $-4.59$ & $-3.75$ & $-2.91$ & $-1.66$ & $-1.03$ & $-0.82$ & $-0.40$ & $-0.10$ & $0.20$  & $0.70$  & $0.96$  & $3.23$  & $4.23$  & $14.96$ & $16.30$ \\
 $\log I_j$       & $-2.74$ & $-2.46$ & $-2.67$ & $-2.07$ & $-2.92$ & $-1.76$ & $-1.83$ & $-1.25$ & $-1.70$ & $0.38$  & $-1.12$ & $-1.48$ & $-1.57$ & $-1.35$ & $-1.91$ & $-3.73$ & $-4.92$ \\
 $\log I_j$(16\%) & $-3.36$ & $-2.63$ & $-2.86$ & $-2.22$ & $-3.13$ & $-1.88$ & $-1.98$ & $-1.41$ & $-2.03$ & $-0.75$ & $-1.41$ & $-1.59$ & $-1.67$ & $-1.42$ & $-1.98$ & $-4.10$ & $-5.59$ \\
 $\log I_j$(84\%) & $-2.19$ & $-2.31$ & $-2.49$ & $-1.93$ & $-2.73$ & $-1.64$ & $-1.69$ & $-1.11$ & $-1.41$ & $1.47$  & $-0.86$ & $-1.37$ & $-1.48$ & $-1.29$ & $-1.85$ & $-3.41$ & $-4.35$ \\
\hline
\end{tabular}
\caption{Stream density measurements at the interpolation nodes (medians) together with 16\%, 84\% percentiles.}
\label{tab:meas_dens}
\end{table*}
\begin{figure*}
\includegraphics{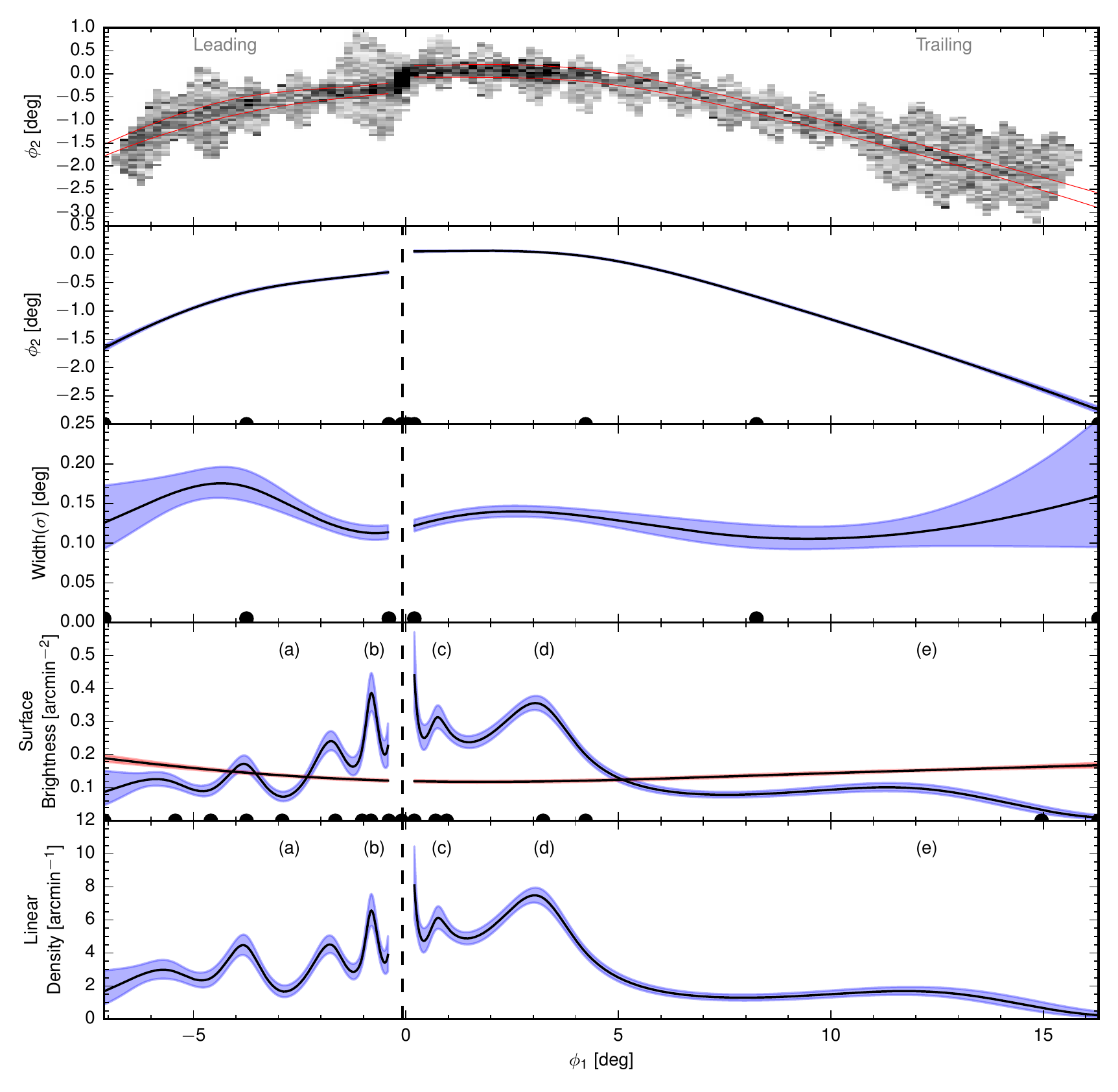}
\caption{Summary of the measured Pal~5 stream properties, each shown
  as a function of the along-stream longitude $\phi_1$. The dashed line indicates the location of the Pal~5 cluster. 
  {\it Top}:
  Two-dimensional density map of the stream stars with the red lines
  delineating the stream track $\pm$ stream width. {\it 2nd panel}:
  Distribution of the posterior samples of the stream track. In this
  and the rest of the panels below, the black line shows the median of the
  parameter in consideration, while the blue band gives the 16\% and
  84\% credible intervals. Filled black circles at the bottom of each
  panel mark the locations of the interpolation nodes chosen. {\it 3rd
    panel}: Stream width evolution. Note that in the trailing tail,
  the width of the debris distribution remains approximately
  constant. This is in stark contrast with the leading tail, which
  broadens significantly at $\phi_1\sim -3\degr$. {\it 4th panel}:
  Stream central surface brightness. For comparison, the surface
  brightness of foreground/background stars is shown in red. Note
  several prominent features in the trailing arm: i) the rise of the
  surface brightness near the progenitor, ii) a striking overdensity
  at $\phi_1 \sim 3\degr$ (labeled by $(d)$), iii) the density 
  depletion near $\phi_1
  \sim 8\degr$, and iv) another lower amplitude overdensity at $\phi_1
  \sim 12\degr$ (labeled by $(e)$) followed by apparent complete or 
  almost complete
  density drop off at $\phi_1\gtrsim 15\degr$. Similarly, in the
  leading tail, there are several notable features as well: i) a rise
  of the star counts towards the progenitor, ii) a narrow peak at
  $\phi_1\sim-0\fdg8$ (labeled by $(b)$) and iii) a pronounced dip at
  $\phi_1\sim-3\degr$ (labeled by $(a)$) surrounded by low-significance bumps on
  either side. {\it 5th (or bottom) panel}: Linear density along the
  stream. Since the stream width is quite constant with the exception
  of the edge of the leading stream, the linear density mimics closely
  the surface brightness profile along the stream.}
\label{fig:stream_meas}
\end{figure*}

\subsection{Results: a sharper view of the Pal~5 tails} \label{sec:results}

This section presents the results of modelling the Pal~5 stream
data. First, we focus on the two-dimensional stellar density maps
showing the area in and around the stream, comparing the observed
features and the behavior of our model. We also discuss the
measurements of the individual stream parameters, such as the debris
track on the sky, its width and its density. The overall quality of
the stream reconstruction can be gleaned from
Figure~\ref{fig:model2d}. The top panel of the Figure shows the
two-dimensional stellar density distribution in the stream-aligned
coordinate system, $(\phi_1, \phi_2)$. The middle panel gives the
reconstructed 2D maximum likelihood model of the stream and the
Galactic background. In the bottom panel of the Figure, the map of the
best-fit model residuals is presented. As the Figure clearly
demonstrates, the model does indeed correctly reproduce the vast
majority of the observed features of the Pal~5 stream leaving no
significant residuals.

A more nuanced view of the Pal~5 stream and the model performance can
be obtained by glancing at Figure~\ref{fig:model2dsub}. Here, the
observed density distribution and its model representation are given
in coordinates corrected for the curvature of the average stream
track. The transformed coordinates are $(\phi_1, \hat{\phi}_2)$, where
$\hat{\phi}_2 = \phi_2 - \Phi_2(\phi_1)$, so that the stream would be
located at the $\hat{\phi}_2=0\degr$ line in the case of a perfect
fit. According to the Figure, the stream appears to follow the
$\hat{\phi}_2=0\degr$ line without deviations. Additionally, the plot
highlights some of the important stream features discussed below. In
particular, a dramatic density variation along the trailing tail is
visible, as well as a significant width evolution along the
stream. More precisely, the debris distribution appears much broader
in the shorter leading tail compared to the long and narrow trailing
tail. Finally, the bottom panel of the Figure gives the 1D histogram
of the model residuals within 0.2 degrees of the stream
center. Impressively, no significant small-scale over- or
under-densities are discernible, confirming that the density behavior
of the stream is correctly reconstructed by the model.

The summary of the behavior of the stream model parameters as a
function of the longitude $\phi_1$ is presented in
Figure~\ref{fig:stream_meas}. Extracted from the posterior samples and
shown here as medians, the 16\% and the 84\% percentiles are: the run
of the stream track on the sky ($\Phi_2(\phi_1)$; second from the
top), the stream width ($\Sigma(\phi_1)$; middle panel), the stream
central surface brightness ($I(\phi_1)$; second from the bottom
panel), as well as the linear stellar density evolution
($\sqrt{2\pi}I(\phi_1)\Sigma(\phi_1)$; bottom panel). As evidenced in
the Figure, the above properties of the Pal~5 tails are measured with
unprecedented precision, thus enabling us to make the following key
inferences.

\begin{enumerate}

\item The stream track displayed in the top two panels of
  Figure~\ref{fig:stream_meas} exhibits a significant large-scale
  deviation from the great circle as measured with unprecedentedly
  high precision: the median $\phi_2$ error is spectacularly low at
  $0.014$ degrees.  The stream itself seems to flow extremely smoothly
  without major small-scale irregularities. The only two broad
  features visible in the track is the expected misalignment between
  the leading and the trailing tails and a significant change in the
  stream curvature at $\phi_1\sim 4\degr$. We also note that,
  expectedly, the stream track uncertainties are truly minuscule in
  the central, most populated, parts of the stream and start to expand
  closer to either end at high values of $|\phi_1|$, where the tails
  are traced with fewer stars.

\item The recovered stream width profile demonstrates the power of the
  adaptive non-parametric approach employed. While the longer trailing
  tail seems to have an almost constant width of 0\fdg12, the shorter leading tail possesses
  a very quick and significant width increase, from 0\fdg12 to $\sim$
  0\fdg18 on moving away from the progenitor. Given the CFHT coverage
  available, it is not clear whether the leading tail width stays high
  further out at $\phi_1<-5\degr$ or returns back to the original
  narrow values of $\sim$ 0\fdg12. We also note that the width of the
  stream at the very edge of the trailing tail at $\phi_1 \sim
  16\degr$ appears to grow up to somewhat higher values, i.e. nearly
  0\fdg15. However, the stream width error-bars also increase
  significantly due to the low stream density far away from the
  progenitor (see the bottom panel of the Figure). Therefore, we
  believe that there is no strong evidence for the width increase in
  the trailing tail.

\item The most interesting and surprising measurement is perhaps that
  of the Pal~5 stream surface brightness (shown on the second from the
  bottom panel of Fig.~\ref{fig:stream_meas}). First, we would like to
  draw attention to the stream's average surface brightness
  levels. Even though the CFHT data reaches an impressive depth of
  $r\sim 23.5$, and in spite of the fact that the Pal~5's debris is
  the most prominent globular cluster stream in the Milky Way halo,
  typically, there is still only 1 stream star per 10 square
  arc-minutes! This means that despite the application of a matched
  filter, the Galactic foreground/background density is at a similar
  or even higher level (as illustrated by the red band in the second
  panel from the bottom) across the lion's share of the surveyed
  area. This truly emphasizes the extreme challenges associated with
  the analysis of the low surface brightness halo sub-structures.

\item Notwithstanding the rather humble star counts in the stream, the
  evolution of the debris density is captured with high precision and
  is shown to exhibit prominent variations. Let us start by looking at
  the trailing tail ($\phi_1>0\degr$), where the most noticeable
  feature is located, namely the very clear surface brightness peak at
  $\phi_1 \sim 3\degr$ (labeled $(d)$). The surface brightness in this
  debris pile-up is more than three times that of the typical level in
  the rest of the tail.  Most importantly, this striking trailing tail
  feature does not seem to have a detectable counterpart on the
  leading side of the stream. Curiously, beyond the bump, i.e. at
  higher $\phi_1$, the stream density decreases but then briefly
  recovers at $\phi_1 \sim 12 \degr$ ($(e)$ label in the Figure),
  where another, albeit lower amplitude bump can be seen.  In fact,
  the bumpy behaviour of the trailing tail can already be glimpsed
  from the two-dimensional density maps shown in
  Figure~\ref{fig:model2d} and particularly
  Figure~\ref{fig:model2dsub}.  This strong density variation present
  in the trailing tail and missing from the leading tail will be the
  main focus of the rest of the paper.  Farther, past the second
  shallow density peak ($\phi_1 \gtrsim 12\degr$), the star counts in
  the stream appear to drop. It is unclear whether the stream stops,
  i.e. the density decays to zero, or simply reaches the levels below
  our detection limit.
  
\item Pal~5's leading tail ($\phi_1<0\degr$) also boasts a number of
  interesting features in its density profile. In order to quantify
  the significance of these density fluctuations, we introduce the
  following statistic. The density excursion parameter, $S$, is simply
  the ratio of the density level at the location $\phi_1$ to the
  density linearly interpolated between two fiducial background points
  $(\phi_{1,l},\phi_{1,r})$:
  $$ S =
  \frac{(\phi_{1,r}-\phi_{1,l})D(\phi_1)}{(\phi_{1,r}-\phi_{1})\,D(\phi_{1,l})+(\phi_1-\phi_{1,l})D(\phi_{1,r})} $$
  Therefore, $S$ approximates the amplitude of the density drop or
  enhancement at given $\phi_1$. If the density changes linearly the
  $S$ statistic will be equal to 1, but if there is a x\% density
  drop, its value will change to $1-x/100$.  Samples from the
  posterior (see Section~\ref{sec:posterior}) are used to evaluate
  the uncertainty in the $S$ statistic and thus assign significance to
  a possible depletion or overdensity in the stream star counts by
  looking at the tail probabilities $\mathcal{P}(S<1|\mathcal{D})$ or $\mathcal{P}(S>1|\mathcal{D})$ 
  respectively. Note that for depletions, the fiducial points are chosen to be the peaks
  closest to the gap, while for overdensities, the fiducial points are chosen
  to be the nearest troughs\footnote{We note that the tail probabilities
  based on the S-statistic measure the local significance, so the Bonferroni
  correction \citep{Gross2010} with the number of tests equal to $\sim$ 15 (the number of 
  parameters of the density model) may need to be applied to 
   evaluate the global significance.}.  Equipped with the $S$-statistic, we examine the
  structures visible in the leading tail.  Closest to the
  progenitor is the sharp peak at $\phi_1\sim -0\fdg8$ (labeled as
  $(b)$ in the Figure). This density spike is quite narrow (width of
  $\sim 0\fdg2$) and has a respectable significance of
  $\sim3.3\,\sigma$ using fiducial points at $-1\fdg2$ and $-0\fdg5$.  The likely explanation for this compact
  overdensity located right next to Pal~5 itself is the so-called
  epicyclic ``bunching", a phenomenon commonly observed in simulations
  of globular cluster disruption\citep[see
    e.g.][]{combes1999,capuzzo2005,kuepper_et_al_2010,amorisco_2014}.
  Indeed, in N-body simulations presented in
  \Secref{sec:unperturbed_pal5}, an epicyclic overdensity can be seen
  at almost exactly the same location in the leading arm, i.e. at
  $\phi_1 = -0\fdg7$. Interestingly, on the other side of the
  progenitor, the trailing tail also seems to have a small peak at
  $\phi_1\sim 0\fdg9$ (labeled $(c)$). We can speculate that this feature
  is also related to the epicyclic bunching, however, in the current
  dataset it is not very statistically significant ($< 2\,\sigma$).
  
\item Additionally, the leading tail shows a considerable density
  decrease at $\phi_1=-3\degr$ (labeled $(a)$ in the Figure). The
  significance of this drop is higher than $5\,\sigma$ using fiducial points at $-4\fdg$ and $-1\fdg8$ (the true
  significance cannot be computed as none of our posterior samples has
  $S>1$). The $S$ statistic value is $0.3 \pm 0.1$ indicating that the
  density decrease in this feature is about 70\%. In fact, this strong
  debris depletion can also be spotted in the two-dimensional density
  maps of Figure~\ref{fig:model2dsub}. We will discuss the possible
  cause of this feature in latter Sections. As a note of caution on
  this and other features in the leading tail, we emphasize that while
  the detection of the density variations at $-4\degr<\phi_1<-0\fdg5$
  is unambiguous, the classification of the observed features as
  over-densities and under-densities is somewhat model-dependent and
  as such is open to interpretation.
  
\item Finally, the stream density in the leading tail has a small
  scale density drop visible at $\phi_1\sim -5\degr$. However, since
  its significance is only $\sim 2.5\,\sigma$ using fiducial points at $-5\fdg7$ and $-3\fdg9$, its nature remains
  uncertain. Furthermore, by examining the extinction maps in this
  area, we noticed that a filament of increased reddening ($E(B-V)\sim
  0.16$) crossing the stream roughly at the location of this
  feature. In other words, it is possible that this particular
  density feature is spurious and is related to the effects of
  obscuration by the inter-stellar dust. 
  To demonstrate this point quantitatively we show the average reddening within one stream width of Pal 5 in \Figref{fig:ebv} which shows there is only significant reddening near $\phi_1\sim -5^\circ$.

\end{enumerate}

The amount of structure visible in the Pal~5 debris distribution is
remarkable. However, while hints of some of the features had been seen
previously, others are revealed here for the first time. Therefore, to
verify the robustness of the stream measuring machinery, several
consistency checks have been performed. The results of two of these checks are
presented in the Appendix.

The first check (see Appendix~\ref{sec:decals_appendix}) compares the
model constrained on the CFHT data to the stream density profile
obtained with the data from the DECam Legacy Survey (DECaLS)
\citep{Blum2016}. Although DECaLS data clearly suffers from similar
Poisson sampling errors, the instrumental effects are expected to be
different. Re-assuringly, as shown in Fig.~\ref{fig:decals}, the
independent DECaLS dataset appears to exhibit very similar density
fluctuations to the ones extracted from CFHT data. The second
consistency check is reported in
Appendix~\ref{sec:sim_compare_appendix}. Here, the same adaptive
non-parametric density measurement algorithm is used on simulated
stream data. The streams with and without the expected $\Lambda$CDM
substructure are analyzed. Importantly, in all cases, the
reconstructed density profile matches well the underlying density,
correctly identifying the presence (or lack) of the density features
over a large range of angular scales. Finally, we have also checked 
that the features are not due to a varying detection efficiency of 
stars across the field of view \citep[as seen in Fig. 6 of][]{ibata_et_al_pal5}.
More precisely, we mask out the bottom half of the bottom right CCD 
and the top left corner of the top left CCD and repeat the analysis. 
Reassuringly, this gives an almost identical stream track, width and 
density, as well as depth of each stream over/under-density (i.e. the S-statistic of 
the most significant density drop at $\phi_1=-3^\circ$ is $S=0.35\pm 0.1$,
indistinguishable from the measurement without masking of $S=0.3\pm 0.1$).

Now let us briefly compare our results with the analyses of
the Pal~5 stream reported previously. For example, in one of the very
first studies of the Pal~5 tails, \citet{odenkirchen2003} show the
stream star count behavior in their Figure~4. There, one can already
get a glimpse of a strong density peak at $\phi_1\sim 3\degr$ and the
resulting leading/trailing tail asymmetry. The later work by
\citet{carlberg_pal5_2012} presents the results of a search for gaps
in the trailing tail of Pal~5 using the SDSS data. One of the most
significant gaps they found was at the distance of $8.45\degr$ from
the progenitor, with a size of $7.7^\circ$ which agrees quite well
with the gap-like feature visible in the bottom panel of
\Figref{fig:stream_meas}. This is an under-density which continues
from $\phi_1 \sim 3\degr$ to $\phi_1 \sim 12\degr$ (between peaks
labeled as $(d)$ and $(e)$ on the Figure). In their analysis of the
SDSS data, \cite{kuepper_et_al_pal5} found many possible
overdensities, some of which could be matched to the stream features
reported here. For example, their most prominent overdensity (labeled
$T4$) can be matched to the large density peak that we observe at
$\phi_1\sim 3\degr$. Furthermore, their overdensity near the
progenitor labeled $L1$ could be identified with the epicyclic
overdensity that we see at $\phi_1\sim-0.8$. 

Lastly, we can compare our results with those of \citet{ibata_et_al_pal5}. In the Pal~5 panorama
presented in their Fig. 7, one can observe a density peak
at $\phi_1\sim3\degr$ (located at $\xi \sim 2.5\degr$ in their
coordinate system), as well as a hint of the gap at $\phi_1\sim-3$ (at
$\xi \sim -2\degr$), best seen in the bottom two left panels of their Fig.~8. One of the main differences between this work and \citet{ibata_et_al_pal5}
is that they limited their search to features below $1^\circ$ in size while we searched for features on a range of scales motivated
by the gap size predictions of \cite{number_of_gaps} and by the small-scale features expected from epicyclic overdensities. Indeed, in the left panel of their Fig. 8, where they show the results of their match filtering technique on different scales, the 
gap at $\xi \sim -2^\circ$ is growing in significance as the search is performed on successively larger scales up to $1^\circ$. If their search had continued to larger scales, $\sim 2^\circ$, it is likely that they would have found a statistically significant gap. We note that \cite{ibata_et_al_pal5} caution that background subtraction can introduce gaps and overdensities in the stream due to inhomogeneities in the survey. As a check that this is not the case for our claimed features, we show the background subtracted density from DECaLS in Appendix~\ref{sec:decals_appendix} which shows precisely the same features as seen with the CFHT data.

\section{Streams in a smooth and static potential} \label{sec:unperturbed_pal5}

In order to understand the significance of the features observed in
the tidal tails of Pal~5, let us first review the mechanism of the
stellar stream formation in the simplest case, namely in a smooth and
static potential.

\subsection{Mechanics of tidal disruption} \label{sec:mechanics}

\begin{figure}
\centering
\includegraphics{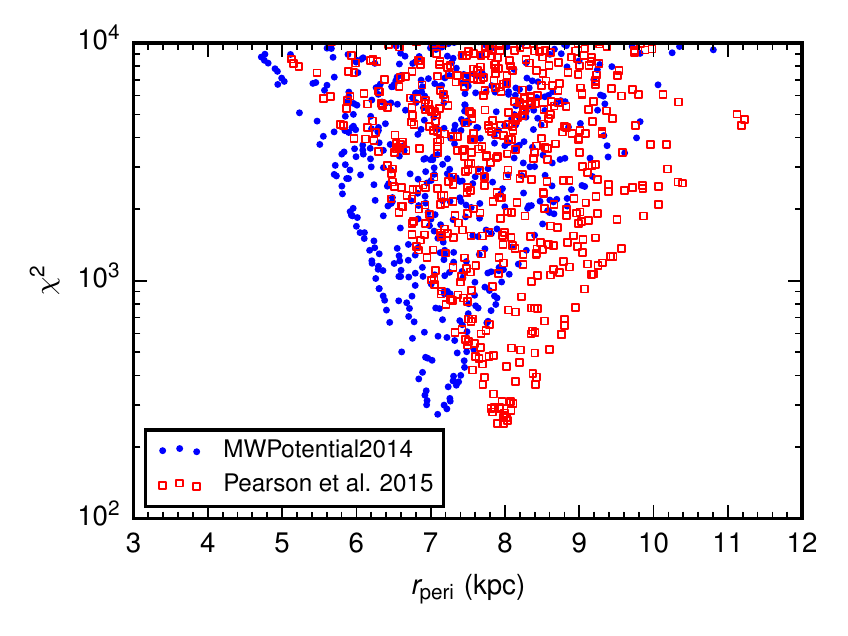}
\caption{Goodness-of-fit $\chi^2$ as a function of $r_{\rm peri}$ for
  Pal~5 stream models with the cluster proper motions sampled near the
  best fit within each potential. The filled blue circles show the
  result for the \texttt{MWPotential2014} potential from
  \protect\cite{bovy_galpy}. The empty red squares show the result for
  the potential from \protect\cite{pearson_et_al_pal5} with a spherical halo. As
  evidenced from this plot, there is a substantial uncertainty in the
  pericentric radius of the Pal 5 cluster due to the uncertainty in
  the MW potential.}
\label{fig:chi2_comparison}
\end{figure}

The basics of the process that leads to the emergence of narrow tidal
tails around in-falling satellites are now well established
\citep[e.g.][]{johnston_1998,helmi_white_1999,eyre_binney_2011}. The
studies above and references therein have presented a simple picture
where stars escape from the Lagrange points of the disrupting
progenitor. The stripped stars have slightly different energies and
angular momenta compared to the parent host, and hence the orbits of
the debris and the satellite will diverge in time. More precisely,
stars ejected from the outer (at larger Galactocentric distance)
Lagrange point have higher energies than the progenitor and hence will
have a longer orbital period and form the trailing tail. Likewise, the
leading tail comprises of lower energy stars ejected from the inner
(lower Galactocentric distance) Lagrange point. The Lagrange points
are typically taken to be equidistant from the progenitor's center,
which is justified if the progenitor is significantly smaller than the
scale over which the host gravitational potential changes. For
globular clusters in particular, it is difficult to imagine a
situation where this assumption can be broken. Since the Lagrange
points are symmetric and assuming the progenitor is roughly spherical, it follows that a similar number of stars should
be leaving each Lagrange point, thus yielding comparable number of
stars pumped into each tail.

To compare stellar densities at increasing distance intervals along
the tails, we have to consider the rate at which the stripped stars
move away from the progenitor. This rate is governed by the offsets in
energy and angular momentum of the debris which are equal in amplitude
and opposite in sign
\citep[e.g.][]{johnston_1998,helmi_white_1999,eyre_binney_2011}. Thus,
we expect the density in the leading and trailing arms to be roughly
symmetric about the progenitor. Note that the exposition above
neglects the stretching and compressing of the stream at pericenter
and apocenter respectively. Therefore, the assertion of a symmetric
density distribution along the tails only holds when the leading and
trailing arm have roughly the same Galactocentric distances. This result 
was also found in \cite{niederste-ostholt_et_al_2012}
where they used simulations to show that the leading and trailing
tails of the Sagittarius dwarf should be symmetric at apocenter and
pericenter.

The picture based on Lagrange point stripping not only predicts the
stream density, but also provides an expectation for the stream track
on the sky and the stream width. If the progenitor is orbiting in a static and spherical
potential, the tidal debris will remain in the same plane and an
observer in the center of the galaxy would see the stream which is
confined to a great circle on the sky. For a heliocentric observer,
the stream track should deviate smoothly from a great circle due to
the Galactic parallax. In an aspherical potential, angular momentum is
no longer conserved and the orbits of the tidal debris will precess
and nutate
\citep[e.g.][]{ibata_et_al_2001,helmi_2004,johnston_et_al_2005,belokurov_sgr_precess,stream_width}. The
effect of the aspherical host potential on the tidal debris
distribution is two-fold.  First, it will cause a further divergence
of the stream track from a great circle albeit producing a gradual and
smooth deviation. Moreover, given slightly different initial
conditions of the stripped stars, they will experience differential
orbital plane precession, leading to an increase in the width of the
stream on the sky. Thus, in a representative Galaxy, the stream track
is expected to deviate slowly from a great circle but have no
small-scale features. In addition, the stream width is not expected to stay
constant and must evolve smoothly with distance from the progenitor.

\subsection{Orbit of Pal 5} \label{sec:pal5orbit}

In what follows, we are interested in discerning between various
mechanisms capable of producing a significant asymmetry in the Pal~5
tails. For the two of these, namely the rotating bar and the giant
molecular clouds, the exact value of the cluster's pericenter plays
an important role. Therefore, while not striving to produce the
absolute best fit to the stream observables, we would like to identify
the closest match to the cluster's orbit in a realistic host
potential.

The radial velocity\footnote{Note that here and throughout the rest of this paper, the radial velocities are always measured in the heliocentric frame.} of Pal~5 itself was measured in
\cite{odenkirchen_et_al_2002_pal5_vel} to be $v_r = -58.7\pm0.2$\kms. Similar results have been found in \cite{kuzma_et_al_2015} with $v_r
= -57.4\pm0.3$\kms\ (and a systematic uncertainty of 0.8\kms\ in the
radial velocity zero-point), as well as \cite{ishigaki_et_al_2016} who measured $v_r = -58.1 \pm 0.7$\kms. The proper motion of the Pal~5
cluster was recently measured in \cite{fritz_kallivayalil_pal5} who
give $\mu_\alpha = -2.296\pm0.186$\masyr\ and $\mu_\delta =
-2.257\pm0.181$\masyr. \cite{kuzma_et_al_2015} also measured the
radial velocity of 17 stars in the leading arm and 30 stars in the
trailing arm. Finally, \cite{dotter_et_al_2011} measured the distance
to Pal~5 using isochrone fitting and found a value of $23.6\pm 0.9$
kpc. Equipped with these measurements, as well as the stream track
measured in the previous Section and shown in
\Figref{fig:stream_meas}, we can explore the range of pericentric
distances within a given potential.

In particular, we consider two choices of the gravitational potential
for the Milky Way host. The first one is the \texttt{MWPotential2014}
model from \cite{bovy_galpy} which consists of a spherical NFW halo, a
Miyamoto-Nagai disk \citep{miyamoto_nagai_1975}, and a power-law
density bulge with an exponential truncation. Specifically, the NFW
halo has a mass of $M_{\rm vir}=8\times 10^{11} M_\odot$, $c=15.3$,
and a scale radius of 16 kpc, the Miyamoto-Nagai disk has a mass of
$M=6.8\times 10^{10} M_\odot$, $a=3$ kpc, and $b=280$ pc, and the
bulge has a mass of $M=5\times 10^9 M_\odot$, a power-law exponent of
$\alpha=-1.8$, and an exponential truncation radius of $1.9$ kpc. The
second model is the spherical halo potential described in
\cite{pearson_et_al_pal5}, which consists of a spherical logarithmic
halo, a Miyamoto-Nagai disk, and a Hernquist bulge. The logarithmic
halo has a circular velocity of $172.39$\kms\ and a scale radius of 12
kpc, the Miyamoto-Nagai disk has a mass of $M=10^{11} M_\odot$,
$a=6.5$ kpc, and $b=260$ pc, and the Hernquist bulge has a mass of
$M=3.4\times 10^{10} M_\odot$ and a scale radius of 700 pc. For our
coordinates, we have the $X$ axis pointing towards the Galactic
center, $Y$ is aligned with the Galactic rotation, and $Z$ pointing
towards the Northern Galctic pole. Following a combination of
\cite{schoenrich_et_al_2010} and \cite{bovy_et_al_2012}, we take the
Sun's velocity relative to the local circular velocity to be
$(U_\odot,V_\odot,W_\odot)$ = (11.1,26,7.3)\kms\ and place the Sun at
$(-8.3,0,0)$\,kpc. Within each potential, we then compute the
tangential motion of the Sun by adding $V_\odot$ to the circular
velocity, $V_c$, at the Sun's location. In the
\texttt{MWPotential2014} potential from \cite{bovy_galpy} we find $V_c
= 219.0$\kms, and in the spherical halo potential from
\cite{pearson_et_al_pal5} we find $V_c = 220.8$\kms.

To zoom-in onto the most appropriate orbit in these potentials, we
sample the proper motion of the Pal~5 cluster from the allowed range
of the observed values and generate model streams using the modified
Lagrange Cloud Stripping (mLCS) method described in
\cite{gibbons_et_al_2014}. The progenitor is modelled as a $2\times
10^4 M_\odot$ Plummer sphere with a scale radius of $15$ pc. Particles
are released from the Lagrange points with a velocity dispersion given
by the Plummer profile. The particles are stripped near each
pericenter following a Gaussian stripping rate with $\sigma = 10$
Myr. At the present epoch, the progenitor is fixed to be at a distance
of 23.6 kpc and a radial velocity of $-57.4$\kms. Given the chosen
proper motion value, we then rewind the cluster's orbit for 5 Gyr and
produce a stream. With the suite of stream simulations in hand, we
explore how the goodness of the model fit depends on Pal~5's
pericentric distance.

In order to assess how well each realization fits the data, we define
a likelihood for the stream track and the run of radial
velocities. For the stream track, we perform a linear fit to the
simulated stream in bins of 1$^\circ$ in $\phi_1$ and determine the
value and uncertainty of $\phi_2$ in the center of each bin. For the
data, we take the mean track with uncertainties from
\Secref{sec:results}. The likelihood for the track is then defined by
\eq{ \mathcal{L} = \prod_i \frac{1}{\sqrt{2\pi \sigma_i^2}} \exp\left( -\frac{(d_i-m_i)^2}{2 \sigma_i^2} \right) ,}
where the index $i$ runs over the $\phi_1$ bins, $d_i$ is the measured
$\phi_2$ from the data, $m_i$ is the mean of $\phi_2$ for the model,
and $\sigma_i$ is the sum in quadrature of the observational and model
error. The likelihood is similarly defined for the radial velocity
except the index $i$ now runs over the radial velocity data
points. For each data point, a Gaussian fit is performed for the model
points within 0.5$^\circ$ in $\phi_1$ to get the mean and error on the
mean of the radial velocity. The likelihoods for the stream track and
radial velocity are then multiplied to get the total likelihood.

For the \texttt{MWPotential2014} model from \cite{bovy_galpy}, we find
a best fit proper motion of $\mu_\alpha=-2.23$\masyr\ and
$\mu_\delta=-2.22$\masyr. This proper motion gives a pericenter of
7.1 kpc. For the model from \cite{pearson_et_al_pal5}, we find a
best-fit proper motion of $\mu_\alpha=-2.30$\masyr\ and
$\mu_\delta=-2.29$\masyr\ which gives a pericenter of 8.0 kpc. Note
that this is slightly different than the best-fit proper motion
reported in \cite{pearson_et_al_pal5}, in part due to our use of a
different cluster's radial velocity and solar motion, and in part due to our updated stream track. In order to
showcase the range of possible pericentric distances in these
potentials, the proper motion is sampled 1000 times around the
best-fit values with a spread of $0.1$\masyr. 
\Figref{fig:chi2_comparison} displays the $\chi^2$ of these
models as a function of the pericentric distance. We see that within
a given potential, the range of pericenters consistent with the stream
data available is quite small. However, the systematic error, i.e. the
uncertainty due to the choice of the potential, is large. Namely, the
allowed range of pericenters for Pal~5 is from 7 to 8 kpc. Further
modelling of the Pal~5 disruption with a more flexible potential is
needed to produce a more robust measurement of the cluster's
pericenter.

\begin{figure*}
\centering
\includegraphics{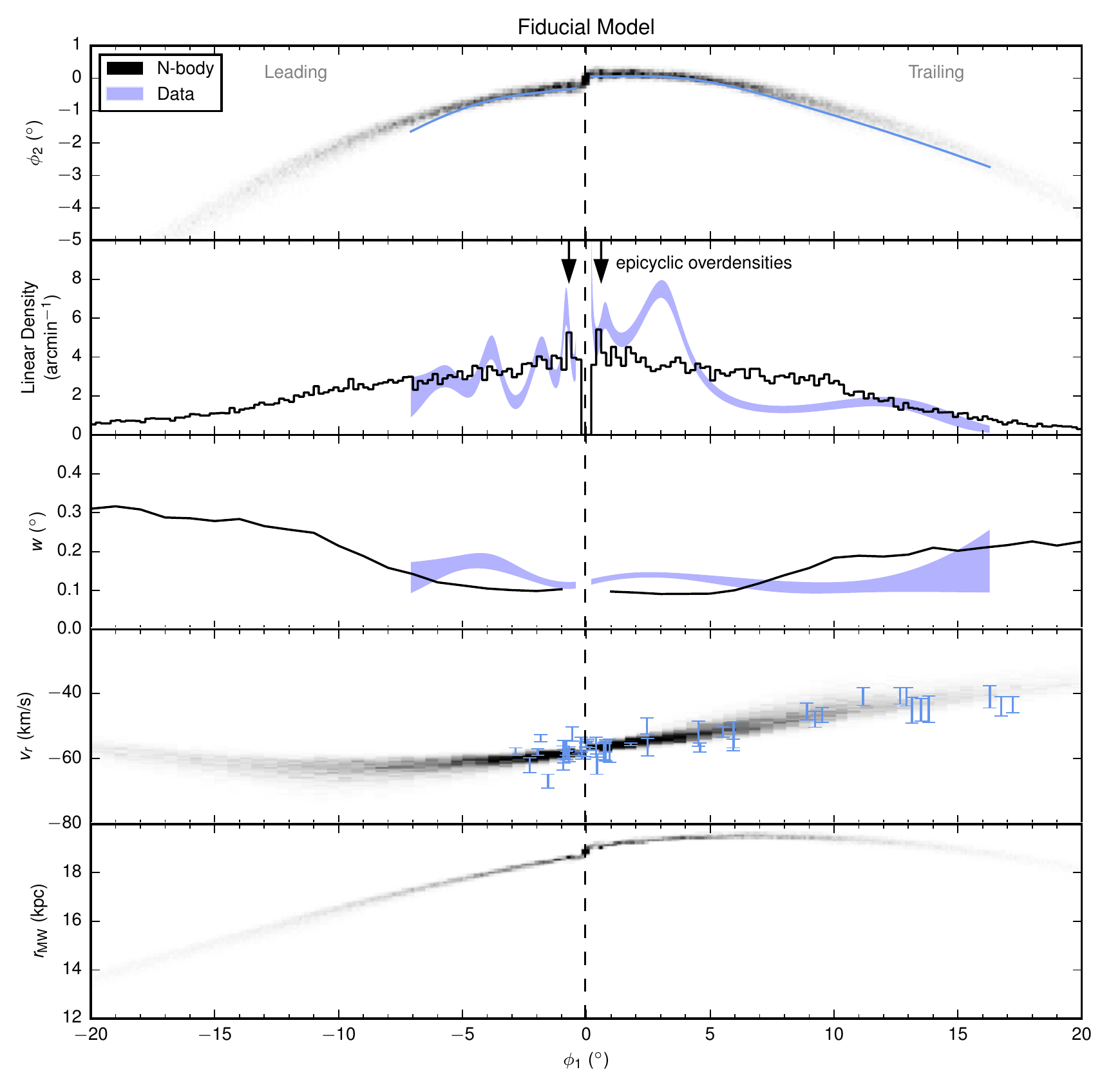}
\caption{Comparison between measurements of the Pal 5 stream and
  the fiducial (unperturbed) N-body model of the cluster
  disruption. Similar to Figure~\ref{fig:stream_meas}, this shows the
  evolution of the stream centroid (top), density of particles (2nd
  panel) and the width of the debris distribution (3rd panel).
  Additionally, the stream's radial velocity (4th panel), and the
  Galactocentric radius (5th panel) are shown.  In all panels, the
  black histograms show the results of our N-body simulations. In the
  top panel, the solid blue line shows the median from
  \protect\figref{fig:stream_meas}. In the second and third panels,
  the filled blue region shows the 16-84\% credible interval from
  \protect\figref{fig:stream_meas}. In the fourth panel, the observed
  radial velocities from \protect\cite{kuzma_et_al_2015} are shown as blue error bars. The
  vertical dashed line shows the location of the progenitor at
  $\phi_1=-0.05^{\circ}$ which separates the leading (left) and trailing (right) tails of the stream. While the simulated stream track is remarkably
  close to the observed track, the simulated density is broadly
  symmetric, as expected, and in stark contrast to the observed Pal 5
  density. The two arrows in the second panel mark the approximate
  locations of the two most prevalent epicycles at $\phi_1 =
  -0.7\degr,+0.6\degr$. This slight asymmetry is due to the fact that
  the simulated progenitor is located at $\phi_1 = -0.05\degr$. The
  epicycle in the leading arm also lines up well an overdensity in the
  data at roughly the same location, $\phi_1 = -0.8\degr$. There is
  also a reasonable match in the trailing arm albeit the measured
  overdensity is not as significant. Note that the second panel, the linear density of particles in the simulation has been scaled down by a factor of 8.3 to approximately match the observed linear density in the leading arm.}
\label{fig:pal5_unpert}
\end{figure*}

\subsection{Model of Pal 5 in a static and smooth potential} \label{sec:smooth_pal5}

Our fiducial model of the Pal~5 disruption is shown
\Figref{fig:pal5_unpert}.  This simulation is run using the N-body
part of \textsc{gadget-3} which is similar to \textsc{gadget-2}
\citep{springel_2005}. For the potential, we use
\texttt{MWPotential2014} from \cite{bovy_galpy} which satisfies a wide
range of constraints, as opposed to the potential model of
\cite{pearson_et_al_pal5} which was selected specifically to reproduce
the Pal~5 stream. In this potential, the globular is represented with
a King profile with a mass of $2\times 10^4 M_\odot$, a core radius of
15 pc, and $w=2$. It is modelled with $10^5$ equal mass particles and a
softening length of $1$ pc. The final phase-space coordinates of the
progenitor are as follows: a proper motion of
$(\mu_\alpha,\mu_\delta)=(-2.235,-2.228)$\masyr, a radial velocity of
$v_r=-57.4$\kms, and a distance of 23.6 kpc. From the position and
velocity of the cluster today, the orbit is rewound for 5 Gyr and the
disruption is initiated. 

\Figref{fig:pal5_unpert} is a clear demonstration of the expectations
of a typical stream behavior as described in
Section~\ref{sec:mechanics}. The Figure shows the same set of stream
properties as in \Figref{fig:stream_meas} as seen by a heliocentric
observer, as well as the run of the stream's line-of-sight velocity
and Galactocentric distance with $\phi_1$. As expected, we see that
the debris density is broadly symmetric about the progenitor with no significant small scale features except for the epicycles, in stark
contrast to the data from Pal 5 which is overplotted. As demonstrated
in the Figure, the centroid of the debris smoothly deviates from a
great circle in a manner very similar to that of the observed stream
track although the match is not perfect. Reassuringly, the simulated stream has a width similar to
that observed in Pal~5. The fourth panel of the Figure (second from
the bottom) shows the radial velocity variation along the stream in
comparison to the measurements provided by
\citet{kuzma_et_al_2015}. Finally, the fifth (bottom) panel gives the
Galactocentric distance evolution along the tails. Note that, in the
vicinity of the progenitor, the leading and trailing arms have very
similar distances, thus demonstrating that the stretching and
compressing of the stream cannot be responsible for the asymmetry
discussed in Section \ref{sec:results}. This is in agreement with
\cite{ibata_et_al_pal5} who also find similar distances to the leading
and trailing arms, with a small variation of $\sim 3$ kpc along the
observed portion of the stream. To conclude, overall, the fiducial
model faithfully reproduces the behavior of the centroid of each of
the tails in the phase-space, but not the details of the stellar
density along the stream.

\subsection{Epicyclic overdensities}

The simple picture of tidal disruption described in
\Secref{sec:mechanics} glosses over the small-scale details of the
stellar stream formation, namely the density variations due to
epicyclic motion
\citep{kuepper_et_al_2008,just_et_al_2009,kuepper_et_al_2010}. As
mentioned before, these overdensities arise because the initial
conditions of the stripped stars are similar and they complete the
motion around the epicyclic ellipse at comparable times. The similarity
of the perturbed motion of the debris (compared to the stream track)
is emphasized at the vertices of the ellipse where the stars spend the
most time. In the fiducial stream displayed in
\Figref{fig:pal5_unpert}, the epicycles are included
self-consistently. The locations of the two most prominent epicyclic
overdensities are marked with arrows and as discussed in
\cite{kuepper_et_al_2008,just_et_al_2009,kuepper_et_al_2010}, appear
equally spaced from the progenitor.

\begin{figure}
\centering
\includegraphics{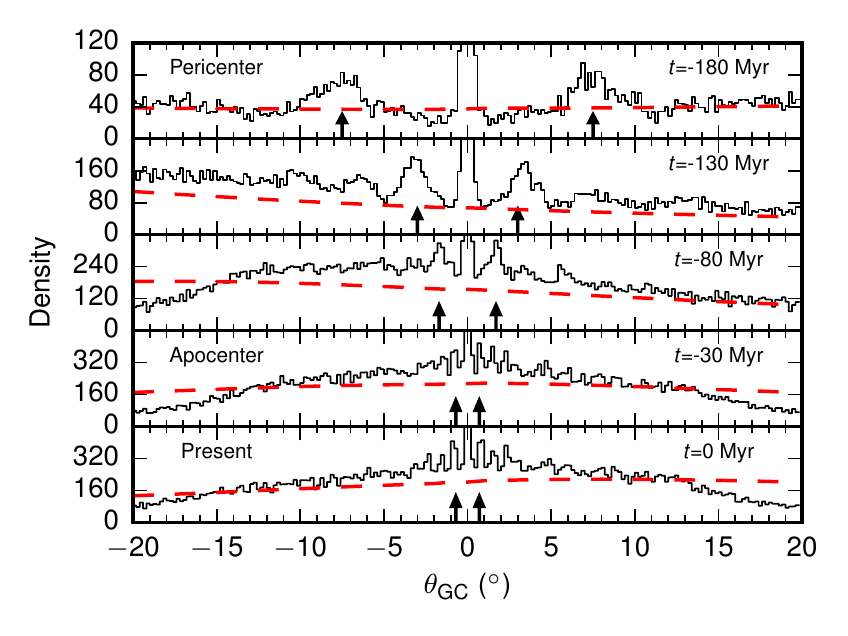}
\caption{Evolution of the stream density in the fiducial (unperturbed)
  Pal 5 simulation at various progenitor's orbital phases for an
  observer at the Galactic center. The solid black histogram shows the
  linear density computed in 0.2$^\circ$ bins. The dashed red curve is
  proportional to $r^2$ and has a maximum equal to the median of the
  density. At each time, the stream plane is defined by the angular
  momentum of the progenitor. The most pronounced epicyclic
  overdensities can be seen at the pericenter (top panel) where they
  appear to be highly symmetric. The second and third panels show the
  density at intermediate times between pericenter and apocenter. The
  density is asymmetric at these epochs due to the differing radii of
  the leading and trailing arms. This asymmetry roughly follows the
  expected scaling of $r^2$. Note that the epicyclic overdensities
  become less pronounced as the progenitor approaches apocenter. The
  fourth panel shows the stream at apocenter where it is broadly
  symmetric and the epicyclic overdensities are barely visible. The
  final panel shows the stream at the present time where the tails are
  slightly asymmetric and there are two epicyclic overdensities
  visible at $\theta_{\rm GC} = \pm 0.7^\circ$.  }
\label{fig:pal5_orbital_comparison}
\end{figure}

Epicyclic feathering is unambiguously abundant in the numerical
simulations of the globular cluster disruption referenced above. Is it
surprising that the overdensities appear so underwhelming in
\Figref{fig:pal5_unpert}? To answer this question, let us follow the
evolution of the epicyclic overdensities as a function of the
satellite's orbital phase. \Figref{fig:pal5_orbital_comparison}
presents the debris density at five different times, starting from the
previous pericenter and ending at the present. Note that this plot
shows the stream star counts as viewed from the Galactic center, hence
we denote the angle along the stream as $\theta_{\rm GC}$ rather than
$\phi_1$. As evidenced in the Figure, the clumping is most visible at
pericenter and is barely detectable at apocenter. Some of this
behavior might be due to the stretching and compressing of the stream
as it goes from pericenter to apocenter. However, if the stream is
thought of as a train of particles following the same orbit, then
conservation of the angular momentum dictates that $r^2
\frac{d\theta}{dt}$ is constant in time, where $d\theta$ and $dt$ are respectively the angle
and the time delay between the two particles. The angle between two particles trailing each other will
thus vary as $d\theta \propto r^{-2}$. This implies that if the growth
of the stream is ignored, the debris density goes like $r^2$. Thus,
the angular distances between the stars in the epicyclic clumps are
expected to compress by a factor of $(r_{\rm peri}/r_{\rm apo})^2$ on
going from pericenter to apocenter. Similarly, we expect their density
to go up by a factor of $(r_{\rm apo}/r_{\rm peri})^2$. Perhaps, the
the strength of the epicyclic overdensities at pericenter is simply
due to the fact that most of the stripping actually happens at
pericenter. Given that the clumps form at integer multiples of the
radial period after they are stripped, their amplitudes are pronounced
near peri crossings. This is illustrated in the top panel of
\Figref{fig:pal5_orbital_comparison} where the strong epicyclic
bunchings are composed of the particles stripped during the previous
pericentric passage.

Note that some of the panels in \Figref{fig:pal5_orbital_comparison}
exhibit a small asymmetry in the density between the leading and
trailing arm. Also included in the Figure is a dashed red curve which
is proportional to $r^2$ as measured in $1^\circ$ bins along the
stream. The slope of the curve reveals that the asymmetry seen in the
second and third panels of \Figref{fig:pal5_orbital_comparison} is
mostly due to the difference in the Galactocentric radius along the
stream. Most importantly, we see that at the present day (bottom
panel), the debris density near the progenitor is roughly symmetric
since the two tails are at approximately the same distance from the
Galactic center. The small asymmetry discussed above agrees well with
the findings of \cite{just_et_al_2009} and \cite{kuepper_et_al_2010}
who both note that the amplitudes of the epicyclic overdensities
could differ slightly. We note, however, that this difference is
nowhere near the dramatic mismatch between the trailing and leading
tail densities as measured here. Finally, \cite{zotos_2015} found some
evidence that the rate at which the stars leave the inner and outer
Lagrange points can be different. Yet, again, the reported asymmetry,
if real, is at much lower level, i.e. some 10\% rather than a factor
of two as detected in the Pal~5 stream.

\section{Mechanisms to produce observed features} \label{sec:asymmetric_mechanisms}

Having established the expected shape and density behavior of a Pal~5
stream in a smooth and static potential, this Section considers two
distinct mechanisms which can plausibly produce the observed
discrepant features. Namely, a fly-by of substructure, either in the
form of a dark matter subhalo or a giant molecular cloud, and the
effect of the Milky Way bar.

\subsection{Interaction with subhaloes} \label{sec:subhalo_interaction}

Dark matter (DM) subhaloes affect streams locally
\citep[e.g.][]{ibata_et_al_2002,johnston_et_al_2002} and hence ought
to be able to induce the features discussed above. In fact, models of
subhalo flybys generically predict an underdense region (around the
point of closest approach) surrounded by overdensities,
\citep[e.g.][]{carlberg_2012,three_phases,subhalo_properties}, similar
to what can be seen in the trailing tail of Pal~5. Thus, interactions with dark perturbers are an obvious candidate
mechanism, which is considered in this subsection.

To demonstrate that perturbations caused by DM subhaloes can look
very similar to the features detected, we run a suite of semi-analytic
simulations of subhalo flybys. We start with the fiducial Pal~5
stream described in \Secref{sec:smooth_pal5} and shown in
\Figref{fig:pal5_unpert}. Using the effective N-body model described
in \cite{number_of_gaps}, we can rapidly simulate the effect of a
flyby by taking a snapshot of the simulation at an earlier time and
then perturbing the particles using the form of the impulse
approximation of \cite{sanders_bovy_erkal_2015}. As in
\cite{number_of_gaps}, the subhaloes are assumed to be Plummer
spheres. The perturbed particles are then evolved using a
kick-drift-kick leapfrog integrator to the end of the simulation
where they are combined with the unperturbed particles.

With a small amount of trial and error, we found a two flyby setup whose
effect roughly matches the observed stream density. The first flyby, which creates the feature in the trailing tail, occurs 1.4 Gyr in the past with a mass of $5 \times 10^7
M_\odot$, a scale radius of $1.15$ pc, and a velocity of
$200$\kms\ relative to the stream. The flyby
occurs 5.7 kpc from the progenitor, roughly 45\% of the way along the trailing
stream at that time, and has an impact parameter of 1 kpc. The second flyby, which creates the feature in the leading tail, occurs 500 Myr in the past with a mass of $10^6 M_\odot$, a scale radius of $162$ pc, and a velocity of $100$\kms\ relative to the stream. The flyby occurs 2.0 kpc from the progenitor, roughly 17\% of the way along the leading stream at that time, and is a direct impact. Interestingly, almost all of the simulated particles within the gap at $\phi_1 \sim -3\degr$ were stripped within the last 2 Gyr indicating that the flyby must have occurred within that time period. 

\begin{figure*}
\centering
\includegraphics{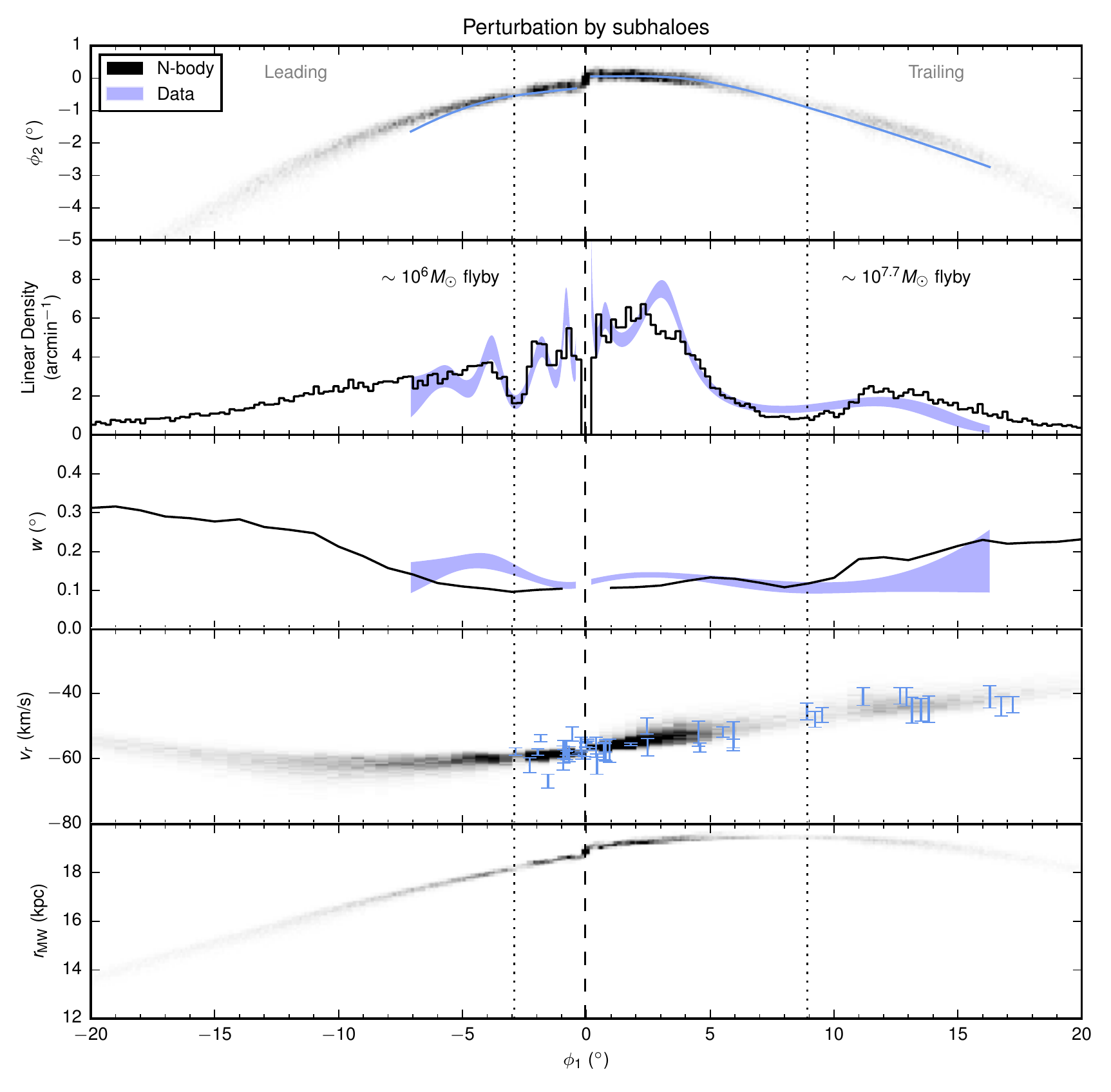}
\caption{Comparison between the measured stream properties and the model
  of the Pal 5 stream in the \texttt{MWPotential2014} potential from
  \protect\cite{bovy_galpy} perturbed by two subhalo fly-bys. As in
  \protect\figref{fig:pal5_unpert}, the stream track, debris density,
  stream width, radial velocity and Galactocentric radius are
  shown. Details of the panels are described in the caption of \protect\figref{fig:pal5_unpert}. The vertical dotted lines show the approximate locations of
  the centers of the two gaps produced as a result of interaction with
  dark matter subhaloes. The gap in the leading arm (left) is created
  by a $10^6 M_\odot$ subhalo while the gap in the trailing arm
  (right) is created by a $5\times 10^7 M_\odot$ subhalo. In an
  apparent contrast to \protect\figref{fig:pal5_unpert}, there is a
  good match between the measured and modelled stream density. In addition, the trailing tail now has a roughly constant width until $\phi_1 \sim 10^\circ$, providing a better match to the observations.}
\label{fig:pal5_dark_flyby}
\end{figure*}

\Figref{fig:pal5_dark_flyby}
compares the realization of Pal 5 which has interacted with both
subhaloes to the stream data and shows a qualitative match to the density profile. In addition,
the width in the trailing arm is now almost constant until $\phi_1 \sim 10^\circ$. This provides
a better match to the observed width, unlike the unperturbed stream (see \figref{fig:pal5_unpert}), which has fanned appreciably by then. Our search for a convincing DM fly-by
configuration has been far from exhaustive. The quick success of the
parameter exploration exercise is easy to understand. As discussed in
\cite{subhalo_properties}, there is a large degeneracy in going from
the shape of the gap to the subhalo properties and hence a wide range
of flybys can give rise to the same density profile. Note, however,
that this degeneracy is almost entirely broken by looking at other
observables, such as the stream track and the radial velocity
profile. Yet, before a more comprehensive modelling of the Pal~5 stream
is carried out, we must stress that the subhalo properties presented
above are merely chosen by hand to generate an approximately similar
density profile.

Because the debris gap properties are solely controlled by the
velocity kick imparted on the stream stars, similar sized kicks can be
produced by a slowly moving low-mass perturber nearby and a
fast-flying massive satellite farther away
\citep[see][]{three_phases}. Therefore, it is prudent to consider the two most obvious
candidate perturbers in the Milky Way: the Sagittarius dwarf galaxy and the Large
Magellanic Cloud. Both have in-fall masses in excess of $10^{10}$
M$_{\odot}$ \citep[see e.g.][]{Jorge2016,Gibbons2016} and therefore
may have affected a number of objects throughout the Galaxy. However, the size of the gap induced in the stream
is also related to the impact parameter \citep[see][]{three_phases} so a distant flyby, $\gtrsim 10$\,kpc, will produce gaps significantly larger than even the $9\degr$ feature seen in \Figref{fig:stream_meas}. Thus, unless Sagittarius had a passage significantly closer than this, or the gap in Pal 5 is actually much larger than $\sim 9^\circ$,  it is unlikely that either could produce the gap-like features seen in Pal 5. As an additional argument, \cite{number_of_gaps} studied the gaps created in a Pal 5-like stream by subhaloes in the range $10^5-10^9 M_\odot$ and found that subhaloes in the range $10^8-10^9 M_\odot$ made a significantly lower contribution to the number of gaps than subhaloes in the range $10^6-10^8 M_\odot$ due to their low rate of encountering the stream. We note that while these massive perturbers are unlikely to have caused the features seen in Pal 5, the streams of dark matter stripped from the massive satellites can also produce gaps as argued in \cite{bovy_2016}.

In addition to creating features in the density, the perturbations from subhalo flybys also cause the stream track to oscillate about the unperturbed track \citep{three_phases,subhalo_properties}. This oscillation takes place over roughly the same scale as the gap size with an amplitude governed by the properties of the flyby \citep[e.g.][]{subhalo_properties}. For the flyby near the trailing arm shown in \Figref{fig:pal5_dark_flyby}, the amplitude of the oscillation is $0.5^\circ$ and, by chance, the oscillation happens to be close to zero at the present time. Thus, the lack of a stream track variation does not necessarily imply there was no subhalo perturbation. Furthermore, in order to detect such a small oscillation of $\sim 0.5^\circ$ over the gap size, $\sim 10^\circ$, one would need an accurate model of the Milky Way and the unperturbed stream track. Thus, the lack of an obvious stream track oscillation in the data does not rule out a subhalo flyby.

\subsection{Interaction with Giant Molecular Clouds} \label{sec:GMCs}

Recent work by \cite{amorisco_et_al_2016_gmcs} has also found that the
Pal 5 stream could be affected by giant molecular clouds (GMCs). They found that since Pal 5 is on a prograde orbit, these gaps are
somewhat enhanced compared to those produced by subhaloes due to the smaller relative velocity between the
stream and the GMCs. However, due to the lower mass of the GMCs,
$M<10^7 M_\odot$, the gaps they produce span a smaller range of sizes. The distribution of gap sizes is explored
in Fig. 4 of \cite{amorisco_et_al_2016_gmcs} where they find that deep
gaps have sizes between $0.4-3$ kpc. At a distance of 23 kpc, this would
correspond to $1-7.5\degr$. We also note that since \cite{amorisco_et_al_2016_gmcs} did not require their Pal 5-analogues to be near apocenter at the present, the gap sizes they predict are likely an overestimate. Thus, while the $2^\circ$ feature in the leading arm at $\phi_1 \sim -3^\circ$ is consistent with a GMC perturbation, the larger feature at $\phi_1 \sim 8^\circ$ with a size of $9^\circ$ is unlikely to have been caused by a GMC. This also agrees with the result of \Secref{sec:subhalo_interaction} where we found that a subhalo mass of $10^6 M_\odot$ could explain the feature in the leading arm.

While we cannot distinguish between the effects of a DM subhalo and a
GMC with the current data, in \cite{subhalo_properties} it was shown
that the properties of the flyby, notably the time since impact, can
be extracted from the shape and density of the stream
perturbation. Thus, with additional measurements of the radial
velocity, and, perhaps, proper motion, one should be able to
reconstruct the flyby and identify the likely perturber. This should
be possible because interactions with GMCs only occur in the disk, unlike
those with DM subhaloes which can occur anywhere in the halo.

\subsection{Rotating bar} \label{sec:rotating_bar}

An interaction with an intervening satellite provides a short
timescale change of the local gravitational potential and thus can
affect only a portion of a long stellar stream. Similarly, a rotating
Milky Way bar adds a varying component to the force field of the
Galaxy. \cite{kohei_rotating_bar} studied the consequences of the
presence of a bar on the Ophiuchus stream \citep{ophiuchus_disc} and
found a dramatic effect where the bar can induce different changes in
the energy to different sections of a stream. Thus, since these
sections will now orbit with different periods, the bar can cause
variations in the debris distribution, resulting in both
under- and overdensities. The effect of the bar on streams was also
considered in \cite{price_whelan_et_al_chaotic_fanning} where they
focused on the importance of orbital chaos, a topic which
will be briefly discussed in \Secref{sec:other}.

Here, since we expect that the bar can only induce relatively large scale features compared to the $\sim$ few degree scale features possible from substructure), we focus solely on the asymmetry in the density near the
progenitor to ascertain whether the bar can plausibly create this. The Galaxy is represented with the NFW halo and
Miyamoto-Nagai disk of \texttt{MWPotential2014} from \cite{bovy_galpy}
and the bar is the prolate bar model of \cite{long_murali_1992}
described in \cite{kohei_rotating_bar}. Specifically, we use a mass of
$M=5\times 10^9 M_\odot$, a half-mass size of $3$ kpc, and a Plummer
softening of 1 kpc. These values give a bar which is broadly
consistent with the mass constraints reported in
\cite{portail_et_al_2015}. The current bar angle is taken to be
$-30^\circ$ to match observed constraints
\citep[e.g.][]{lopez_et_al_2005}. In order to explore the broad enough
range of bar effects, we consider three different pattern speeds and a
non-rotating bar. We use pattern speeds of $\Omega_{\rm bar} =
-30,-50,-70$ km/s/kpc which span the gamut consistent with the Milky
Way's bar observations \citep[see][for a review]{gerhard_review}. In
this potential, 1000 model streams are evolved using the mLCS
technique described in \Secref{sec:pal5orbit}. As in
\Secref{sec:pal5orbit}, the current distance and radial velocity of
Pal 5 are fixed at 23.6 kpc and $-57.4$\kms. The proper motions are
sampled from a normal distribution centered on
$(\mu_\alpha,\mu_\delta) = (-2.23,-2.22)$\masyr\ with a spread of 
$0.1$\masyr\ in each component. Each simulated stream is evolved for 5 Gyr
and the minimum pericentric distance during this time is recorded. To
quantify the bar's influence on the stream, the asymmetry statistic,
$A$, describing the density difference between the leading and
trailing arm is calculated. The asymmetry statistic is defined as
follows:
\eq{ A=\sum_i (\rho^{\rm lead}_i - \rho_i^{\rm trail})^2 , }
where the index $i$ runs over bins in $\phi_1$ between $1-5^\circ$ in
steps of 0.5$^\circ$, and $\rho^{\rm lead}_i,\rho^{\rm trail}_i$ are
the normalized densities in the leading and trailing arm
respectively. For the observed Pal~5 density, a value of $A=0.059$ is
measured which can now be compared against the asymmetry seen in each
model realization of the stream. Of course, this asymmetry statistic may overestimate the impact of the bar alone since the small-scale feature in the data near $\phi_1 \sim -3^\circ$ will contribute to the asymmetry but was likely created by a different mechanism.

\begin{figure}
\centering
\includegraphics{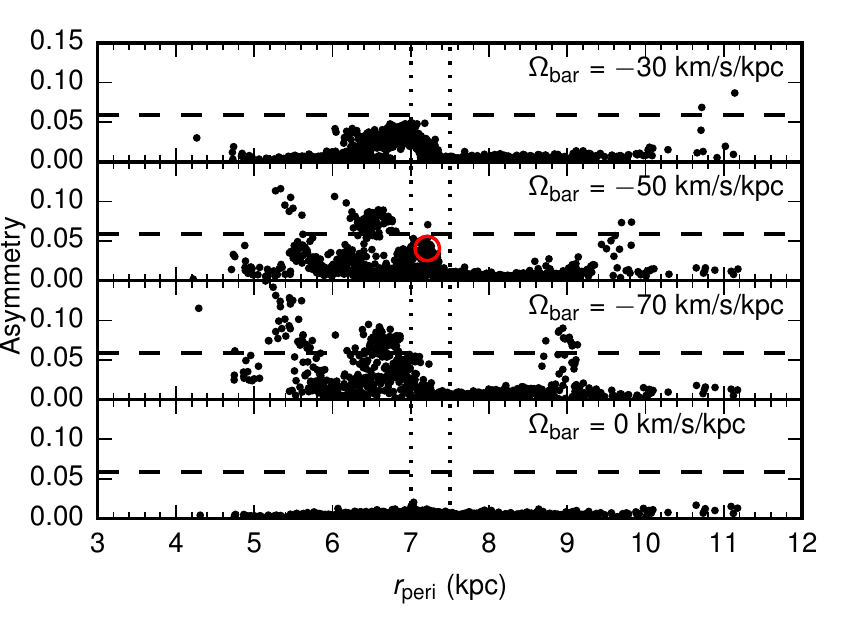}
\caption{Asymmetry statistic, $A$ (see main text for details), for a
  sample of streams with varying proper motion in the presence of a
  rotating bar with three different pattern speeds as a function of
  Pal 5's pericenter.  The horizontal dashed line shows the
  observed asymmetry in Pal 5 of $A=0.059$. Interestingly, the bar can
  produce a similar amount of asymmetry as in the observed Pal 5's
  tails even if its orbit has a pericenter larger than $7$ kpc. The
  red circle in the top panel shows the stream we have resimulated
  with \textsc{gadget-3} (see
  \protect\figref{fig:pal5_bar_example}). This particular point was
  chosen since the stream track and radial velocity also provide a
  good match to that of Pal 5. The dotted vertical lines show the
  range of pericentric radii where we get the best fits to the Pal 5
  stream.}
\label{fig:bar_asymmetry}
\end{figure}

\Figref{fig:bar_asymmetry} shows this asymmetry versus the
pericentric distance for three different bar pattern speeds and for a
non-rotating bar. As evidenced in the Figure, a rotating bar can
create a substantial asymmetry in the stream even if the pericenter is
sufficiently large, i.e. $\sim 7-8$ kpc. The exact location of the
maximal asymmetry depends on the pattern speed of the bar. The Figure
also shows that a non-rotating bar can only create symmetric
streams. Thus, it is the rotation of the bar that is crucial, as
expected from the earlier results of \cite{kohei_rotating_bar}. As a test, we also ran the streams in a bar with a present-day angle of $-20^\circ$ and found a similar number of asymmetric streams. Note
that, importantly, while a reasonable fraction of the model streams
studied here exhibit a substantial asymmetry, many of these do not
provide a good match to the stream track and the radial velocity
profile of Pal 5. 

\begin{figure}
\centering
\includegraphics{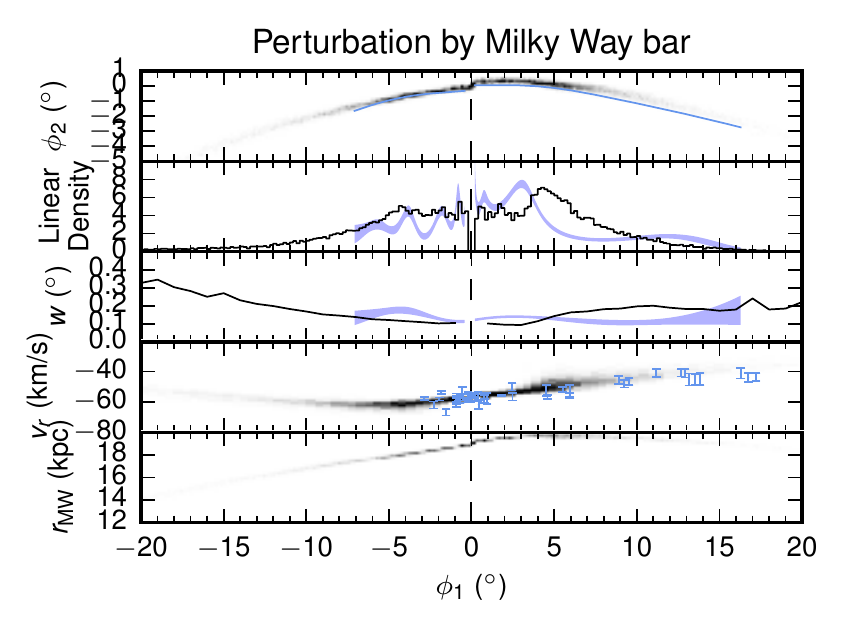}
\caption{Influence of a rotating bar on Pal 5's tidal
  debris. The panels respectively show the stream track, density, width, radial velocity, and
  Galactocentric radius of a Pal 5 realization in a potential with
  $\Omega_{\rm bar} = -50$ km/s/kpc. See the caption of \protect\figref{fig:pal5_unpert} for more details on the panels. This realization has an asymmetry
  of $A=0.041$ and demonstrates that the bar can produce significant
  asymmetries in the stream density. This particular realization
  corresponds to the red circle in \protect\figref{fig:bar_asymmetry}
  with a pericenter of $7.1$ kpc. We note that the asymmetry is
  extremely sensitive to the pattern speed and a change of only 0.5
  km/s/kpc will yield a different stream perturbation with a
  significantly different value of $A$. }
\label{fig:pal5_bar_example}
\end{figure}

As a further piece of evidence, one of the streams shown in
\Figref{fig:bar_asymmetry} (the red-circle) is re-simulated with an
N-body disruption. This particular stream was chosen since it has a
significant asymmetry and the stream track and radial velocity run are
a relatively good fit to Pal 5. The simulation setup is the same as in
\Secref{sec:smooth_pal5}, except the bulge has been replaced by the
rotating bar described above. The resulting stream properties are
shown in \Figref{fig:pal5_bar_example}. The density has an asymmetry
of $A=0.041$, $\sim 30\%$ less than that in Pal 5 and the perturbation
is slightly off-set along $\phi_1$ compared to what is
observed. However, overall, the size and amplitude of the feature
produced are similar to what is seen in the trailing arm. During the progenitor's orbit, its minimum
pericentric distance is $7.1$ kpc, confirming the result of
\Figref{fig:bar_asymmetry} that the bar is important as such large
distances. Finally, we note that this asymmetry is extremely sensitive
to the pattern speed. For this particular set of final velocities, the
asymmetry is erased if the pattern speed is changed by just 0.5
km/s/kpc. This is likely because the asymmetry depends on a precise
alignment of the bar and the stream's pericenter at an earlier time,
as was the case in \cite{kohei_rotating_bar}. Thus, while we have
demonstrated that the bar can in principle create the density
asymmetry seen in the Pal 5 tails, the uncertainty in both the pattern
speed and the orbit of Pal 5 means we cannot be sure that the bar is the culprit for the feature in the trailing arm.
Future work is needed to determine whether the effect of the bar can produce the precise features seen in Pal 5.

Note that only a single bar model is examined in detail in this work,
i.e. that described in \cite{long_murali_1992}. As noted in
\cite{kohei_rotating_bar}, this bar is slightly lighter than the
constraints in \cite{portail_et_al_2015}. Thus, it is possible that we
have somewhat underestimated the effect of the bar. In addition, it is
possible that bars obeying different density laws may produce
different features in the stream. Furthermore, note that the effect of
the bar on the Pal 5 cluster itself was considered in
\cite{allen_et_al_2006} where they studied a single realization of Pal
5's orbit and found that their version of the rotating bar did not
substantially affect it. However, as was shown above and in
\cite{kohei_rotating_bar}, the presence of a bar induces slight
changes in the orbital periods of stars in a stream and hence can
create structure while not dramatically altering the orbit of the
progenitor.

\section{Other mechanisms}
\label{sec:other}

This Section considers other mechanisms that could lead to the
features seen in the density of the Pal~5 tidal
tails, some of which can be ruled out. In addition, several routes
are suggested to help distinguish between the plausible mechanisms.

\subsection{Rotating Pal 5}

Observations of globular clusters around the Milky Way suggest that a
large fraction of them may possess internal rotation
\citep[e.g.][]{Bellazzini_et_al_2012,fabricius_et_al_2014}. Non-zero
net angular momentum will affect the velocities with which stars leave
the globular cluster and therefore can naturally affect the stream
properties. The rate at which particles move away from the progenitor
is controlled by their energy and hence mainly by the component of
their velocity aligned with the progenitor's systemic velocity. As
long as the progenitor remains roughly spherical, the stars at the
Lagrange points will receive opposite but equal boosts of their
velocity from the rotation. Thus, they will move away from the
progenitor at the same rate, maintaining a symmetric stream
density. In addition, this mechanism should not create significant small-scale features in the stream. According to this picture, if the progenitor's rotation is
aligned with the orbital angular momentum, the debris will spread out
at a faster rate. Likewise, if the progenitor's rotation is
anti-aligned, we should expect a shorter stream.

\begin{figure}
\centering
\includegraphics{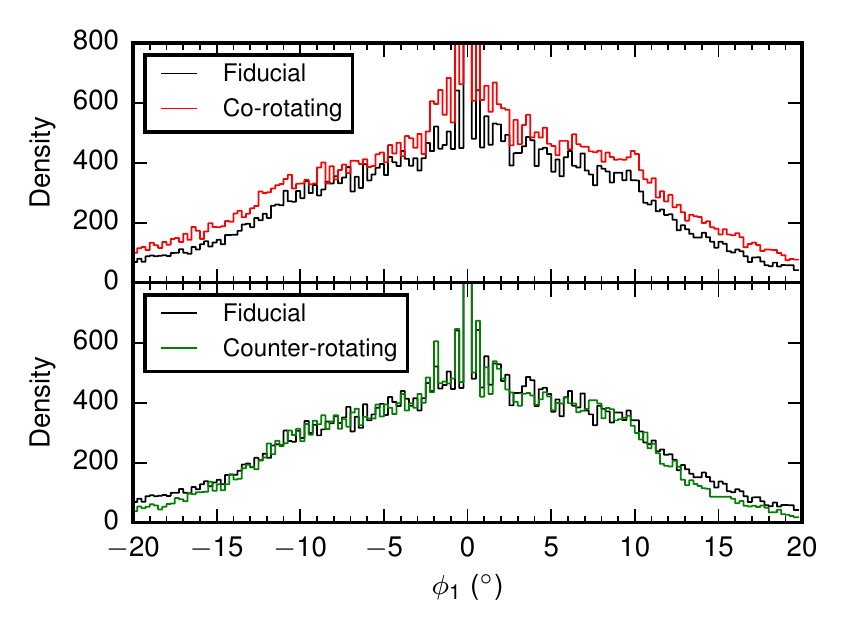}
\caption{Stream density for rotating progenitors. Comparison of the tidal tail density
  for a non-rotating (black) Pal 5 with a co-rotating (red, top) and
  counter-rotating (green, bottom) Pal 5. As expected, the
  non-rotating Pal 5 has an intermediate extent with the co-rotating
  (counter-rotating) being more (less) extended. More importantly,
  however, the rotation does not lead to any significant asymmetry in
  the debris distribution or significant small-scale density
  features. }
\label{fig:rotation}
\end{figure}

In order to study this effect, we ran a simulation of a rotating
globular cluster based on the unperturbed Pal 5 model described in \Secref{sec:smooth_pal5} and presented in
\Figref{fig:pal5_unpert}. The particle initial conditions in the
fiducial simulation depend only on the magnitude of the velocity
relative to the cluster's center of mass. Thus, we can flip the sign
of any individual velocity and still have a stable system. In order to
have a net rotation around a given direction, we compute the component
of the angular momentum along this axis and require each particle to
have a positive angular momentum in this direction. If the angular
momentum is negative, we simply flip the sign of the particle's
velocity. This results in a spherical and stable progenitor with a net
rotation. \Figref{fig:rotation} presents the stream density at the
present day for a non-rotating Pal 5, a Pal 5 whose internal rotation
is aligned with the initial orbital angular momentum (co-rotating),
and a Pal 5 whose internal rotation is anti-aligned with the initial
orbital angular momentum (counter-rotating). As expected, the
co-rotating realization produces the most extended stream and the
counter-rotating realization produces the shortest debris
distribution. Critically, however, the cluster rotation does not
introduce a significant density mismatch at levels similar to those
measured in Pal 5 tails or any significant small-scale features. Thus, it appears that rotation can not be
responsible for the observed features.

\subsection{Chaos}

Another important mechanism worth considering is
chaos. \cite{pearson_et_al_pal5} simulated Pal 5 in the triaxial
potential of \cite{law_majewski_2010} and found that the cluster's
tidal debris experienced a large amount of dispersal, which resulted in
a dramatic dramatic drop in surface density along the stream. The exact nature of this
stream fanning is uncertain. However, the same potential was studied
in more detail in \cite{price_whelan_et_al_chaos} who found the
presence of weak chaos in the orbital behavior and concluded that even
weak chaos could lead to substantial stream fanning. Note that at least
part of this fanning is due to differential orbital plane precession which
occurs in axisymmetric and triaxial potentials \citep{stream_width}. However, this effect
treats each tail equally and hence cannot create asymmetries. We also note that the most dispersed stream models, e.g. Fig. 5 of
\cite{pearson_et_al_pal5}, look very different from the observed Pal 5
tails. Namely, the centroid of the simulated debris distribution is
dramatically misaligned with respect to the observed stream track and
the tails display a rapidly increasing width (inconsistent with our
measurement presented above). Moreover, the simulation in the triaxial
potential can not reproduce the run of the radial velocity along the
stream. It remains unclear if a potential which was mildly
chaotic could ameliorate these discrepancies and also exhibit an
asymmetry in the density. We conclude that further investigation is
needed to understand the importance of chaos for the Pal 5 stream although
it does not appear to be a likely mechanism.

\subsection{Other baryonic effects}

As far as the localised stream perturbations are concerned, in
addition to DM subhaloes and GMCs, there are several other baryonic
substructures which could potential re-shape the Pal~5 tails. Three
obvious candidates are globular clusters, the disk of the Milky Way, and the spiral arms of the Milky Way.

In order to estimate the importance of interactions with (other)
globular clusters, we can compare their numbers and masses to those of
subhaloes and use the formalism of \cite{number_of_gaps}. Using the
table of globular cluster properties around the Milky Way from
\cite{gnedin_ostriker_1997}, we find there are 3 globular clusters in
the mass range $10^6 M_\odot < M < 10^7 M_\odot$ with Galactocentric
radii between 7 and 20 kpc (e.g. spanning the range of Pal 5's orbit),
with the most massive being $1.45\times10^6 M_\odot$. In contrast,
on average there are expected to be roughly 10 subhaloes in the same mass and
radial range for Milky Way-mass hosts \citep[e.g.][and references
  therein]{number_of_gaps}. If we assume that the 3 globular clusters seen at present are representative of the average number of globular clusters in this radial range, there will be roughly three times as many impacts by subhaloes as by globular clusters in this mass range. Furthermore, many of the subhaloes will have substantially higher masses so we
expect the subhaloes to produce significantly more prominent gaps. Thus, we conclude that subhaloes are more likely to have created gaps in Pal 5. However, future orbital analysis of the most massive globular clusters will also shed light on whether they interacted with Pal 5.

The disk itself cannot create an asymmetry since each tail will pass
through nearly the same region of the disk and will experience the
same forces. This was also demonstrated in \Secref{sec:rotating_bar}
where we showed that only a rotating bar can create an asymmetry in
the stream. If the bar is static, each part of the stream receives
almost the same perturbation and the stream remains symmetric. In
addition, the fiducial model presented in \Figref{fig:pal5_unpert}
included the effect of a disk and produced a symmetric stream with no
significant small-scale features. However, as we will discuss in the
next subsection, disk shocks can vary the stripping rate of the
progenitor and introduce small-scale, symmetric structure in the
stream. We leave the importance of spiral arms for future work.

\subsection{Variable stripping rate}

Another mechanism which can produce structure in the density is a
variable stripping rate. For globular cluster streams, the majority of
the stripping occurs at pericenter, especially those corresponding to
disk crossings \citep[e.g.][]{johnston_et_al_1999_2,dehnen2004}, where
the tidal forces are the highest. Each stripping episode sends a
packet of material into the stream which then broadens due to the
debris energy dispersion. Depending on the amount stripped in each
subsequent passage and the rate at which each packet broadens, this
can introduce structure into the stream density which will appear
symmetric near the progenitor. Indeed, this effect is naturally
included in our N-body simulations and we saw evidence of this in
\Figref{fig:pal5_orbital_comparison} where we presented the density of
our fiducial model at various times. Although the density variations
at the present time are not remotely as significant as what is seen in
the data (e.g. \figref{fig:pal5_unpert}), it is possible that the
effect can be enhanced in a different potential with, for example, a
significant flattening in the halo. This will result in a larger
variation in the tidal force at each pericentric passage which can
cause a more substantial variation in the amount of stripped
material. This could in principle explain the density minimum seen at
$\phi_1 \sim -3^\circ$ although not the inter-tail asymmetry. However,
estimates of the halo flattening based on recent modelling efforts
suggest it is not significantly flattened
\citep[e.g.][]{Koposov2010,bowden_et_al_gd1,kuepper_et_al_pal5} so we
do not think this is a likely explanation. Of course, the stripping
rate also depends on the properties of the progenitor
\citep[e.g.][]{dehnen2004}. Clearly, additional work is needed to
explore the range of density variations produced in different
potentials for progenitors consistent with the observations of Pal 5
and its tails.

\subsection{Distinguishing the various mechanisms}

The stream-fanning due to weak orbital chaos could in principle cause
density variations along the tidal tails. However, the most obvious
feature of the debris dispersal, the fast stream width growth, is
inconsistent with the results of our analysis. Similarly, although the
variable stripping rate could in principle produce substantial density
variations along the stream, our N-body streams evolved in a realistic
potential exhibit no such features (e.g. \figref{fig:pal5_unpert}),
and we argued that a highly flattened which could enhance the effect
is not supported by observations. While we cannot completely rule
these effects out, we argue that there are two main plausible
mechanisms which could create the features seen in Pal 5: namely, an
impact by substructure (e.g. \figref{fig:pal5_dark_flyby}) and the
effect of the Milky Way bar (e.g. \figref{fig:bar_asymmetry}). In
\cite{subhalo_properties} it was shown that the flyby of a subhalo
produces an almost unique signature which can be used to infer the
entire set of subhalo properties. This signature is imprinted in the
6D phase-space structure of the stream and only by combining
observations (e.g. the stream density, radial velocity profile, and
stream track) is it possible to make the inference. Presumably, the
effect of the bar also produces its own unique signatures which can be
distinguished from those caused by substructure. Additional efforts
are clearly needed to map out these signatures. Going forward, the
large catalogues of radial velocities expected from WEAVE \citep{weave}, 4MOST \citep{4most} and
DESI \citep{desi}, as well as the proper motions from Gaia \citep{gaia}, will help distinguish
these mechanisms or perhaps even show that they are working in
concert.

\begin{figure}
\centering
\includegraphics{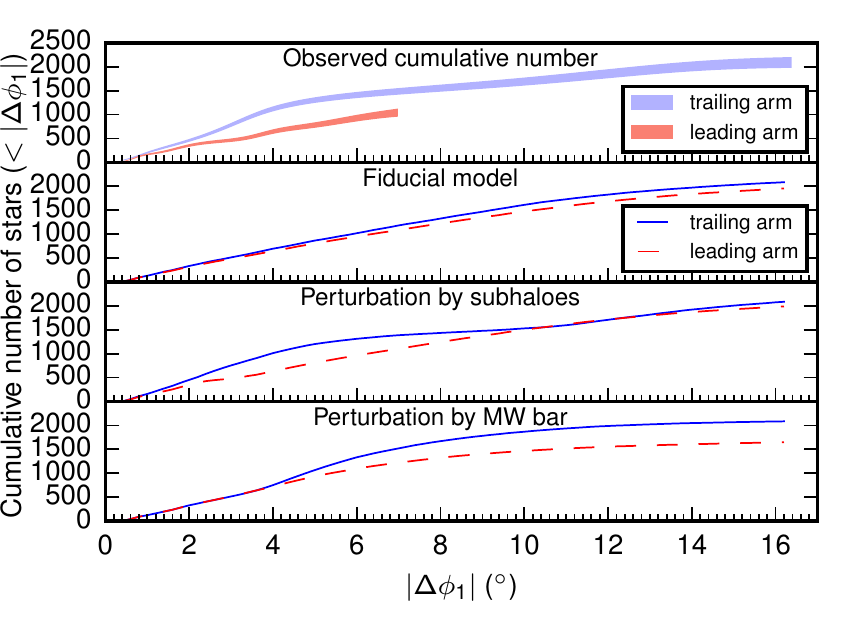}
\caption{Comparison of the cumulative number of stars for the leading and
  trailing arms for the observed Pal 5 stream and three different
  stream models. Since the progenitor of Pal 5 is slightly offset from
  the origin, we show the
  cumulative number in terms of $\Delta \phi_1 = \phi_1 -
  \phi_{1,{\rm Pal\, 5}}$. {\it Top}: 16-84\% confidence interval of
  the observed cumulative number distribution. This
  demonstrates that there is significantly less material in the
  observed section of the leading arm compared to the observed section
  of the trailing arm. {\it 2nd panel}: Fiducial (unperturbed) N-body
  simulation of the Pal 5 disruption. As expected, the arms appear
  almost completely symmetric. {\it 3rd panel}: Example of a two subhalo
  flyby from \protect\secref{sec:subhalo_interaction} which looks similar to the observations by construction. The
  leading and trailing arms have a similar cumulative number of stars for
  $|\Delta \phi_1| > 12^\circ$. {\it 4th panel}: Example of a
  perturbation by the Milky Way bar from \protect\secref{sec:rotating_bar}. In contrast to the subhalo
  interaction, here the difference in the debris density persists
  beyond the region shown here, eventually becoming symmetric only at
  very large distances from the progenitor. Thus, it is not
  immediately apparent what observations of the leading arm will
  reveal. We note that the cumulative numbers of stars in the simulations are
  scaled to match the observed cumulative number of the trailing arm in the
  right most bin.}
\label{fig:pal5_cumulative_density}
\end{figure}

In this vein, we propose an additional diagnostic of the asymmetry
mechanisms. \Figref{fig:pal5_cumulative_density} compares the observed
cumulative number of stars along the leading and trailing Pal 5 tails against
our fiducial N-body simulation, as well as the perturbed simulations
including a subhalo flyby and the effect of the Milky Way bar. The
observed cumulative number (top panel) shows that there is
substantially more material in the studied section of the trailing arm
compared to the measured portion of the leading arm. The fiducial
model (second panel) re-iterates that an unperturbed stream would
possess an almost symmetric cumulative number of stars. The slight difference
between the two arms here is due to the fact that leading arm is
heading towards pericenter and is being stretched out. The third and
fourth panels show the cumulative number behavior for a stream
perturbed by a subhalo flyby and by the Milky Way bar respectively. In
the case of the subhalo flyby, the asymmetry only continues until
$|\Delta \phi_1| \sim 12^\circ$ after which the cumulative numbers
are symmetric. In contrast, the perturbation by the Milky Way bar
produces an asymmetry which persists beyond the region shown here,
although it does become symmetric sufficiently far from the
progenitor. Further observations of the leading arm beyond what is
measured to date could help distinguish between these cases although
significant additional modelling efforts are also needed to understand
the features the bar (and other mechanisms) can produce.

\section{Discussion}

\label{sec:discussion}

\subsection{Width of Pal 5}

In this work we have focused on the stream's centroid and the debris
density. However there is also information that can be gleaned from
the stream width. Ignoring the perturbations we have discussed above,
the change in the stream width is due to the stream fanning out in a
non-spherical potential
\citep[e.g.][]{ibata_et_al_2001,helmi_2004,johnston_et_al_2005,stream_width}. A
constant width indicates a spherical potential and a rapidly
increasing width indicates a flattened (or a triaxial) potential. In
this light, we can re-visit \Figref{fig:pal5_unpert} which compares
the observed properties of Pal 5 with that of an N-body
simulation. The N-body simulation has widths which are broadly
symmetric near the progenitor with both tails exhibiting a modest
fanning out. In contrast, the observed width is nearly constant for
the trailing arm ($\phi_1 > 0^\circ$) but with a more rapid change in
the leading arm ($\phi_1 < 0^\circ$). As with the density, such an
asymmetry is likely a hallmark of a perturbation to Pal 5's stellar
debris which affects each tail differently. If the leading tail is
unperturbed, the large change in the width of the leading arm could
indicate that the potential in the region inhabited by Pal 5 is
flatter than suggested by \texttt{MWPotential2014}. Alternatively, if
the trailing tail is unperturbed, the constant width would be an
indication that the potential is in fact more spherical. In order to
extract information from the width, we would first need determine which tail
has been perturbed. Intriguingly, in \Figref{fig:pal5_dark_flyby} which shows a simulated
stream impacted by two subhaloes, we see that the subhalo impact on the 
trailing arm makes the width constant until $\phi_1 \sim 10^\circ$, suggesting
that perhaps the relatively constant observed width is due to perturbations. 
Interestingly, in the presence of a bar the stream width broadens rapidly (see \figref{fig:pal5_bar_example}) 
suggesting that the width may also help distinguish a subhalo impact from the effect of the bar.

\subsection{Constraining the subhalo properties} \label{sec:subhalo_inference}

If the asymmetry seen in Pal 5 is indeed due to DM substructure, we
can ask what mass range subhalo was responsible for the
perturbation. In \cite{subhalo_properties}, a simplified model of a
stream on a circular orbit was used to show that given three
observables of the stream, e.g. the stream track, the stream density,
and the radial velocity along the stream, there is enough information
to constrain the properties of the subhalo flyby down to a degeneracy
between the subhalo mass and the flyby velocity. In
\Figref{fig:pal5_dark_flyby} we compare the observed properties of Pal
5 with those of an N-body simulation including the effect of two
flybys. We include two dotted vertical lines which show the locations of the gaps. If the Milky Way's gravitational
potential was well understood there should in principle be enough
information to infer the subhalo properties. This inference would also
give the time since impact, allowing us to determine whether either
interaction occurred near or far from the disk. This would allow us to
more confidently attribute the effect to a subhalo or a giant
molecular cloud. However, given the current level of uncertainty in
the Milky Way's mass distribution, (as well as the influence of the
bar/chaos, and the properties of the Pal 5 progenitor), we will leave
this for future work.

To provide a preliminary ball-park estimate, we can use the results of
\cite{number_of_gaps} which computed the characteristic gap size
produced by several different mass subhaloes for the Pal 5
stream. From Fig. 8 of that work, we see that subhaloes with a mass of $10^6 M_\odot, 10^7 M_\odot, 10^8
M_\odot$ would produce gaps with characteristic sizes of $\sim 3.5^\circ,10^\circ,30^\circ$ in Pal 5 respectively, assuming that Pal 5 has an
age of 3.4 Gyr. While the two possible gaps in Pal 5 shown in
\Figref{fig:pal5_dark_flyby} have sizes of $\sim 2^\circ$ and $\sim 9^\circ$, the
analysis in \cite{number_of_gaps} neglects the stretching and
contraction of the stream gap due to the eccentric orbit of Pal
5. Since Pal 5 is currently near apocenter, the gaps will be
contracted. This suggests that the subhalo mass for the gap in the leading arm would be in the $10^6-10^7 M_\odot$ range while the subhalo mass for the gap in the trailing arm would be in the $10^7-10^8 M_\odot$ range.

\subsection{Comparison with expected number of gaps}

There are arguably two non-epicyclic features in the Pal 5 stream,
namely an underdensity between $\phi_1 \sim 3-12^\circ$ and a
$2^\circ$ underdensity at $\phi_1 \sim -3^\circ$. Although we have
argued that there are other possible explanations for these features,
we can ask if they are consistent with expectations from
substructure. In \cite{number_of_gaps}, it was argued that the
expected background of subhaloes should produce $0.7$ gaps deeper than
$75\%$ in the Pal 5 stream, where the depth is relative to the
unperturbed stream. In terms of the $S$ statistic discussed in \Secref{sec:results}, this would approximately correspond to $S$ smaller than 0.75. Note that we cannot measure the depth of the gaps observed in Pal 5 this way since we do not know the unperturbed stream density. However, we argue that a gap deeper than $75\%$ should be relatively easy to detect so the prediction should be thought of as 0.7 gaps which are detectable. The gaps from subhaloes can span a wide range of
sizes from a few degrees up to a few tens of degrees with a
characteristic size of roughly $8^\circ$ \citep{number_of_gaps}. Note,
however, that the prediction from \cite{number_of_gaps} overestimates
the gap size since it assumes that the stream is on a circular orbit
while Pal 5 is near apocenter which compresses the stream and the
gap. Thus, both features are consistent with the expected number and
size of gaps from subhaloes.

In \cite{amorisco_et_al_2016_gmcs}, the expected number of gaps from
GMCs was studied and they found 0.65 gaps deeper than $\sim 71\%$. As
discussed in \Secref{sec:GMCs}, due to their lower masses, the GMCs
create smaller gaps. So, while a perturbation from a GMC could explain
the small gap seen near $\phi_1 \sim -3^\circ$, it is very unlikely
that a GMC could produce the feature seen at $\phi_1 \sim
8^\circ$. Thus, the number and sizes of the features appear consistent
with the combined effect of both subhaloes and GMCs. We note that
\cite{amorisco_et_al_2016_gmcs} assumed a pericenter of 8 kpc for Pal
5. However, as we saw in \Secref{sec:pal5orbit}, the pericenter is
uncertain due to underlying uncertainty about the Milky Way
potential. Since the number density of GMCs depends on distance from
the Milky Way's center \citep[e.g.][]{roman-duval_et_al_2010}, the
predicted number of gaps can change if the pericenter of Pal 5 is
substantially different from 8 kpc.

Finally, in \cite{bovy_erkal_sanders_2016} the authors showed that
the power spectrum and bi-spectrum of stream observables are sensitive
to the amount of substructure. Using the stream density reported in \cite{ibata_et_al_pal5}
they found that the density fluctuations in Pal 5 are consistent with a population of subhaloes 1.5-9 times 
more numerous than expected in $\Lambda$CDM \citep[accounting for a factor of 3 depletion of the subhaloes by the Milky Way disk, e.g.][]{donghia_et_al_2010}.
This is in agreement with the direct gap counting where we expected 0.7 gaps from \cite{number_of_gaps} but found 2 gaps which suggests $\sim 3$ 
times the population of subhaloes. However, since the uncertainties are large both estimates are still consistent with $\Lambda$CDM. The density
and stream track in this work can also be used with the technique of \cite{bovy_erkal_sanders_2016} however care should be taken
to apply the same cubic spline technique on the simulated streams since the measured density and stream track have been somewhat smoothed over and hence will lack
power on small scales.

\subsection{Putting the puzzle together} 

Thanks to the staggering progress in the last ten years, there exist
clear predictions as to what might have produced the density
variations we have detected in the Pal~5 stream. Interestingly, a
different phenomenon (or even a set of phenomena) is involved in the
production of density fluctuations at each angular scale. For example,
the sharp, small-scale density peak at $\phi_1\sim-0\fdg8$ (as well as
its less significant counterpart in the trailing tail) is likely the
result of the epicyclic bunching. The epicyclic overdensities have
been studied extensively in the literature \citep[see
  e.g.][]{combes1999,capuzzo2005,kuepper_et_al_2008,kuepper_et_al_2010,Ma2012,amorisco_2014}
and are the result of the similarity of the orbits of the unbound
stars. As further confirmation, these observed peaks are closely
aligned with the peaks seen in our simulation at $\phi_1 =
-0.7^\circ,+0.6^\circ$ (\figref{fig:pal5_unpert}). Additionally, we
have also studied the prominence of the epicyclic overdensities as a
function of the progenitor's orbital phase and found that near the
apocentre, i.e. close to the current position of the cluster, the
epicycles in the N-body simulations are strongly suppressed. Moreover,
when transforming from counts of particles in the simulations to the
observable number of stars, the small-amplitude epicyclic bunches are
further reduced in significance, as demonstrated recently by
\citet{thomas_et_al_2016}. It is therefore unsurprising that even in
data of such depth, only the epicycles nearest to the progenitor are
detected.

While the small-scale bunching due to the epicyclic feathering is a
generic feature of the stream production in any gravitational
potential, the other larger scale density perturbations detected here,
and in particular the asymmetry between the trailing and leading tails
are highly unlikely in a smooth and static potential.

Dark matter subhaloes have long been predicted to cause damage to the
stellar streams \citep[see
  e.g.][]{ibata_et_al_2002,johnston_et_al_2002}. More recently, the
expectation as to the shape and size of stream perturbations caused by
subhaloes has finally crystallized. First, it is now quite clear that
the flyby does not only produce a depletion in the stream density
around the impact point, but also causes stars to pile up at either
side of the gap \citep[see e.g.][]{carlberg_2009,
  three_phases}. Typically, in the numerical experiments, the
perturbed parts of the stream are chosen to be sufficiently far away
from the progenitor. These sections of tidal tails are characterized
by relatively uniform debris densities. Therefore, the density bumps
on either side of the gap have broadly similar strength with slight asymmetries in peak height possible for flybys with an impact parameter \citep[e.g.][]{three_phases}. Closer to the
progenitor, however, non-zero density gradients are expected, with the
number of stars normally decaying away with distance from the Lagrange
points. Accordingly, for stream perturbations in the vicinity of the
parent satellite, significantly more stars might gather on the side of
the gap which is nearest to the progenitor. Second, there exists a
lower bound to the size of the stream gap a subhalo can tear. It takes
time for the affected stars to move away from the impact point, and by
the time the stream density drops to detectable levels, the gap has
already grown substantially. \citet{number_of_gaps} carefully
considered the combination of factors affecting the gap evolution and
gave gap sizes expected for perturbers of different mass. In their
picture, the characteristic gap size created by DM subhaloes of $10^7
M_{\odot}$ is $\sim10^{\circ}$.

Guided by these expectations, we put forward a hypothesis that the
dramatic density asymmetry between the trailing and leading debris is
the result of a flyby of a DM subhalo. More precisely, the projection
of the perturber's impact point lies somewhere between 8 and 10
degrees, which corresponds to the center of the density depletion of an
appreciable extent. The gap is accompanied by a substantial pile-up of
stars at $\phi_1=3^{\circ}$ on the side of the progenitor. We thus
estimate that the gap size is approximately $9^{\circ}$, which is
supported by the fact that the stream density recovers to normal
levels at around $\phi_1=12^{\circ}$. To further demonstrate the
feasibility of our conjecture, we run a series of numerical
simulations of the disruption of a Pal~5-like cluster. The unperturbed
density of a stream produced in a static and smooth potential is
always symmetric (see \figref{fig:pal5_unpert}), thus emphasizing both the peak at $\phi_1=3^{\circ}$
and the dip between $\phi_1=5^{\circ}$ and
$\phi_1=11^{\circ}$. Additional confirmation is provided by a
simulation of a DM flyby which appears to re-shape the stream density
profile into one closely resembling the observations (see \figref{fig:pal5_dark_flyby}). Based on the
analytic predictions of \citet{number_of_gaps} and the numerical
experiments reported here, we believe that the mass of the perturber
interaction which could have produced the observed density
fluctuations must be in the range of $10^7-10^8 M_\odot$.

We considered several other mechanisms that could possibly produce the
observed Pal~5 stream asymmetry. For example, we studied the effect of
the Milky Way's rotating bar whose importance for stellar streams has
been highlighted in \cite{kohei_rotating_bar} and
\cite{price_whelan_et_al_chaotic_fanning}.  In particular, we
simulated the disruption of a large number of Pal 5 realizations by
sampling the cluster's proper motion and found that the presence of a
bar can indeed create significant asymmetries in the stream (e.g.
\figref{fig:pal5_bar_example}). We have established that this effect
depends sensitively on the pattern speed of the bar and on the orbit of
Pal 5, similar to what was found in \cite{kohei_rotating_bar}.  Thus,
better modelling of the bar and the Milky Way potential is needed to
conclusively determine if the features in Pal 5 are due to the
bar. Another important effect is that of chaos which can be arise in both a
smooth static potential as well as the rotating bar
\citep[e.g.][]{price_whelan_et_al_chaotic_fanning}.  \cite{pearson_et_al_pal5} evolved a Pal 5
analogue in a triaxial potential and found that a seemingly mild chaos
could give rise to a significantly perturbed stream.  However, the
particular model they considered did not match Pal 5. Most
importantly, the best fit to the Pal 5 stream data available at the
time was found in an axisymmetric potential which exhibited no
chaos. Thus, it is unclear whether chaos could create the observed density variations or asymmetry while maintaining a thin stream. To conclude, based on the evidence in hand, the conspicuous
features in the trailing tail of Pal~5 could easily be produced by a
flyby of a DM subhalo or by the Galaxy's bar. Less likely, albeit impossible to rule
out at the present, is the possibility that the stream asymmetry was caused by chaos in the Milky Way's
gravitational potential.

We stress that this substructure is not necessarily a dark subhalo,
but could also be a giant molecular clouds as suggested in
\cite{amorisco_et_al_2016_gmcs}. However, we also note that the gap
sizes found in that work are smaller than the proposed gap in the Pal
5's trailing tail. More precisely, given that the typical masses of
the GMCs are less than $10^6 M_{\odot}$, the characteristic size of
the gaps they tend to produce are less than 3 degrees. Interestingly,
this is a good match to the size of the density depletion we detect in
the leading tail at $\phi_1=-3^{\circ}$. This 70\% dip in the star
counts (as measured from trough to peak) is clearly less prominent than the spectacular ripple in the
trailing arm, but, nonetheless, carries a significance of at least
5$\sigma$. We conjecture that either a GMC or a DM subhalo, both with
a mass around $10^6-10^7 M_{\odot}$ could be responsible for this
gap. However, unfortunately, a straightforward interpretation of the
nature of this $\sim 2^{\circ}$ wide gap is not possible at the
moment. This is because an alternative theory can be put forward to
explain it. Namely, a substantial change in the Pal~5's pericentric
distance could cause the stripping rate to vary significantly between
the bouts of disruption. The varying stripping rate would induce
features in the debris distribution, that might look similar to the
feature discussed. Of course, the debris density waves due to variable
stripping efficiency must by symmetric with respect to the
progenitor's position, which is not observed. However, the presence of
a large density fluctuation in the trailing tail prevents us from
testing whether there exists a counterpart feature there.

\section{Conclusions}

\label{sec:conclusion}

In this paper we have analyzed new, high quality photometry of the
Palomar~5 stellar stream published recently by
\citet{ibata_et_al_pal5}. In order to fully take advantage of this
superb data we have developed a novel non-parametric method to extract
the stream properties such as the density along the tails, the
centroid track of the debris distribution on the sky, as well as the
stream width. Our probabilistic method is adaptive in the sense that
the model complexity is not fixed {\it a priori}, but rather is driven
by the data in hand. The combination of the quality of the data and
the power of the modelling technique yields an exquisite determination
of the stream properties.

For the first time, we measure significant changes in the stream width
and show that the debris cross-section varies differently along each
tail. We also detect dramatic stream density fluctuations on a variety
of angular scales. First, on the scale of a fraction of a degree,
density spikes are detected very close to the progenitor. Second,
further away from the Pal~5 cluster, at $\phi_1=-3^{\circ}$ in the
leading tail, a density depletion approximately 2 degrees across is
measured. It is accompanied by two low-level bumps on either
side. Finally, the trailing tail exhibits a prominent density
enhancement at a distance of $\sim$ 3 degrees from the progenitor,
followed by a smooth drop in star counts, observable for some 8
degrees along the tail, and then a mild bump at $\phi_1 \sim
12^\circ$. As we demonstrate with utmost clarity, the remarkable rise
and fall of the trailing debris density do not have counterparts in
the leading tail.

We interpret the small-scale debris pile-ups in the vicinity of the
progenitor as epicyclic overdensities, and conjecture that the two
larger scale density perturbations are induced by interactions with
small substructure. If dark matter subhaloes were the cause of these
stream wrinkles, then their masses are of order of $10^6-10^7$ and
$10^7-10^8 M_{\odot}$. Impressively, the size of the larger gap
discovered here agrees well with the characteristic gap scale expected
in the presence of $\Lambda$CDM sub-structure with masses between
$10^5$ and $10^9 M_{\odot}$ as predicted in \citet{number_of_gaps}. It
is not easy to over-emphasize the importance of this discovery if the
sub-structure that wrought havoc in the stream was non-baryonic. In
fact, a subhalo in the $10^6-10^7 M_\odot$ range would increase the
lower bound on the warm dark matter particle mass to $>9$-$18$ keV
\citep{viel_et_al_2013}.

Note, however, that currently we cannot rule out other plausible
explanations, such as the effect of the rotating bar, impacts by GMCs,
or other complexities in the gravitational potential of the Galaxy,
that would lead to mild orbital chaos or would induce substantial
variations in the stripping rate of the cluster. In order to
distinguish between these mechanisms, we must predict the features
each of these can create in the Pal~5 tails.  For example, as was
shown in \cite{subhalo_properties}, the flyby of a subhalo produces an
almost unique signature which can be used to infer the subhalo
properties.  The precise signatures of the bar, chaos and the
pericentre wobble are likely very different, so with additional data,
especially the improved radial velocity measurements expected from
WEAVE, 4MOST and DESI, it should be possible to determine the culprit.

\section*{Acknowledgements}

We are grateful to \citet{ibata_et_al_pal5} for making their catalogs
publicly available. We thank the Streams group at Cambridge for valuable discussions. The research leading to these results has received
funding from the European Research Council under the European Union's
Seventh Framework Programme (FP/2007-2013)/ERC Grant Agreement
no. 308024. SK thanks the United Kingdom Science and Technology
Council (STFC) for the award of Ernest Rutherford fellowship (grant
number ST/N004493/1). Finally, we thank the referee for a constructive report and useful suggestions.

\bibliographystyle{mnras}
\bibliography{citations_pal5_asym,citations_pal5_asym1}

\begin{thebibliography}{}
\makeatletter
\relax
\def\mn@urlcharsother{\let\do\@makeother \do\$\do\&\do\#\do\^\do\_\do\%\do\~}
\def\mn@doi{\begingroup\mn@urlcharsother \@ifnextchar [ {\mn@doi@}
  {\mn@doi@[]}}
\def\mn@doi@[#1]#2{\def\@tempa{#1}\ifx\@tempa\@empty \href
  {http://dx.doi.org/#2} {doi:#2}\else \href {http://dx.doi.org/#2} {#1}\fi
  \endgroup}
\def\mn@eprint#1#2{\mn@eprint@#1:#2::\@nil}
\def\mn@eprint@arXiv#1{\href {http://arxiv.org/abs/#1} {{\tt arXiv:#1}}}
\def\mn@eprint@dblp#1{\href {http://dblp.uni-trier.de/rec/bibtex/#1.xml}
  {dblp:#1}}
\def\mn@eprint@#1:#2:#3:#4\@nil{\def\@tempa {#1}\def\@tempb {#2}\def\@tempc
  {#3}\ifx \@tempc \@empty \let \@tempc \@tempb \let \@tempb \@tempa \fi \ifx
  \@tempb \@empty \def\@tempb {arXiv}\fi \@ifundefined
  {mn@eprint@\@tempb}{\@tempb:\@tempc}{\expandafter \expandafter \csname
  mn@eprint@\@tempb\endcsname \expandafter{\@tempc}}}

\bibitem[\protect\citeauthoryear{{Akaike}}{{Akaike}}{1974}]{Akaike1974}
{Akaike} H.,  1974, IEEE Transactions on Automatic Control,
  \href{http://adsabs.harvard.edu/abs/1974ITAC...19..716A}{19, 716}

\bibitem[\protect\citeauthoryear{{Allen}, {Moreno}  \& {Pichardo}}{{Allen}
  et~al.}{2006}]{allen_et_al_2006}
{Allen} C.,  {Moreno} E.,   {Pichardo} B.,  2006, \mn@doi [\apj]
  {10.1086/508676},
  \href{http://adsabs.harvard.edu/abs/2006ApJ...652.1150A}{652, 1150}

\bibitem[\protect\citeauthoryear{{Amorisco}}{{Amorisco}}{2015}]{amorisco_2014}
{Amorisco} N.~C.,  2015, \mn@doi [\mnras] {10.1093/mnras/stv648},
  \href{http://adsabs.harvard.edu/abs/2015MNRAS.450..575A}{450, 575}

\bibitem[\protect\citeauthoryear{{Amorisco}, {G{\'o}mez}, {Vegetti}  \&
  {White}}{{Amorisco} et~al.}{2016}]{amorisco_et_al_2016_gmcs}
{Amorisco} N.~C.,  {G{\'o}mez} F.~A.,  {Vegetti} S.,   {White} S.~D.~M.,  2016,
  \mn@doi [\mnras] {10.1093/mnrasl/slw148},
  \href{http://adsabs.harvard.edu/abs/2016MNRAS.463L..17A}{463, L17}

\bibitem[\protect\citeauthoryear{{Bellazzini}, {Bragaglia}, {Carretta},
  {Gratton}, {Lucatello}, {Catanzaro}  \& {Leone}}{{Bellazzini}
  et~al.}{2012}]{Bellazzini_et_al_2012}
{Bellazzini} M.,  {Bragaglia} A.,  {Carretta} E.,  {Gratton} R.~G.,
  {Lucatello} S.,  {Catanzaro} G.,   {Leone} F.,  2012, \mn@doi [\aap]
  {10.1051/0004-6361/201118056},
  \href{http://adsabs.harvard.edu/abs/2012A%26A...538A..18B}{538, A18}

\bibitem[\protect\citeauthoryear{{Belokurov} et~al.,}{{Belokurov}
  et~al.}{2014}]{belokurov_sgr_precess}
{Belokurov} V.,  et~al., 2014, \mn@doi [\mnras] {10.1093/mnras/stt1862},
  \href{http://adsabs.harvard.edu/abs/2014MNRAS.437..116B}{437, 116}

\bibitem[\protect\citeauthoryear{{Bernard} et~al.,}{{Bernard}
  et~al.}{2014}]{ophiuchus_disc}
{Bernard} E.~J.,  et~al., 2014, \mn@doi [\mnras] {10.1093/mnrasl/slu089},
  \href{http://adsabs.harvard.edu/abs/2014MNRAS.443L..84B}{443, L84}

\bibitem[\protect\citeauthoryear{{Binney} \& {Tremaine}}{{Binney} \&
  {Tremaine}}{2008}]{bt2008}
{Binney} J.,  {Tremaine} S.,  2008, {Galactic Dynamics: Second Edition}.
Princeton University Press

\bibitem[\protect\citeauthoryear{{Blum} et~al.,}{{Blum}
  et~al.}{2016}]{Blum2016}
{Blum} R.~D.,  et~al., 2016, in American Astronomical Society Meeting
  Abstracts. p. 317.01

\bibitem[\protect\citeauthoryear{{Bovy}}{{Bovy}}{2015}]{bovy_galpy}
{Bovy} J.,  2015, \mn@doi [\apjs] {10.1088/0067-0049/216/2/29},
  \href{http://adsabs.harvard.edu/abs/2015ApJS..216...29B}{216, 29}

\bibitem[\protect\citeauthoryear{{Bovy}}{{Bovy}}{2016}]{bovy_2016}
{Bovy} J.,  2016, \mn@doi [Physical Review Letters]
  {10.1103/PhysRevLett.116.121301},
  \href{http://adsabs.harvard.edu/abs/2016PhRvL.116l1301B}{116, 121301}

\bibitem[\protect\citeauthoryear{{Bovy} et~al.,}{{Bovy}
  et~al.}{2012}]{bovy_et_al_2012}
{Bovy} J.,  et~al., 2012, \mn@doi [\apj] {10.1088/0004-637X/759/2/131},
  \href{http://adsabs.harvard.edu/abs/2012ApJ...759..131B}{759, 131}

\bibitem[\protect\citeauthoryear{{Bovy}, {Erkal}  \& {Sanders}}{{Bovy}
  et~al.}{2017}]{bovy_erkal_sanders_2016}
{Bovy} J.,  {Erkal} D.,   {Sanders} J.~L.,  2017, \mn@doi [\mnras]
  {10.1093/mnras/stw3067},
  \href{http://adsabs.harvard.edu/abs/2017MNRAS.466..628B}{466, 628}

\bibitem[\protect\citeauthoryear{{Bowden}, {Belokurov}  \& {Evans}}{{Bowden}
  et~al.}{2015}]{bowden_et_al_gd1}
{Bowden} A.,  {Belokurov} V.,   {Evans} N.~W.,  2015, \mn@doi [\mnras]
  {10.1093/mnras/stv285},
  \href{http://adsabs.harvard.edu/abs/2015MNRAS.449.1391B}{449, 1391}

\bibitem[\protect\citeauthoryear{Bronshtein, Semendyayev, Musiol  \&
  M{\"u}hlig}{Bronshtein et~al.}{2007}]{Bronshtein2007}
Bronshtein I.,  Semendyayev K.,  Musiol G.,   M{\"u}hlig H.,  2007, Handbook of
  Mathematics.
Springer Berlin Heidelberg, \url
  {https://books.google.co.uk/books?id=gCgOoMpluh8C}

\bibitem[\protect\citeauthoryear{{Capuzzo Dolcetta}, {Di Matteo}  \&
  {Miocchi}}{{Capuzzo Dolcetta} et~al.}{2005}]{capuzzo2005}
{Capuzzo Dolcetta} R.,  {Di Matteo} P.,   {Miocchi} P.,  2005, \mn@doi [\aj]
  {10.1086/426006},
  \href{http://adsabs.harvard.edu/abs/2005AJ....129.1906C}{129, 1906}

\bibitem[\protect\citeauthoryear{{Carlberg}}{{Carlberg}}{2009}]{carlberg_2009}
{Carlberg} R.~G.,  2009, \mn@doi [\apjl] {10.1088/0004-637X/705/2/L223},
  \href{http://adsabs.harvard.edu/abs/2009ApJ...705L.223C}{705, L223}

\bibitem[\protect\citeauthoryear{{Carlberg}}{{Carlberg}}{2012}]{carlberg_2012}
{Carlberg} R.~G.,  2012, \mn@doi [\apj] {10.1088/0004-637X/748/1/20},
  \href{http://adsabs.harvard.edu/abs/2012ApJ...748...20C}{748, 20}

\bibitem[\protect\citeauthoryear{{Carlberg}}{{Carlberg}}{2013}]{carlberg_2013}
{Carlberg} R.~G.,  2013, \mn@doi [\apj] {10.1088/0004-637X/775/2/90},
  \href{http://adsabs.harvard.edu/abs/2013ApJ...775...90C}{775, 90}

\bibitem[\protect\citeauthoryear{{Carlberg}, {Grillmair}  \&
  {Hetherington}}{{Carlberg} et~al.}{2012}]{carlberg_pal5_2012}
{Carlberg} R.~G.,  {Grillmair} C.~J.,   {Hetherington} N.,  2012, \mn@doi
  [\apj] {10.1088/0004-637X/760/1/75},
  \href{http://adsabs.harvard.edu/abs/2012ApJ...760...75C}{760, 75}

\bibitem[\protect\citeauthoryear{{Combes}, {Leon}  \& {Meylan}}{{Combes}
  et~al.}{1999}]{combes1999}
{Combes} F.,  {Leon} S.,   {Meylan} G.,  1999, \aap,
  \href{http://adsabs.harvard.edu/abs/1999A%26A...352..149C}{352, 149}

\bibitem[\protect\citeauthoryear{{D'Onghia}, {Springel}, {Hernquist}  \&
  {Keres}}{{D'Onghia} et~al.}{2010}]{donghia_et_al_2010}
{D'Onghia} E.,  {Springel} V.,  {Hernquist} L.,   {Keres} D.,  2010, \mn@doi
  [\apj] {10.1088/0004-637X/709/2/1138},
  \href{http://adsabs.harvard.edu/abs/2010ApJ...709.1138D}{709, 1138}

\bibitem[\protect\citeauthoryear{{Dalton} et~al.,}{{Dalton}
  et~al.}{2012}]{weave}
{Dalton} G.,  et~al., 2012, in Ground-based and Airborne Instrumentation for
  Astronomy IV. p. 84460P, \mn@doi{10.1117/12.925950}

\bibitem[\protect\citeauthoryear{{Dehnen}, {Odenkirchen}, {Grebel}  \&
  {Rix}}{{Dehnen} et~al.}{2004}]{dehnen2004}
{Dehnen} W.,  {Odenkirchen} M.,  {Grebel} E.~K.,   {Rix} H.-W.,  2004, \mn@doi
  [\aj] {10.1086/383214},
  \href{http://adsabs.harvard.edu/abs/2004AJ....127.2753D}{127, 2753}

\bibitem[\protect\citeauthoryear{{Diemand}, {Kuhlen}, {Madau}, {Zemp}, {Moore},
  {Potter}  \& {Stadel}}{{Diemand} et~al.}{2008}]{Diemand2008}
{Diemand} J.,  {Kuhlen} M.,  {Madau} P.,  {Zemp} M.,  {Moore} B.,  {Potter} D.,
    {Stadel} J.,  2008, \mn@doi [\nat] {10.1038/nature07153},
  \href{http://adsabs.harvard.edu/abs/2008Natur.454..735D}{454, 735}

\bibitem[\protect\citeauthoryear{{Dotter}, {Sarajedini}  \&
  {Anderson}}{{Dotter} et~al.}{2011}]{dotter_et_al_2011}
{Dotter} A.,  {Sarajedini} A.,   {Anderson} J.,  2011, \mn@doi [\apj]
  {10.1088/0004-637X/738/1/74},
  \href{http://adsabs.harvard.edu/abs/2011ApJ...738...74D}{738, 74}

\bibitem[\protect\citeauthoryear{{Erkal} \& {Belokurov}}{{Erkal} \&
  {Belokurov}}{2015a}]{three_phases}
{Erkal} D.,  {Belokurov} V.,  2015a, \mn@doi [\mnras] {10.1093/mnras/stv655},
  \href{http://adsabs.harvard.edu/abs/2015MNRAS.450.1136E}{450, 1136}

\bibitem[\protect\citeauthoryear{{Erkal} \& {Belokurov}}{{Erkal} \&
  {Belokurov}}{2015b}]{subhalo_properties}
{Erkal} D.,  {Belokurov} V.,  2015b, \mn@doi [\mnras] {10.1093/mnras/stv2122},
  \href{http://adsabs.harvard.edu/abs/2015MNRAS.454.3542E}{454, 3542}

\bibitem[\protect\citeauthoryear{{Erkal}, {Sanders}  \& {Belokurov}}{{Erkal}
  et~al.}{2016a}]{stream_width}
{Erkal} D.,  {Sanders} J.~L.,   {Belokurov} V.,  2016a, \mn@doi [\mnras]
  {10.1093/mnras/stw1400},
  \href{http://adsabs.harvard.edu/abs/2016MNRAS.461.1590E}{461, 1590}

\bibitem[\protect\citeauthoryear{{Erkal}, {Belokurov}, {Bovy}  \&
  {Sanders}}{{Erkal} et~al.}{2016b}]{number_of_gaps}
{Erkal} D.,  {Belokurov} V.,  {Bovy} J.,   {Sanders} J.~L.,  2016b, \mn@doi
  [\mnras] {10.1093/mnras/stw1957},
  \href{http://adsabs.harvard.edu/abs/2016MNRAS.463..102E}{463, 102}

\bibitem[\protect\citeauthoryear{{Eyre}}{{Eyre}}{2010}]{Eyre2010}
{Eyre} A.,  2010, \mn@doi [\mnras] {10.1111/j.1365-2966.2009.16234.x},
  \href{http://adsabs.harvard.edu/abs/2010MNRAS.403.1999E}{403, 1999}

\bibitem[\protect\citeauthoryear{{Eyre} \& {Binney}}{{Eyre} \&
  {Binney}}{2011}]{eyre_binney_2011}
{Eyre} A.,  {Binney} J.,  2011, \mn@doi [\mnras]
  {10.1111/j.1365-2966.2011.18270.x},
  \href{http://adsabs.harvard.edu/abs/2011MNRAS.413.1852E}{413, 1852}

\bibitem[\protect\citeauthoryear{{Fabricius} et~al.,}{{Fabricius}
  et~al.}{2014}]{fabricius_et_al_2014}
{Fabricius} M.~H.,  et~al., 2014, \mn@doi [\apjl]
  {10.1088/2041-8205/787/2/L26},
  \href{http://adsabs.harvard.edu/abs/2014ApJ...787L..26F}{787, L26}

\bibitem[\protect\citeauthoryear{{Fritz} \& {Kallivayalil}}{{Fritz} \&
  {Kallivayalil}}{2015}]{fritz_kallivayalil_pal5}
{Fritz} T.~K.,  {Kallivayalil} N.,  2015, \mn@doi [\apj]
  {10.1088/0004-637X/811/2/123},
  \href{http://adsabs.harvard.edu/abs/2015ApJ...811..123F}{811, 123}

\bibitem[\protect\citeauthoryear{Gelman \& Rubin}{Gelman \&
  Rubin}{1992}]{Gelman1992}
Gelman A.,  Rubin D.~B.,  1992, \mn@doi [Statist. Sci.]
  {10.1214/ss/1177011136}, 7, 457

\bibitem[\protect\citeauthoryear{Gelman, Hwang  \& Vehtari}{Gelman
  et~al.}{2014}]{Gelman2014}
Gelman A.,  Hwang J.,   Vehtari A.,  2014, Statistics and Computing, 24, 997

\bibitem[\protect\citeauthoryear{{Gerhard}}{{Gerhard}}{2011}]{gerhard_review}
{Gerhard} O.,  2011, Memorie della Societa Astronomica Italiana Supplementi,
  \href{http://adsabs.harvard.edu/abs/2011MSAIS..18..185G}{18, 185}

\bibitem[\protect\citeauthoryear{{Gibbons}, {Belokurov}  \& {Evans}}{{Gibbons}
  et~al.}{2014}]{gibbons_et_al_2014}
{Gibbons} S.~L.~J.,  {Belokurov} V.,   {Evans} N.~W.,  2014, \mn@doi [\mnras]
  {10.1093/mnras/stu1986},
  \href{http://adsabs.harvard.edu/abs/2014MNRAS.445.3788G}{445, 3788}

\bibitem[\protect\citeauthoryear{{Gibbons}, {Belokurov}  \& {Evans}}{{Gibbons}
  et~al.}{2017}]{Gibbons2016}
{Gibbons} S.~L.~J.,  {Belokurov} V.,   {Evans} N.~W.,  2017, \mn@doi [\mnras]
  {10.1093/mnras/stw2328},
  \href{http://adsabs.harvard.edu/abs/2017MNRAS.464..794G}{464, 794}

\bibitem[\protect\citeauthoryear{{Gnedin} \& {Ostriker}}{{Gnedin} \&
  {Ostriker}}{1997}]{gnedin_ostriker_1997}
{Gnedin} O.~Y.,  {Ostriker} J.~P.,  1997, \apj,
  \href{http://adsabs.harvard.edu/abs/1997ApJ...474..223G}{474, 223}

\bibitem[\protect\citeauthoryear{{G{\'o}rski}, {Hivon}, {Banday}, {Wandelt},
  {Hansen}, {Reinecke}  \& {Bartelmann}}{{G{\'o}rski}
  et~al.}{2005}]{Gorski2005}
{G{\'o}rski} K.~M.,  {Hivon} E.,  {Banday} A.~J.,  {Wandelt} B.~D.,  {Hansen}
  F.~K.,  {Reinecke} M.,   {Bartelmann} M.,  2005, \mn@doi [\apj]
  {10.1086/427976},
  \href{http://adsabs.harvard.edu/abs/2005ApJ...622..759G}{622, 759}

\bibitem[\protect\citeauthoryear{{Grillmair}}{{Grillmair}}{2009}]{Grillmair2009}
{Grillmair} C.~J.,  2009, \mn@doi [\apj] {10.1088/0004-637X/693/2/1118},
  \href{http://adsabs.harvard.edu/abs/2009ApJ...693.1118G}{693, 1118}

\bibitem[\protect\citeauthoryear{{Grillmair} \& {Dionatos}}{{Grillmair} \&
  {Dionatos}}{2006}]{gd_pal5}
{Grillmair} C.~J.,  {Dionatos} O.,  2006, \mn@doi [\apjl] {10.1086/503744},
  \href{http://adsabs.harvard.edu/abs/2006ApJ...641L..37G}{641, L37}

\bibitem[\protect\citeauthoryear{{Gross} \& {Vitells}}{{Gross} \&
  {Vitells}}{2010}]{Gross2010}
{Gross} E.,  {Vitells} O.,  2010, \mn@doi [European Physical Journal C]
  {10.1140/epjc/s10052-010-1470-8},
  \href{http://adsabs.harvard.edu/abs/2010EPJC...70..525G}{70, 525}

\bibitem[\protect\citeauthoryear{{Hattori}, {Erkal}  \& {Sanders}}{{Hattori}
  et~al.}{2016}]{kohei_rotating_bar}
{Hattori} K.,  {Erkal} D.,   {Sanders} J.~L.,  2016, \mn@doi [\mnras]
  {10.1093/mnras/stw1006},
  \href{http://adsabs.harvard.edu/abs/2016MNRAS.460..497H}{460, 497}

\bibitem[\protect\citeauthoryear{{Helmi}}{{Helmi}}{2004}]{helmi_2004}
{Helmi} A.,  2004, \mn@doi [\mnras] {10.1111/j.1365-2966.2004.07812.x},
  \href{http://adsabs.harvard.edu/abs/2004MNRAS.351..643H}{351, 643}

\bibitem[\protect\citeauthoryear{{Helmi} \& {White}}{{Helmi} \&
  {White}}{1999}]{helmi_white_1999}
{Helmi} A.,  {White} S.~D.~M.,  1999, \mn@doi [\mnras]
  {10.1046/j.1365-8711.1999.02616.x},
  \href{http://adsabs.harvard.edu/abs/1999MNRAS.307..495H}{307, 495}

\bibitem[\protect\citeauthoryear{{Ibata}, {Lewis}, {Irwin}, {Totten}  \&
  {Quinn}}{{Ibata} et~al.}{2001}]{ibata_et_al_2001}
{Ibata} R.,  {Lewis} G.~F.,  {Irwin} M.,  {Totten} E.,   {Quinn} T.,  2001,
  \mn@doi [\apj] {10.1086/320060},
  \href{http://adsabs.harvard.edu/abs/2001ApJ...551..294I}{551, 294}

\bibitem[\protect\citeauthoryear{{Ibata}, {Lewis}, {Irwin}  \& {Quinn}}{{Ibata}
  et~al.}{2002}]{ibata_et_al_2002}
{Ibata} R.~A.,  {Lewis} G.~F.,  {Irwin} M.~J.,   {Quinn} T.,  2002, \mn@doi
  [\mnras] {10.1046/j.1365-8711.2002.05358.x},
  \href{http://adsabs.harvard.edu/abs/2002MNRAS.332..915I}{332, 915}

\bibitem[\protect\citeauthoryear{{Ibata}, {Lewis}  \& {Martin}}{{Ibata}
  et~al.}{2016}]{ibata_et_al_pal5}
{Ibata} R.~A.,  {Lewis} G.~F.,   {Martin} N.~F.,  2016, \mn@doi [\apj]
  {10.3847/0004-637X/819/1/1},
  \href{http://adsabs.harvard.edu/abs/2016ApJ...819....1I}{819, 1}

\bibitem[\protect\citeauthoryear{{Ishigaki}, {Hwang}, {Chiba}  \&
  {Aoki}}{{Ishigaki} et~al.}{2016}]{ishigaki_et_al_2016}
{Ishigaki} M.~N.,  {Hwang} N.,  {Chiba} M.,   {Aoki} W.,  2016, \mn@doi [\apj]
  {10.3847/0004-637X/823/2/157},
  \href{http://adsabs.harvard.edu/abs/2016ApJ...823..157I}{823, 157}

\bibitem[\protect\citeauthoryear{{Johnston}}{{Johnston}}{1998}]{johnston_1998}
{Johnston} K.~V.,  1998, \mn@doi [\apj] {10.1086/305273},
  \href{http://adsabs.harvard.edu/abs/1998ApJ...495..297J}{495, 297}

\bibitem[\protect\citeauthoryear{{Johnston}, {Sigurdsson}  \&
  {Hernquist}}{{Johnston} et~al.}{1999}]{johnston_et_al_1999_2}
{Johnston} K.~V.,  {Sigurdsson} S.,   {Hernquist} L.,  1999, \mn@doi [\mnras]
  {10.1046/j.1365-8711.1999.02200.x},
  \href{http://adsabs.harvard.edu/abs/1999MNRAS.302..771J}{302, 771}

\bibitem[\protect\citeauthoryear{{Johnston}, {Spergel}  \& {Haydn}}{{Johnston}
  et~al.}{2002}]{johnston_et_al_2002}
{Johnston} K.~V.,  {Spergel} D.~N.,   {Haydn} C.,  2002, \mn@doi [\apj]
  {10.1086/339791},
  \href{http://adsabs.harvard.edu/abs/2002ApJ...570..656J}{570, 656}

\bibitem[\protect\citeauthoryear{{Johnston}, {Law}  \& {Majewski}}{{Johnston}
  et~al.}{2005}]{johnston_et_al_2005}
{Johnston} K.~V.,  {Law} D.~R.,   {Majewski} S.~R.,  2005, \mn@doi [\apj]
  {10.1086/426777},
  \href{http://adsabs.harvard.edu/abs/2005ApJ...619..800J}{619, 800}

\bibitem[\protect\citeauthoryear{{Just}, {Berczik}, {Petrov}  \&
  {Ernst}}{{Just} et~al.}{2009}]{just_et_al_2009}
{Just} A.,  {Berczik} P.,  {Petrov} M.~I.,   {Ernst} A.,  2009, \mn@doi
  [\mnras] {10.1111/j.1365-2966.2008.14099.x},
  \href{http://adsabs.harvard.edu/abs/2009MNRAS.392..969J}{392, 969}

\bibitem[\protect\citeauthoryear{{Koposov}, {Rix}  \& {Hogg}}{{Koposov}
  et~al.}{2010}]{Koposov2010}
{Koposov} S.~E.,  {Rix} H.-W.,   {Hogg} D.~W.,  2010, \mn@doi [\apj]
  {10.1088/0004-637X/712/1/260},
  \href{http://adsabs.harvard.edu/abs/2010ApJ...712..260K}{712, 260}

\bibitem[\protect\citeauthoryear{{K{\"u}pper}, {MacLeod}  \&
  {Heggie}}{{K{\"u}pper} et~al.}{2008}]{kuepper_et_al_2008}
{K{\"u}pper} A.~H.~W.,  {MacLeod} A.,   {Heggie} D.~C.,  2008, \mn@doi [\mnras]
  {10.1111/j.1365-2966.2008.13323.x},
  \href{http://adsabs.harvard.edu/abs/2008MNRAS.387.1248K}{387, 1248}

\bibitem[\protect\citeauthoryear{{K{\"u}pper}, {Kroupa}, {Baumgardt}  \&
  {Heggie}}{{K{\"u}pper} et~al.}{2010}]{kuepper_et_al_2010}
{K{\"u}pper} A.~H.~W.,  {Kroupa} P.,  {Baumgardt} H.,   {Heggie} D.~C.,  2010,
  \mn@doi [\mnras] {10.1111/j.1365-2966.2009.15690.x},
  \href{http://adsabs.harvard.edu/abs/2010MNRAS.401..105K}{401, 105}

\bibitem[\protect\citeauthoryear{{K{\"u}pper}, {Balbinot}, {Bonaca},
  {Johnston}, {Hogg}, {Kroupa}  \& {Santiago}}{{K{\"u}pper}
  et~al.}{2015}]{kuepper_et_al_pal5}
{K{\"u}pper} A.~H.~W.,  {Balbinot} E.,  {Bonaca} A.,  {Johnston} K.~V.,  {Hogg}
  D.~W.,  {Kroupa} P.,   {Santiago} B.~X.,  2015, \mn@doi [\apj]
  {10.1088/0004-637X/803/2/80},
  \href{http://adsabs.harvard.edu/abs/2015ApJ...803...80K}{803, 80}

\bibitem[\protect\citeauthoryear{{Kuzma}, {Da Costa}, {Keller}  \&
  {Maunder}}{{Kuzma} et~al.}{2015}]{kuzma_et_al_2015}
{Kuzma} P.~B.,  {Da Costa} G.~S.,  {Keller} S.~C.,   {Maunder} E.,  2015,
  \mn@doi [\mnras] {10.1093/mnras/stu2343},
  \href{http://adsabs.harvard.edu/abs/2015MNRAS.446.3297K}{446, 3297}

\bibitem[\protect\citeauthoryear{{Lang}, {Hogg}  \& {Mykytyn}}{{Lang}
  et~al.}{2016}]{Lang2016}
{Lang} D.,  {Hogg} D.~W.,   {Mykytyn} D.,  2016, {The Tractor: Probabilistic
  astronomical source detection and measurement}, Astrophysics Source Code
  Library (\mn@eprint {ascl} {1604.008})

\bibitem[\protect\citeauthoryear{{Law} \& {Majewski}}{{Law} \&
  {Majewski}}{2010}]{law_majewski_2010}
{Law} D.~R.,  {Majewski} S.~R.,  2010, \mn@doi [\apj]
  {10.1088/0004-637X/714/1/229},
  \href{http://adsabs.harvard.edu/abs/2010ApJ...714..229L}{714, 229}

\bibitem[\protect\citeauthoryear{{Levi} et~al.,}{{Levi} et~al.}{2013}]{desi}
{Levi} M.,  et~al., 2013, preprint,
  \href{http://adsabs.harvard.edu/abs/2013arXiv1308.0847L}{} (\mn@eprint
  {arXiv} {1308.0847})

\bibitem[\protect\citeauthoryear{{Long} \& {Murali}}{{Long} \&
  {Murali}}{1992}]{long_murali_1992}
{Long} K.,  {Murali} C.,  1992, \mn@doi [\apj] {10.1086/171764},
  \href{http://adsabs.harvard.edu/abs/1992ApJ...397...44L}{397, 44}

\bibitem[\protect\citeauthoryear{{L{\'o}pez-Corredoira}, {Cabrera-Lavers}  \&
  {Gerhard}}{{L{\'o}pez-Corredoira} et~al.}{2005}]{lopez_et_al_2005}
{L{\'o}pez-Corredoira} M.,  {Cabrera-Lavers} A.,   {Gerhard} O.~E.,  2005,
  \mn@doi [\aap] {10.1051/0004-6361:20053075},
  \href{http://adsabs.harvard.edu/abs/2005A%26A...439..107L}{439, 107}

\bibitem[\protect\citeauthoryear{{Majewski}, {Skrutskie}, {Weinberg}  \&
  {Ostheimer}}{{Majewski} et~al.}{2003}]{sag}
{Majewski} S.~R.,  {Skrutskie} M.~F.,  {Weinberg} M.~D.,   {Ostheimer} J.~C.,
  2003, \mn@doi [\apj] {10.1086/379504},
  \href{http://adsabs.harvard.edu/abs/2003ApJ...599.1082M}{599, 1082}

\bibitem[\protect\citeauthoryear{{Mastrobuono-Battisti}, {Di Matteo},
  {Montuori}  \& {Haywood}}{{Mastrobuono-Battisti} et~al.}{2012}]{Ma2012}
{Mastrobuono-Battisti} A.,  {Di Matteo} P.,  {Montuori} M.,   {Haywood} M.,
  2012, \mn@doi [\aap] {10.1051/0004-6361/201219563},
  \href{http://adsabs.harvard.edu/abs/2012A%26A...546L...7M}{546, L7}

\bibitem[\protect\citeauthoryear{{Miyamoto} \& {Nagai}}{{Miyamoto} \&
  {Nagai}}{1975}]{miyamoto_nagai_1975}
{Miyamoto} M.,  {Nagai} R.,  1975, \pasj,
  \href{http://adsabs.harvard.edu/abs/1975PASJ...27..533M}{27, 533}

\bibitem[\protect\citeauthoryear{Neal}{Neal}{2011}]{Neal2012}
Neal R.,  2011, Handbook of Markov Chain Monte Carlo, pp 113--162

\bibitem[\protect\citeauthoryear{{Niederste-Ostholt}, {Belokurov}  \&
  {Evans}}{{Niederste-Ostholt} et~al.}{2012}]{niederste-ostholt_et_al_2012}
{Niederste-Ostholt} M.,  {Belokurov} V.,   {Evans} N.~W.,  2012, \mn@doi
  [\mnras] {10.1111/j.1365-2966.2012.20602.x},
  \href{http://adsabs.harvard.edu/abs/2012MNRAS.422..207N}{422, 207}

\bibitem[\protect\citeauthoryear{{Odenkirchen} et~al.,}{{Odenkirchen}
  et~al.}{2001}]{pal5disc}
{Odenkirchen} M.,  et~al., 2001, \mn@doi [\apjl] {10.1086/319095},
  \href{http://adsabs.harvard.edu/abs/2001ApJ...548L.165O}{548, L165}

\bibitem[\protect\citeauthoryear{{Odenkirchen}, {Grebel}, {Dehnen}, {Rix}  \&
  {Cudworth}}{{Odenkirchen} et~al.}{2002}]{odenkirchen_et_al_2002_pal5_vel}
{Odenkirchen} M.,  {Grebel} E.~K.,  {Dehnen} W.,  {Rix} H.-W.,   {Cudworth}
  K.~M.,  2002, \mn@doi [\aj] {10.1086/342287},
  \href{http://adsabs.harvard.edu/abs/2002AJ....124.1497O}{124, 1497}

\bibitem[\protect\citeauthoryear{{Odenkirchen} et~al.,}{{Odenkirchen}
  et~al.}{2003}]{odenkirchen2003}
{Odenkirchen} M.,  et~al., 2003, \mn@doi [\aj] {10.1086/378601},
  \href{http://adsabs.harvard.edu/abs/2003AJ....126.2385O}{126, 2385}

\bibitem[\protect\citeauthoryear{{Pe{\~n}arrubia}, {G{\'o}mez}, {Besla},
  {Erkal}  \& {Ma}}{{Pe{\~n}arrubia} et~al.}{2016}]{Jorge2016}
{Pe{\~n}arrubia} J.,  {G{\'o}mez} F.~A.,  {Besla} G.,  {Erkal} D.,   {Ma}
  Y.-Z.,  2016, \mn@doi [\mnras] {10.1093/mnrasl/slv160},
  \href{http://adsabs.harvard.edu/abs/2016MNRAS.456L..54P}{456, L54}

\bibitem[\protect\citeauthoryear{{Pearson}, {K{\"u}pper}, {Johnston}  \&
  {Price-Whelan}}{{Pearson} et~al.}{2015}]{pearson_et_al_pal5}
{Pearson} S.,  {K{\"u}pper} A.~H.~W.,  {Johnston} K.~V.,   {Price-Whelan}
  A.~M.,  2015, \mn@doi [\apj] {10.1088/0004-637X/799/1/28},
  \href{http://adsabs.harvard.edu/abs/2015ApJ...799...28P}{799, 28}

\bibitem[\protect\citeauthoryear{{Perryman} et~al.,}{{Perryman}
  et~al.}{2001}]{gaia}
{Perryman} M.~A.~C.,  et~al., 2001, \mn@doi [\aap]
  {10.1051/0004-6361:20010085},
  \href{http://adsabs.harvard.edu/abs/2001A%26A...369..339P}{369, 339}

\bibitem[\protect\citeauthoryear{{Portail}, {Wegg}, {Gerhard}  \&
  {Martinez-Valpuesta}}{{Portail} et~al.}{2015}]{portail_et_al_2015}
{Portail} M.,  {Wegg} C.,  {Gerhard} O.,   {Martinez-Valpuesta} I.,  2015,
  \mn@doi [\mnras] {10.1093/mnras/stv058},
  \href{http://adsabs.harvard.edu/abs/2015MNRAS.448..713P}{448, 713}

\bibitem[\protect\citeauthoryear{Press, Teukolsky, Vetterling  \&
  Flannery}{Press et~al.}{2007}]{Press2007}
Press W.~H.,  Teukolsky S.~A.,  Vetterling W.~T.,   Flannery B.~P.,  2007,
  Numerical Recipes 3rd Edition: The Art of Scientific Computing, 3 edn.
Cambridge University Press, New York, NY, USA

\bibitem[\protect\citeauthoryear{{Price-Whelan}, {Johnston}, {Valluri},
  {Pearson}, {K{\"u}pper}  \& {Hogg}}{{Price-Whelan}
  et~al.}{2016a}]{price_whelan_et_al_chaos}
{Price-Whelan} A.~M.,  {Johnston} K.~V.,  {Valluri} M.,  {Pearson} S.,
  {K{\"u}pper} A.~H.~W.,   {Hogg} D.~W.,  2016a, \mn@doi [\mnras]
  {10.1093/mnras/stv2383},
  \href{http://adsabs.harvard.edu/abs/2016MNRAS.455.1079P}{455, 1079}

\bibitem[\protect\citeauthoryear{{Price-Whelan}, {Sesar}, {Johnston}  \&
  {Rix}}{{Price-Whelan} et~al.}{2016b}]{price_whelan_et_al_chaotic_fanning}
{Price-Whelan} A.~M.,  {Sesar} B.,  {Johnston} K.~V.,   {Rix} H.-W.,  2016b,
  \mn@doi [\apj] {10.3847/0004-637X/824/2/104},
  \href{http://adsabs.harvard.edu/abs/2016ApJ...824..104P}{824, 104}

\bibitem[\protect\citeauthoryear{{Rockosi} et~al.,}{{Rockosi}
  et~al.}{2002}]{rockosi2002}
{Rockosi} C.~M.,  et~al., 2002, \mn@doi [\aj] {10.1086/340957},
  \href{http://adsabs.harvard.edu/abs/2002AJ....124..349R}{124, 349}

\bibitem[\protect\citeauthoryear{{Roman-Duval}, {Jackson}, {Heyer}, {Rathborne}
   \& {Simon}}{{Roman-Duval} et~al.}{2010}]{roman-duval_et_al_2010}
{Roman-Duval} J.,  {Jackson} J.~M.,  {Heyer} M.,  {Rathborne} J.,   {Simon} R.,
   2010, \mn@doi [\apj] {10.1088/0004-637X/723/1/492},
  \href{http://adsabs.harvard.edu/abs/2010ApJ...723..492R}{723, 492}

\bibitem[\protect\citeauthoryear{{Sanders}, {Bovy}  \& {Erkal}}{{Sanders}
  et~al.}{2016}]{sanders_bovy_erkal_2015}
{Sanders} J.~L.,  {Bovy} J.,   {Erkal} D.,  2016, \mn@doi [\mnras]
  {10.1093/mnras/stw232},
  \href{http://adsabs.harvard.edu/abs/2016MNRAS.457.3817S}{457, 3817}

\bibitem[\protect\citeauthoryear{{Schlegel}, {Finkbeiner}  \&
  {Davis}}{{Schlegel} et~al.}{1998}]{schlegel_etal_1998}
{Schlegel} D.~J.,  {Finkbeiner} D.~P.,   {Davis} M.,  1998, \mn@doi [\apj]
  {10.1086/305772},
  \href{http://adsabs.harvard.edu/abs/1998ApJ...500..525S}{500, 525}

\bibitem[\protect\citeauthoryear{{Sch{\"o}nrich}, {Binney}  \&
  {Dehnen}}{{Sch{\"o}nrich} et~al.}{2010}]{schoenrich_et_al_2010}
{Sch{\"o}nrich} R.,  {Binney} J.,   {Dehnen} W.,  2010, \mn@doi [\mnras]
  {10.1111/j.1365-2966.2010.16253.x},
  \href{http://adsabs.harvard.edu/abs/2010MNRAS.403.1829S}{403, 1829}

\bibitem[\protect\citeauthoryear{{Springel}}{{Springel}}{2005}]{springel_2005}
{Springel} V.,  2005, \mn@doi [\mnras] {10.1111/j.1365-2966.2005.09655.x},
  \href{http://adsabs.harvard.edu/abs/2005MNRAS.364.1105S}{364, 1105}

\bibitem[\protect\citeauthoryear{{Springel} et~al.,}{{Springel}
  et~al.}{2008}]{springel_et_al_2008}
{Springel} V.,  et~al., 2008, \mn@doi [\mnras]
  {10.1111/j.1365-2966.2008.14066.x},
  \href{http://adsabs.harvard.edu/abs/2008MNRAS.391.1685S}{391, 1685}

\bibitem[\protect\citeauthoryear{{Theano Development Team}}{{Theano Development
  Team}}{2016}]{Theano2016}
{Theano Development Team} 2016, arXiv e-prints, abs/1605.02688

\bibitem[\protect\citeauthoryear{{Thomas}, {Ibata}, {Famaey}, {Martin}  \&
  {Lewis}}{{Thomas} et~al.}{2016}]{thomas_et_al_2016}
{Thomas} G.~F.,  {Ibata} R.,  {Famaey} B.,  {Martin} N.~F.,   {Lewis} G.~F.,
  2016, \mn@doi [\mnras] {10.1093/mnras/stw1189},
  \href{http://adsabs.harvard.edu/abs/2016MNRAS.460.2711T}{460, 2711}

\bibitem[\protect\citeauthoryear{{Viel}, {Becker}, {Bolton}  \&
  {Haehnelt}}{{Viel} et~al.}{2013}]{viel_et_al_2013}
{Viel} M.,  {Becker} G.~D.,  {Bolton} J.~S.,   {Haehnelt} M.~G.,  2013, \mn@doi
  [\prd] {10.1103/PhysRevD.88.043502},
  \href{http://adsabs.harvard.edu/abs/2013PhRvD..88d3502V}{88, 043502}

\bibitem[\protect\citeauthoryear{{Wilson}}{{Wilson}}{1955}]{wilson1955}
{Wilson} A.~G.,  1955, \mn@doi [\pasp] {10.1086/126754},
  \href{http://adsabs.harvard.edu/abs/1955PASP...67...27W}{67, 27}

\bibitem[\protect\citeauthoryear{{Yoon}, {Johnston}  \& {Hogg}}{{Yoon}
  et~al.}{2011}]{yoon_etal_2011}
{Yoon} J.~H.,  {Johnston} K.~V.,   {Hogg} D.~W.,  2011, \mn@doi [\apj]
  {10.1088/0004-637X/731/1/58},
  \href{http://adsabs.harvard.edu/abs/2011ApJ...731...58Y}{731, 58}

\bibitem[\protect\citeauthoryear{Zhu, Byrd, Lu  \& Nocedal}{Zhu
  et~al.}{1997}]{Zhu1997}
Zhu C.,  Byrd R.~H.,  Lu P.,   Nocedal J.,  1997, ACM Transactions on
  Mathematical Software (TOMS), 23, 550

\bibitem[\protect\citeauthoryear{{Zotos}}{{Zotos}}{2015}]{zotos_2015}
{Zotos} E.~E.,  2015, \mn@doi [\mnras] {10.1093/mnras/stu2129},
  \href{http://adsabs.harvard.edu/abs/2015MNRAS.446..770Z}{446, 770}

\bibitem[\protect\citeauthoryear{{de Jong} et~al.,}{{de Jong}
  et~al.}{2012}]{4most}
{de Jong} R.~S.,  et~al., 2012, in Ground-based and Airborne Instrumentation
  for Astronomy IV. p. 84460T (\mn@eprint {arXiv} {1206.6885}),
  \mn@doi{10.1117/12.926239}

\makeatother
\end{thebibliography}

\appendix

\section{Coordinate Transformation Matrix} \label{sec:coo_appendix}

In \Secref{sec:coordinate_system} we described the rotated coordinate system, $(\phi_1,\phi_2)$, which is approximately aligned with the stream. The transformation from $(\alpha,\delta)$ to $(\phi_1,\phi_2)$ is given by

\begin{eqnarray}
\begin{bmatrix}
\cos(\phi_1) \cos(\phi_2)\\
\sin(\phi_1) \cos(\phi_2)\\
\sin(\phi_2)
\end{bmatrix}&=&\nonumber\\
\begin{bmatrix}
-0.656057 & -0.754711 & 0.000636\\
0.609115 & -0.528995 & 0.590883\\ 
-0.445608 & 0.388045 & 0.806751
\end{bmatrix} 
&\times &
\begin{bmatrix}
\cos(\alpha) \cos(\delta)\\
\sin(\alpha) \cos(\delta)\\
\sin(\delta)
\end{bmatrix}\nonumber
\end{eqnarray}

\section{Pal 5 stream density in the DECaLS dataset} \label{sec:decals_appendix}

\begin{figure*}
\includegraphics{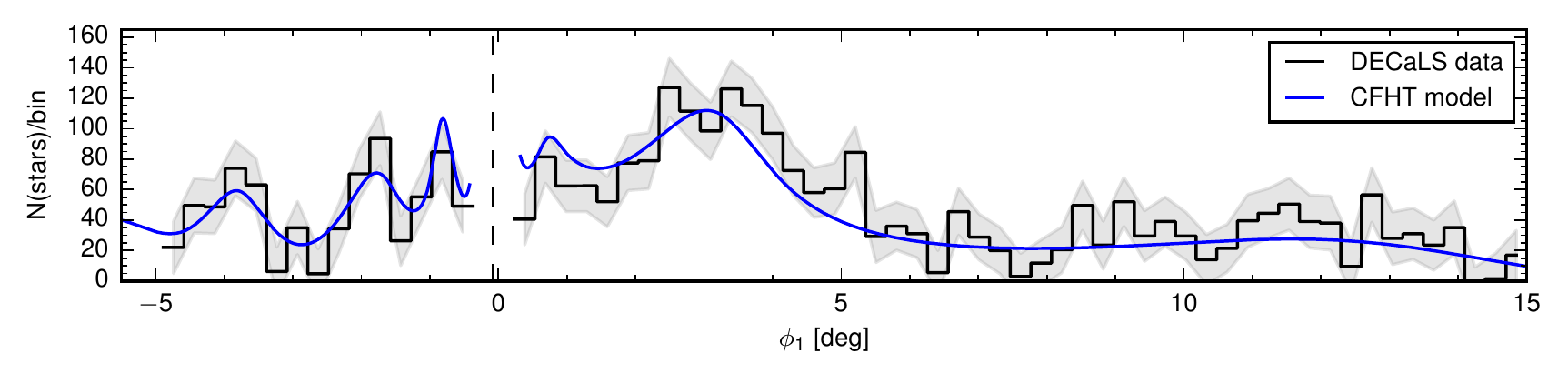}
\caption{Comparison of the model Pal~5 stream density from CFHT data with data
  from the DECaLS survey. The histogram with the grey band shows the
  background subtracted density of the candidate Pal~5 stream stars in
  DECaLS (selected using the optimal color-magnitude mask in $r$,$z$
  bands) within 0.25 degrees of the stream track. The blue curve shows
  the best-fit stream density model that was extracted from the CFHT
  data. The size of the bin is 0.3$\degr$. The grey bands represent
  the Poisson uncertainty of the stream star counts and incorporate the uncertainty from the background subtraction.}
\label{fig:decals}
\end{figure*}

To demonstrate the robustness of the recovered Pal 5 stream density
profile, and to verify that the features observed are not caused by
any issues related to the CFHT observations (such as data quality
variation along the stream or an incomplete footprint), we have compared
the measured stream density with a completely independent dataset,
namely the DECam Legacy Survey (DECaLS) \citep{Blum2016}. The DECaLS
dataset is comprised of $g$-, $r$- and $z$-band photometry covering the stream from $\phi_1\sim-5\degr$ to $\phi_1 \sim 20\degr$ at a depth of 0.5$-$1 magnitude
deeper than the SDSS. Here we use the source catalog from the second
data release of the survey (DR2) and only include objects classified
as point sources \citep[{\tt type="PSF"}; see][]{Lang2016}. In this data release, the continuous coverage along the
stream exists only in $r$ and $z$ bands.  Therefore we use the same
matched filter analysis to select Pal~5 stream stars as described in
Section~\ref{sec:data} but in $r$, $z$ bands (down to a limit of $z<22$). Because
the DECaLS data coverage is not restricted to a narrow region around
the stream we can perform a proper background subtraction of the
stream densities without doing model fitting, but instead using two
background regions above and below the stream.

Figure~\ref{fig:decals} shows the background-subtracted density of the
CMD selected Pal 5 stars within 0.25\, degrees of the stream
track. Over-plotted on top of the density recovered from the DECaLS
data is the scaled best-fit stream density model based on the CFHT
dataset (as measured in Section~\ref{sec:results}).  We also note that
a one-to-one match is not necessary expected given that the DECaLS
data is considerably shallower compared to the CFHT data as well as
less homogeneous. Nonetheless, most density fluctuations extracted
from the CFHT data have clear counterparts in the DECaLS dataset. In
particular, the large overdensity/stream asymmetry at $\phi_1\sim
3\degr$, the epicyclic overdensity at $\phi_1\sim -0.7\degr$ and the
underdensity at $\phi_1\sim -3\degr$ are all present in the DECaLS Pal
5 stream density profile. Accordingly, re-assured by this test, we
conclude that the stream density model presented above gives a fair
representation of the debris distribution along the Pal 5 tidal
tails.

\section{Comparison with simulations} \label{sec:sim_compare_appendix}

\begin{figure*}
\includegraphics{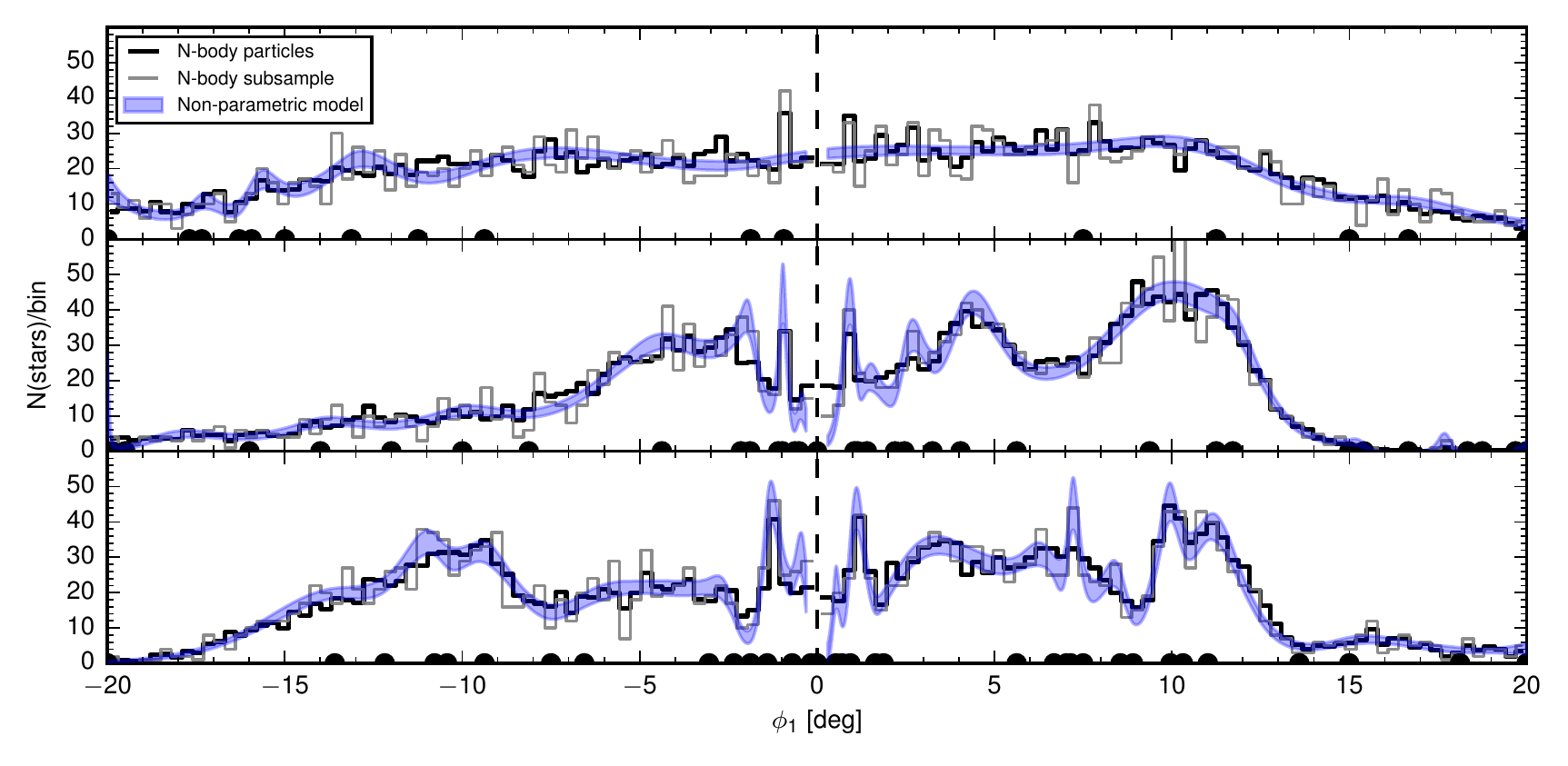}
\caption{Performance of the non-parametric stream modelling method on
  simulated data. This shows the density of stars along simulated
  Pal~5-like streams which were perturbed by different amounts of
  substructure. {\it Top}: Fiducial stream, not perturbed by any
  substructure {\it Middle}: Stream evolved with the expected amount
  of substructure between $10^6-10^9 M_\odot$ matching the
  $\Lambda$CDM predictions. {\it Bottom}: Stream evolved with three
  times the expected amount of $\Lambda$CDM substructure between
  $10^6-10^9 M_\odot$. In each panel, black histogram gives the
  density of the stream particles in 0.3 degree wide bins. The grey
  histogram shows the density of the 20\% subsample of all the
  particles used for stream density extraction. Blue bands show the
  non-parametric density measurement together with 16\%,84\% credible   
  intervals from the posterior samples.  In all three cases the non-parametric method recovers well the majority of the structure in stream densities, including the small scale epicyclic over-densities.
  Note that the number of subhaloes used
  does not account for the effect of the disk of the host which is expected  to deplete the subhaloes by a factor of 3 \protect\citep{donghia_et_al_2010}. 
  }
\label{fig:sim_comparison}
\end{figure*}

To test the performance of the non-parametric stream model described
in \Secref{sec:stream_model}, we have used it to extract density
features from several Pal 5-like stellar streams generated using
N-body simulations and perturbed with varying amounts of
substructure. The simulations are broadly similar to those described
in \Secref{sec:smooth_pal5} and are run with a modified version of
\textsc{gadget-3}.  We use a potential similar to
\texttt{MWPotential2014} from \cite{bovy_galpy} except the bulge is
replaced with a $5\times 10^9 M_\odot$ Hernquist profile with a scale
radius of $500$ pc.  As above, the progenitor is taken to be a King
profile with a mass of $2\times 10^4 M_\odot$, $w=2$, a core radius of
$15$ pc and is modelled with $10^5$ equal mass particles and a
softening length of 1 pc. The progenitor is given the best fit
line-of-sight velocity and proper motions from
\cite{kuepper_et_al_pal5}, rewound for 3.4 Gyr \citep[the
  best-fit age from][]{kuepper_et_al_pal5}, and then allowed to
disrupt for the same amount of time.  In the fiducial simulation, we
do not include any substructure. To perturb the stream, we include the
effect of substructure by taking the reported number density profile
of subhaloes from \cite{springel_et_al_2008}, scaling the profile down
to a host mass of $10^{12} M_\odot$ as described in Sec. 2.4 of
\cite{number_of_gaps} and including the expected population of subhaloes
with masses between $10^6-10^9 M_\odot$.  Each subhalo is modelled as
a single tracer particle which sources a Hernquist profile force given
by the mass and scale radius of the subhalo.  The relation between the
mass and scale radius comes from fits to the $v_{\rm max}-M_{\rm
  tidal}$ relation in the public catalogs of Via Lactea II \citep{
  Diemand2008} and is given by
\eq{ r_s = 1.05 {\rm kpc} \left(\frac{M}{10^8 M_\odot}\right)^{1/2}. }
We then run a simulation in the expected $\Lambda$CDM background and a
simulation in three times the expected background.  Note that the
number of subhaloes used does not account for the effect of the disk
of the host which is known to deplete their number by a factor of 3
\citep[e.g.][]{donghia_et_al_2010}.

For each stream in the simulations described above, we ran our measuring
algorithm to extract a one-dimensional stellar density profile,
$I(\phi_1)$, as described in Section~\ref{sec:model_def}. Importantly,
exactly the same procedure for dynamic spline node placement as
described in Section~\ref{sec:nodes_def} is used. Note that here only
the one-dimensional debris density distribution is determined, i.e. the
track on the sky, stream width and background density are not
modelled. Nonetheless, we believe that even in this somewhat limited
setup, the simulated stream data provides a suitable test case for the
algorithm.  In order to make the simulated streams look closer to the
actual Pal 5 stream as it appears in the CFHT data, we use only $\sim
20\%$ of particles from the simulations. This sub-sampling produces
a particle density similar to the median stellar density observed in
the Pal~5 stream, namely 2 stars per arc-minute (see
Fig.~\ref{fig:stream_meas}).  Figure~\ref{fig:sim_comparison} shows
the comparison of linear densities of the simulated streams as
extracted by our non-parametric model (blue band shows 1$\sigma$
credible intervals from the posterior samples) versus the simple
histogram of the full set of particles in the simulation (thick black
line) and the actual subset used in the fitting (thin grey line). We
see that in all three cases, the density measurement matches very
closely the true stream density, correctly extracting both large-scale
fluctuations produced by the sub-halo flybys and the small-scale
features related to the epicyclic overdensities.

\section{Dust Extinction} \label{sec:dust_appendix}

\begin{figure}
\includegraphics{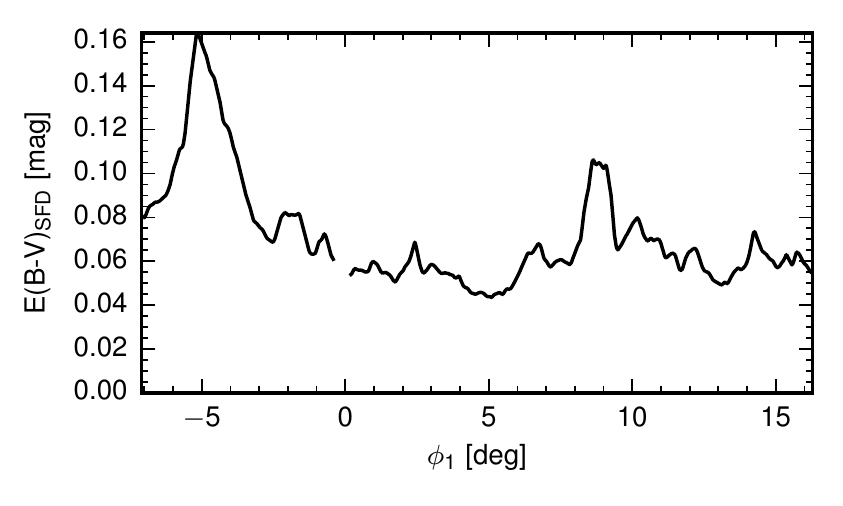}
\caption{Average dust reddening within the measured width of the stream using the maps from \protect\cite{schlegel_etal_1998}. The excess reddening at $-6^\circ < \phi_1 < -4^\circ$ may create a feature in the stream as we discuss in \protect\secref{sec:results} but there are no dust features corresponding to the density variations of interest we find in Pal 5.}
\label{fig:ebv}
\end{figure}

As a check that dust extinction did not create any of the features of interest, we compute the average dust reddening within the width of the stream. Specifically, we take the stream track and width from \Secref{sec:results} and compute the average E(B-V) within one stream width using dust maps from \cite{schlegel_etal_1998}. \Figref{fig:ebv} shows this average with a clear excess of dust at $-5^\circ$. While this excess reddening may create a feature in the stream (as discussed in \secref{sec:results}), there are no dust features corresponding to the gap at $\phi_1\sim-3^\circ$, the over density at $\phi_1 \sim 3^\circ$, or the broad underdensity around $\phi_1 \sim 8^\circ$. Thus, do not believe that extinction can explain the density variations of interest which we find in Pal 5.

\section{Supplementary data} \label{sec:appendix_data}

As supplementary material, we provide the summary of the Pal~5 stream properties shown in Figure~\ref{fig:stream_meas} in machine-readable form with the following columns:

\begin{enumerate}
\item {\tt phi1} Angle along the stream $\phi_1$ ([deg])
\item
{\tt phi2\_16}, {\tt phi2\_50}, {\tt phi2\_84} Angle across the stream [deg] (16\%, 50\%, 84\% percentile) 
\item {\tt sbstream\_16}, {\tt sbstream\_50}, {\tt sbstream\_84} Surface brightness of the stream in stars per square arcminute  (16\%, 50\%, 84\% percentile)
\item {\tt ldens\_16}, {\tt ldens\_50}, {\tt ldens\_84} Linear density of  the stream in stars per arcmin (16\%, 50\%, 84\% percentile)
\item {\tt cumdens\_16}, {\tt cumdens\_50} , { \tt cumdens\_84}  Cumulative density of the stream from the progenitor (16\%, 50\%, 84\% percentile)
\item {\tt width\_16}, {\tt width\_50}, {\tt width\_84} Gaussian stream width (16\%, 50\%, 84\% percentile)
\item {\tt bgdens\_16}, {\tt bgdens\_50}, {\tt bgdens\_84}  Background stellar density (16\%, 50\%, 84\% percentile)
\end {enumerate}

\end{document}